\documentclass[12pt]{article}

\usepackage{chngcntr}  
\usepackage{amsfonts}
\usepackage{amssymb}
\usepackage{amsthm}
\usepackage{amsmath}
\usepackage{bbm}
\usepackage{array}
\usepackage{booktabs,longtable}
\usepackage[para,online,flushleft]{threeparttable}
\usepackage{threeparttablex}
\usepackage{color,tikz}
\usepackage{color}
\usepackage{dsfont}
\usepackage{enumerate}
\usepackage{epstopdf}
\usepackage{float}
\usepackage{graphicx}
\usepackage[hidelinks]{hyperref}
\usepackage{indentfirst}
\usepackage{longtable}
\usepackage{lscape}
\usepackage{multirow}
\usepackage{multicol}
\usepackage{natbib}
\usepackage{rotating}
\usepackage{setspace}
\usepackage{sectsty}
\usepackage{subcaption}
\usepackage{todonotes}
\usepackage{url}
\usepackage{xcolor}
\usepackage{caption}
\usepackage{subcaption}
\usepackage{titlesec}
\usepackage{times}
\usepackage{mathptmx}
\usepackage{chngcntr}
\usepackage{amsmath,amsfonts, amssymb, dsfont, xcolor, comment}
\usepackage{mathtools}
\usepackage{natbib}
\usepackage{epsfig}
\usepackage{setspace}
\usepackage{graphicx, subcaption}
\usepackage{color}
\usepackage{mdframed}
\usepackage{float}
\usepackage[super]{nth}
\usepackage[toc,page]{appendix}
\PassOptionsToPackage{hyphens}{url}
\usepackage{tabularx}
\usepackage{ragged2e}
\usepackage{makecell}
\usepackage{placeins}
\usepackage{natbib}
\usepackage{etoolbox}  
\usepackage{setspace}

\newtheoremstyle{boldprop}  
  {\topsep}   
  {\topsep}   
  {\itshape}  
  {}          
  {\bfseries} 
  {.}         
  { }         
  {\thmname{#1} \thmnumber{#2}} 
\theoremstyle{boldprop}

{\hfill$\blacksquare$\par}

\usepackage{bm}
\renewcommand{\bm}[1]{\boldsymbol{#1}}

\newcommand{\Xomit}[1]{}

\usepackage[margin=1.25in]{geometry}

\usepackage[normalem]{ulem}
\usepackage[mathscr]{eucal}

\long\def\/*#1*/{}

\hypersetup{colorlinks,linkcolor={blue},citecolor={blue},urlcolor={blue}}

\begin{document}
\newgeometry{top=1.2in, bottom=1.2in, left=0.95in, right=0.95in}
\title{\LARGE{\textbf{Interpretable Factors of Firm Characteristics}}
\thanks{We are grateful to seminar participants at Washington University in St. Louis and  conference participants at 2022 SoFiE Financial Econometrics Summer School, 2022 Pacific Basin Finance, Economics, Accounting, and Management (PBFEAM),  the 2023 Asia Meeting of the Econometric Society (AMES), the 2023 China Fintech Research Conference (CFTRC), the 2024 Midwest Finance Association Annual Meeting (MFA) and 2024 American Finance Association Annual Meeting (AFA). 
This work was supported by the Tsinghua University Initiative Scientific Research Program [grant numbers 2022Z04W02016]; and Tsinghua University School of Economics and Management Research Grant [grant number 2022051002].
All errors are our own.}
}
\vspace{-0.5in}
\date {June 2025}

\author{\hspace{2mm}\\ Yuxiao Jiao \footnote{School of Management Science and Engineering, Central University of Finance and Economics, jiaoyuxiao201110@163.com}
\hspace{5mm} Guofu Zhou \footnote{Olin Business School, Washington University in St. Louis, zhou@wustl.edu} \hspace{5mm} Wu Zhu \footnote{School of Economics and Management, Tsinghua University, zhuwu@sem.tsinghua.edu.cn} \hspace{5mm} Yingzi Zhu \footnote{School of Economics and Management, Tsinghua University, zhuyz@sem.tsinghua.edu.cn} }

\maketitle

\vspace{-0.3in}
\maketitle

\begin{abstract}

\singlespacing \normalsize 
We develop a new framework for constructing factors from firm characteristics that balances statistical efficiency and economic interpretability. Instead of using all characteristics equally, our method groups related characteristics and derives one factor per group. The grouping combines economic intuition with data-driven clustering. Applied to the IPCA model by \cite{kelly2019characteristics}, our approach yields economically meaningful factors that match or exceed standard IPCA in pricing performance. Using 94 characteristics from \cite{gu2020empirical}, we show that our parsimonious, transparent factors outperform benchmarks in out-of-sample tests, demonstrating the value of embedding economic structure into statistical modeling.


\medskip




\end{abstract}

\restoregeometry

\newgeometry{top=1.3in, bottom=1.3in, left=1.1in, right=1.1in}
\thispagestyle{empty}
\addtocounter{page}{-1}

\newpage
\setstretch{1.4}

\section{Introduction}
It is well established that a large number of firm characteristics -
possibly hundreds  - are correlated with expected returns in the
cross section of equities (e.g., \cite{harvey2016and,
hou2020replicating, chen2022publication}). However, many of these
characteristics are highly correlated with each other, complicating
the identification of different priced factors and affecting the
economic interpretation of risk premia, as emphasized by
\cite{hou2020replicating}. In his influential presidential address
to the American Finance Association, \cite{cochrane2011presidential}
posed two fundamental questions that remain central to empirical
asset pricing: First, how many factors are truly needed to explain
expected returns? Second, conditional on a set of well-established
factors, such as the five-factor model of \cite{fama2015five}: Do
additional factors provide incremental explanatory power or sharpen
our economic understanding of cross-sectional returns?

Two broad approaches have emerged in the literature to address these fundamental questions. The first is based on
the Principal Component Analysis (PCA). Early contributions, such as \cite{connor1986performance}, employ PCA to extract latent
 factors from asset returns. Although PCA can be extended to panels of firm characteristics, the resulting factors are linear combinations
  of \emph{all} characteristics, making them difficult to interpret economically. For example, if the dataset includes two value-related
  and two growth-related characteristics, any PCA factor will typically load on all four, yielding a factor that reflects neither
  value nor growth. Moreover, as originally formulated, the PCA-based approach assumes static factor loadings over time, which limits
  its ability to capture evolving return dynamics. To address these limitations, \cite{kelly2019characteristics} introduce Instrumented PCA (IPCA),
  which allows factor loadings to vary flexibly as functions of observable firm characteristics. This innovation improves predictive
   performance relative to traditional PCA by incorporating time-varying exposures in a theoretically grounded manner. However, despite
    this methodological advance, the resulting IPCA factors remain often difficult to interpret. Because the loadings tend to be diffusely
    spread across a large number of characteristics, therefore, each factor lacks a dominant economic theme, hindering economic
    interpretation and hindering transparency.

The second approach involves the growing application of machine
learning (ML) techniques to empirical asset pricing. A large and
expanding literature, including  \cite{feng2020taming},
\cite{freyberger2020dissecting}, \cite{gu2020empirical}, and
\cite{kozak2020shrinking},
 develop various ML methods to identify
which firm characteristics are most predictive of future
returns.\footnote{See also \cite{cong2021alphaportfolio},
\cite{demiguel2020transaction}, \cite{gu2020empirical},
\cite{daniel2020cross}, \cite{chen2024deep},
\cite{chordia2020anomalies}, \cite{patton2020you},
\cite{avramov2023integrating}, and \cite{cong2024textual}} These
methods, ranging from penalized regressions to tree-based models and
deep learning, are highly effective in capturing nonlinearities and
interactions in large characteristic spaces. However, they often
struggle to distinguish among highly correlated predictors, leading
to over-identification of priced factors. As a result, ML-based
approaches may detect a large number of statistically significant
signals that reflect overlapping information, thereby limiting their
applications to construct parsimonious and interpretable factor
models that facilitate  economic understanding.

In this paper, we propose a simple and yet effective two-step approach to address both the dimensionality and interpretability challenges in characteristic-based asset pricing. First, we partition the universe of firm characteristics into statistically coherent and economically meaningful clusters. This clustering procedure also allows for the imposition of structure guided by economic theory or interpretability constraints. Second, we extract a single latent factor from each cluster that captures the dominant economic signal driving the characteristics within it. 

This approach offers two key advantages. First, if two
characteristics are driven by the same underlying economic
mechanism—such as exposure to a common risk factor—they are
likely to be highly correlated and hence grouped into the same
cluster. The resulting factor, being derived from a cluster of
characteristics, is therefore economically interpretable with the
same risk source of the cluster. Second, by grouping and
partitioning collinear characteristics into a smaller set of
clusters, our method avoids over-identifying the number of priced
factors. Rather than assigning a distinct factor to each correlated
signal, our approach produces a parsimonious factor structure that
reflects the true underlying dimensionality of the data. In this
way, the estimated factors correspond more closely to economically
meaningful sources of risk,  eliminating  redundancies in the
characteristic space.

The underlying assumption of our approach is that firm
characteristics exhibit a latent clustering structure. If asset
returns are driven by a low-dimensional factor model, then
characteristics that proxy for the same underlying factor should be
statistically related. In this sense, our clustering assumption is
no stronger than the standard factor structure assumption.
Specifically, we posit that characteristics within a cluster share a
common economic origin or represent various noisy measurements of
the same latent factor.




Empirically, we apply this clustering-based approach to construct an economically interpretable version of the IPCA model
proposed by \cite{kelly2019characteristics}. We focus on Instrumented Principal Component Analysis (IPCA) due to its demonstrated
 effectiveness not only in pricing equities,  but also in explaining returns on options and corporate bonds \citep{buchner2022factor, kelly2023modeling}.
 However, like other PCA-based models, IPCA suffers from limited interpretability, as its factors often depend on
 diffuse combinations of firm characteristics. Our modification, named Cluster-IPCA (C-IPCA), addresses this limitation through a
  simple two-step procedure. First, we form clusters of firm characteristics as described above. Second, we restrict the loading on the $k$-th
  C-IPCA factor to depend only on the characteristics of the $k$-th cluster. Consequently, each factor and its exposure are estimated as a linear combination of characteristics within its corresponding cluster, making it directly interpretable in terms of the economic theme captured by that cluster.

Beyond the benefit of enhanced interpretability, our approach has implications for out-of-sample performance. Theoretically, imposing a cluster structure on firm characteristics introduces a classic bias–variance trade-off. On the one hand, restricting factor exposure to depend only on characteristics within a given cluster may introduce bias if the cluster is misspecified. On the other hand, this restriction can substantially reduce the variance of loading estimates by incorporating economically motivated structure. As a result, the relative out-of-sample performance of the C-IPCA model compared to the standard IPCA model ultimately depends on the balance between reduced variance and potential bias.

Empirically, the C-IPCA model exhibits two desirable properties. First, each estimated factor is clearly related to a different source of economic risk, significantly improving both transparency and interpretability. Specifically, we identify 13 interpretable factors, of which the top four, based on Sharpe ratios, are: Operating Illiquidity (OI), Return Volatility (RV), Operating Efficiency (OE), and Size \& Illiquidity (S\&I). This finding is consistent with previous literature. The S\&I factor, for instance, is a key component in several classical factor models, such as the SMB factor (Small Minus Big) in the Fama-French 3-factor model \citep{fama1993common}, the q-factor model \citep{hou2015digesting} and the Fama-French 5-factor model \citep{fama2015five}. Factors such as OI, OE and RV underscore the importance of information asymmetry, financing frictions, and trading frictions in stock pricing, as highlighted in previous studies \citep{palazzo2012cash, sloan1996stock, grullon2012real}.

While the C-IPCA framework generates interpretable factors by construction, not all factors contribute equally to pricing cross-sectional returns. For instance, factors such as Price Delay (PD), Investment (Inv), and Value (Val) exhibit relatively low Sharpe ratios, suggesting limited economic significance. This highlights the potential for a more parsimonious model that retains only a subset of high-performing factors. 

Interestingly, we find that traditional asset pricing models - including the Fama-French three-factor model  \cite{fama1993common}, the Fama-French five-factor model \cite{fama2015five}, and the q-factor model \cite{hou2015digesting} - are only able to explain a subset of C-IPCA factors with low Sharpe ratios. In contrast, for the top-performing C-IPCA factors, these benchmark models yield large and statistically significant alphas. This suggests that our combined economic and data-driven approach successfully extracts novel information that is not captured by standard factor models.

Second, despite the additional structure and restrictions, the C-IPCA model performs comparably to, or even better than, the standard IPCA model in terms of out-of-sample Sharpe ratios. Given previous findings that not all C-IPCA factors are associated with economically significant risk premia, we investigate the Sharpe ratios of tangency portfolios constructed using only a subset of the most informative factors. Specifically, we consider two factor selection methods: ordered model selection and Bayesian model selection. The ordered selection of models ranks the factors based on their Sharpe ratios in the training sample, selecting the top
\(J\) factors to form a J-factor model. In contrast, the selection of the Bayesian model, as proposed by \cite{chib2020comparing}, identifies the model with
 the highest posterior probability among all possible subsets of factors.

For comparison, we apply the same factor selection procedures to the IPCA model, or alternatively, we use the full IPCA model as a benchmark.
Our empirical results indicate that, in most cases, tangency portfolios constructed using a subset of C-IPCA factors outperform those based
on the benchmarks, regardless of the model selection method. These findings suggest that our approach not only preserves the
 statistical power of the IPCA model,  but also enhances its economic interpretability.

Furthermore, we provide evidence on the mechanism underlying the comparable performance (even outperformance) of the C-IPCA and IPCA models. In principle, the performance may arise from three sources: domain knowledge, the clustering information implied by the data and the reduction in parameter estimation resulting from the model structure. Our empirical results indicate that both domain knowledge and the data-driven clustering structure contribute significantly to the superior performance of the C-IPCA model. The combination of these two elements enhances both interpretability and performance. In contrast, we find no evidence that the reduction in the number of parameters plays a significant role in the outperformance. Instead, the special structure imposed by economic intuition, along with the data-driven similarity between characteristics, primarily drives the model's superior performance.


Our approach is also related to \cite{stambaugh2017mispricing}, who are among the first to use clustering methods to isolate time-series factors. In contrast to their method, we apply a clustering algorithm that is well-suited for high-dimensional cases and can determine the optimal number of clusters from the data without imposing it a priori. To our knowledge, this is the first paper in the finance literature to apply clustering algorithms to such a large cross-sectional dataset to construct factors.  
Furthermore, rather than using the average of the factors within a cluster as the new factor, we adopt a data-driven approach that allows the data to identify the most representative factor for each cluster.




The remainder of the paper is organized as follows. Section~\ref{sec:model} describes the construction of factor models based on cluster analysis. Section~\ref{sec:clusters} introduces the data and clustering methodology. Section~\ref{subsec: DC and IC} presents the clustering result and evaluates its effectiveness. Section~\ref{sec: Data and Perf_IC-IPCA} presents empirical results on the performance of various factor models. Section~\ref{sec: Machanism_and_robustness} investigates mechanism underlying the performance advantage of the C-IPCA models and presents some robustness tests. Section~\ref{sec: conclusion} concludes.

\section{Model}\label{sec:model}

In this section, we describe the model framework, which extends the IPCA approach of \cite{kelly2019characteristics} by incorporating characteristic-based clustering to improve interpretability. Our proposed method, which we refer to as Cluster-IPCA (C-IPCA), integrates economic structure into the IPCA framework by restricting factor loadings to depend only on characteristics within statistically and economically coherent clusters.

We begin by briefly reviewing the standard IPCA model and then detail how cluster-based restrictions can be incorporated to yield interpretable and economically grounded factors.

\subsection{IPCA Model}\label{subsec:IPCA}

The IPCA model, proposed by \cite{kelly2019characteristics}, is a conditional factor model that allows for time-varying risk exposures by using firm characteristics as instruments. The key element of IPCA is that it models factor loadings (i.e., risk exposures) as linear functions of observable firm characteristics.

Specifically, the model consists of two equations. The return equation is:
\begin{equation}\label{eq:ret}
\bm{r}_t = \bm{\beta}_{t-1} \bm{f}_t + \bm{e}_t,
\end{equation}
and the exposure equation is:
\begin{equation}
\bm{\beta}_{t-1} = \bm{X}_{t-1} \bm{\Gamma} + \bm{u}_{t-1},
\label{Eq: IPCA_beta}
\end{equation}
where $\bm{r}_t = (r_{1t},\dots,r_{Nt})\in \mathbb{R}^N$ is the vector of excess returns for $N$ assets at time $t$, $\bm{f}_t \in \mathbb{R}^J$ is a vector of $J$ latent factors, $\bm{\beta}_{t-1} \in \mathbb{R}^{N \times J}$ contains the factor loadings (risk exposures) for each asset, $\bm{X}_{t-1} = (\bm{x}'_{1,t-1}, \bm{x}'_{2,t-1}, \ldots, \bm{x}'_{N,t-1}) \in \mathbb{R}^{N \times (I+1)}$ is the matrix of $I$ firm characteristics for each  where the first \(I\) columns represent the firm time-varying characteristics and the final column is the \(\bm 1 = (1,\dots, 1) \in R^N \)  , $\bm{\Gamma} \in \mathbb{R}^{(I+1) \times J}$ contains the characteristic loadings that map characteristics to risk exposures, the \(j^{\text{th}}\) column corresponds to the loadings of the \(j^{\text{th}}\) risk exposure on characteristics,  $\bm{e}_t$ and $\bm{u}_{t-1}$ are idiosyncratic error terms.

Each element $\Gamma_{ij}$ in $\bm{\Gamma}$ reflects the contribution of the $i^{\text{th}}$ firm characteristic to the $j^{\text{th}}$ factor exposure. Specifically, for $i = 1, \ldots, I$ and $j = 1, \ldots, J$, $\Gamma_{ij}$ captures the slope coefficient relating the $i^{\text{th}}$ characteristic to the loading of the $j^{\text{th}}$ factor, while $\Gamma_{(I+1),j}$ corresponds to the intercept term for the $j^{\text{th}}$ factor loading.

For simplicity and following the empirical implementation in \cite{kelly2019characteristics}, we assume that the pricing errors (alphas) are zero. This restriction allows for more tractable estimation and is standard in the literature applying IPCA.

The IPCA model can be estimated by minimizing the value-weighted mean squared error (MSE) of the pricing residuals. Specifically, the objective function is given by:
\begin{equation}
\min_{\bm{\Gamma}, \{\bm{f}_t\}_{t=1}^T} \sum_{t=1}^T \left( \bm{r}_t - \bm{X}_{t-1} \bm{\Gamma} \bm{f}_t \right)' \bm{W}_t \left( \bm{r}_t - \bm{X}_{t-1} \bm{\Gamma} \bm{f}_t \right),
\label{Eq:obj_fun}
\end{equation}
where $\bm{W}_t$ is the weighting matrix at time $t$. Following \cite{kelly2019characteristics}, we use a diagonal value-weighted matrix for $\bm{W}_t$. Different from \cite{kelly2019characteristics}, who employ equal weighting, we use value-weighted portfolios to assign greater importance to larger firms and mitigate concerns that results are driven by micro-cap stocks, as pointed out by \cite{hou2020replicating}. This approach better reflects the economic significance of larger firms and aligns with practical considerations in portfolio management. Our results remain robust when equal-weighted portfolios are used, as reported in Appendix~\ref{app: EW}, confirming that the findings are not sensitive to the weighting scheme.

The first-order conditions for this optimization problem yield the following recursive updating rules. The estimated factors are given by:
\begin{equation}
\hat{\bm{f}}_t = \left( \hat{\bm{\Gamma}}' \bm{X}_{t-1}' \bm{W}_{t-1} \bm{X}_{t-1} \hat{\bm{\Gamma}} \right)^{-1} \hat{\bm{\Gamma}}' \bm{X}_{t-1}' \bm{W}_{t-1} \bm{r}_t,
\label{Eq:f_hat}
\end{equation}
and the estimated characteristic loadings $\hat{\bm{\Gamma}}$ satisfy:
\begin{equation}
\text{vec}(\hat{\bm{\Gamma}}') = \left( \sum_{t=2}^{T} \bm{X}_{t-1}' \bm{W}_{t-1} \bm{X}_{t-1} \otimes \hat{\bm{f}}_t \hat{\bm{f}}_t' \right)^{-1} \left( \sum_{t=2}^{T} \left[ \bm{X}_{t-1} \bm{W}_{t-1} \otimes \hat{\bm{f}}_t' \right]' \bm{r}_t \right).
\label{Eq:gamma_hat}
\end{equation}
These equations are solved recursively using the algorithm by \cite{kelly2019characteristics}. Importantly, equation~\eqref{Eq:f_hat} reveals that the estimated factors $\bm{f}_t$ can be interpreted as portfolio returns, where the weights are given by:
\[
\left( \hat{\bm{\Gamma}}' \bm{X}_{t-1}' \bm{W}_{t-1} \bm{X}_{t-1} \hat{\bm{\Gamma}} \right)^{-1} \hat{\bm{\Gamma}}' \bm{X}_{t-1}' \bm{W}_{t-1},
\]
which is available at period \(t-1\). \cite{kelly2019characteristics} also provide a formal econometric theory to guarantee the consistency and convergence of the IPCA estimators.

\subsection{C-IPCA model}\label{subsec:CIPCA}

We now describe how to incorporate clustering into the IPCA framework to construct the Cluster-IPCA (C-IPCA) model. The standard IPCA model assumes that each risk exposure (i.e., factor loading) is a linear function of \emph{all} firm characteristics, as specified in equation~\eqref{eq:ret}. In contrast, C-IPCA imposes economic structure by assuming that each risk exposure depends only on a subset of firm characteristics—specifically, those within a single cluster representing a shared economic interpretation.

Formally, let $\{P_k\}_{k=1}^K$ denote the set of $K$ clusters of firm characteristics, where each $P_k$ indexes the characteristics grouped into the $k^{\text{th}}$ cluster. The C-IPCA model introduces two restrictions on equation ~\eqref{Eq: IPCA_beta}:

\paragraph{Restriction 1 (Cluster-Based Loadings).} The $k^{\text{th}}$ factor loading is a linear function only of the firm characteristics in the $k^{\text{th}}$ cluster. That is, for each $k = 1, \ldots, K$, we impose:
\begin{equation}
\Gamma_{i k} = 0, \quad \forall i \notin P_k,
\label{Eq:res_CIPCA1}
\end{equation}
where $\Gamma_{i k}$ denotes the $i^{\text{th}}$ row and $k^{\text{th}}$ column of the loading matrix $\bm{\Gamma}$. This restriction ensures that the $k^{\text{th}}$ factor is constructed solely from the characteristics in cluster $P_k$, thereby enhancing interpretability by aligning each factor with a distinct set of economically related variables.

\paragraph{Restriction 2 (Zero-Correlation Factor).} We introduce an additional $(K+1)^{\text{th}}$ factor whose loadings on all firm characteristics are set to zero:
\begin{equation}
\Gamma_{i, K+1} = 0, \quad \forall i = 1, \ldots, I.
\label{Eq:res_CIPCA2}
\end{equation}
This factor is included to capture variation in returns that is not explained by observable characteristics. Specifically, it allows for the possibility of a factor that is orthogonal to all characteristics—such as the market factor—which may otherwise be omitted under the cluster-based structure. Note that such a factor is a special case of the IPCA formulation in which all exposures are constant across assets. In practice, this $(K+1)^{\text{th}}$ factor—referred to as the \emph{zero-correlation (ZC)} factor—is often highly correlated with the market; in our application, its correlation with the market factor exceeds 0.999 as shown in table \ref{Table: Perf_factors}.

These two restrictions jointly define the C-IPCA model, which combines the statistical strength of IPCA with an economically grounded structure based on characteristic clustering. The resulting model enhances interpretability without compromising pricing performance, as we demonstrate in the empirical analysis.

With the two restrictions, the C-IPCA model takes the following form:
\begin{equation}
\bm{r}_t = \bm{\beta}_{1,t-1} f_{1,t} + \bm{\beta}_{2,t-1} f_{2,t} + \cdots + \bm{\beta}_{K+1,t-1} f_{K+1,t} + \bm{e}_t,
\label{Eq:CIPCA_r}
\end{equation}
where the model includes at most $K+1$ factors: one for each of the $K$ characteristic clusters, and one zero-correlation (ZC) factor unrelated to any characteristics. In contrast, the standard IPCA model allows factor loadings to depend on all characteristics, potentially leading to a much larger number of latent factors.

Importantly, not all clusters necessarily contribute significantly to the model. In practice, statistical tests can be employed to eliminate uninformative clusters, resulting in a model with fewer than $K+1$ factors, which will be discussed in the section on model selection \ref{subsec: MaxPort_IC-IPCA}.

\paragraph{An illustrative example.} To better understand the role of the two restrictions, consider a simple illustrative example involving four firm characteristics: two momentum-related variables, \(\text{mom}_1\) and \(\text{mom}_2\), and two value-related variables, \(\text{value}_1\) and \(\text{value}_2\). Suppose the true return-generating process is driven by three latent factors. Under the standard IPCA framework, the model takes the following form (suppressing asset and time subscripts for simplicity):
\begin{equation}
r = \beta_1 f_1 + \beta_2 f_2 + \beta_3 f_3 + e,
\label{Eq:IPCA_r_eg}
\end{equation}
\begin{equation}
\beta_j = \gamma_{1j} \cdot \text{mom}_1 + \gamma_{2j} \cdot \text{mom}_2 + \gamma_{3j} \cdot \text{value}_1 + \gamma_{4j} \cdot \text{value}_2 + \gamma_{5j} + u_j, \quad j = 1,2,3,
\label{Eq:IPCA_beta_eg}
\end{equation}
where \( r \) is the return of a stock, \( f_j \) is the $j^{\text{th}}$ factor, and \( \beta_j \) is the corresponding factor loading. As shown in equation~\eqref{Eq:IPCA_beta_eg}, the IPCA model allows each risk exposure \( \beta_j \) to be a linear combination of all available characteristics, regardless of their economic grouping.

In contrast, the C-IPCA model imposes structure by restricting each exposure to depend only on the relevant cluster. Specifically, it assumes:
\begin{align}
\beta_1 &= \gamma_{11} \cdot \text{mom}_1 + \gamma_{21} \cdot \text{mom}_2 + \gamma_{51} + u_1, \label{Eq:CIPCA_beta_eg1} \\
\beta_2 &= \gamma_{32} \cdot \text{value}_1 + \gamma_{42} \cdot \text{value}_2 + \gamma_{52} + u_2, \label{Eq:CIPCA_beta_eg2} \\
\beta_3 &= \gamma_{53} + u_3. \label{Eq:CIPCA_beta_eg3}
\end{align}

Thus, in the C-IPCA framework, each factor exposure is tied to a single cluster of characteristics. For example, \(\beta_1\) depends only on the momentum characteristics \(\text{mom}_1\) and \(\text{mom}_2\), while \(\beta_2\) depends only on the value characteristics \(\text{value}_1\) and \(\text{value}_2\). The third factor loading \(\beta_3\) is constant and unrelated to any characteristics, capturing the contribution of a latent factor—such as the market—that is not spanned by observable firm-level characteristics. This example illustrates how C-IPCA achieves a cleaner, economically interpretable structure while still accommodating latent variation uncorrelated with known signals.

{Based on this example, we demonstrate how the two restrictions shape the structure of the model. First, 
Restriction 1 in equation~\eqref{Eq:res_CIPCA1} directly leads to equations~\eqref{Eq:CIPCA_beta_eg1} and \eqref{Eq:CIPCA_beta_eg2}. }
Specifically, to derive these two equations from the general specification in equation ~\eqref{Eq:IPCA_beta_eg}, we impose the restriction that each factor exposure depends exclusively on the characteristics within a single cluster.
we restrict each factor exposure to depend only on the characteristics in a single cluster in the general specification in equation. Formally, we impose:
\begin{equation}
\begin{aligned}
&\gamma_{i1} = 0, \quad \text{if } i \notin P_{\text{mom}}, \\
&\gamma_{i2} = 0, \quad \text{if } i \notin P_{\text{value}},
\end{aligned}
\end{equation}
where \( P_{\text{mom}} = \{1, 2\} \) denotes the momentum cluster and \( P_{\text{value}} = \{3, 4\} \) denotes the value cluster. This constraint corresponds to the cluster-based restriction defined in equation~\eqref{Eq:res_CIPCA1}.

{Second, Restriction 2 in equation~\eqref{Eq:res_CIPCA2} directly lead to equations~\eqref{Eq:CIPCA_beta_eg3}.}
 Specifically, to derive the equation \eqref{Eq:CIPCA_beta_eg3} from the general specification in equation ~\eqref{Eq:IPCA_beta_eg}, we assume that one risk exposure is unrelated to any firm characteristics. This implies:
\begin{equation}
\gamma_{i3} = 0, \quad \text{for } i = 1,2,3,4,
\end{equation}
which corresponds to the zero-correlation (ZC) restriction in equation~\eqref{Eq:res_CIPCA2}.

Together, these restrictions yield the structured and interpretable form of the C-IPCA model, aligning each factor with a distinct economic theme while allowing for unobservable but priced sources of risk.

\paragraph{Estimation of the C-IPCA model.} The estimation of the C-IPCA model can be implemented analytically in the same manner as the standard IPCA model. This is because the objective function remains the same as in equation~\eqref{Eq:obj_fun}, but with the additional constraints introduced by equations~\eqref{Eq:res_CIPCA1} and~\eqref{Eq:res_CIPCA2}.

Consider the first-order conditions of the optimization problem. After imposing the restrictions, equation~\eqref{Eq:f_hat} for estimating the factors remains unchanged. However, the estimation of the loading matrix $\bm{\Gamma}$ in equation~\eqref{Eq:gamma_hat} is modified by dropping the conditions associated with parameters restricted to zero. Specifically, equation~\eqref{Eq:gamma_hat} consists of $(I+1) \times J$ first-order conditions, where the $((i-1) \cdot J + j)^{\text{th}}$ row corresponds to the gradient with respect to parameter $\gamma_{ij}$.

Under the C-IPCA restrictions, we omit all rows corresponding to parameters $\gamma_{ij}$ that are set to zero by either cluster assignment (restriction~\eqref{Eq:res_CIPCA1}) or the zero-correlation factor (restriction~\eqref{Eq:res_CIPCA2}). The remaining unrestricted parameters are estimated using the corresponding subset of first-order conditions. The resulting estimator retains the analytical tractability of the IPCA model, while imposing structure that improves interpretability\footnote{As is typical in latent factor models, additional assumptions are required for the identification of IPCA estimators. Specifically, the matrices \(\bm{\Gamma}\) and \(\bm{f_{t+1}}\) are unidentified because any solutions can be rotated into an observationally equivalent form, \(\bm{\Gamma}R^{-1}\) and \(R\bm{f_{t+1}}\), for any non-singular \(K*K\) rotation matrix \(R\). To address this identification issue, \cite{kelly2019characteristics} imposes three restrictions, as detailed in their internet appendix. Since C-IPCA involves additional parameter constraints relative to IPCA, we impose two restrictions to ensure uniqueness of the solution. First, the monthly standard deviation of \(\hat{\bm{f_t}}\) is fixed at 1\%. Second, following  \cite{kelly2019characteristics}, the mean of \(\hat{\bm{f_t}}\) is restricted to be non-negative.
}.

\subsection{OOS estimation}

In the following sections, we evaluate the performance of the IPCA and C-IPCA models based on rolling out-of-sample (OOS) estimates. Specifically, for each period \( t \), we use all available data up to time \( t \) to estimate the model parameters recursively, based on equations~\eqref{Eq:f_hat} and~\eqref{Eq:gamma_hat}.

In particular, starting from month \( t \geq 180 \), we use data through period \( t \) to estimate the parameter matrix \( \hat{\bm{\Gamma}} \) following the procedure described in Section~\ref{subsec:CIPCA}. Given \( \hat{\bm{\Gamma}} \), we then compute the out-of-sample realized factor returns at time \( t+1 \) as:
\begin{equation}
\hat{\bm{f}}_{t+1} = \left( \hat{\bm{\Gamma}}' \bm{X}_{t}' \bm{W}_{t} \bm{X}_{t} \hat{\bm{\Gamma}} \right)^{-1} \hat{\bm{\Gamma}}' \bm{X}_{t}' \bm{W}_{t} \bm{r}_{t+1},
\end{equation}
based on the estimated portfolio weights
\(
\left( \hat{\bm{\Gamma}}' \bm{X}_{t}' \bm{W}_{t} \bm{X}_{t} \hat{\bm{\Gamma}} \right)^{-1} \hat{\bm{\Gamma}}' \bm{X}_{t}' \bm{W}_{t},
\)
which depend only on information available up to time \( t \). Since all parameters used in this calculation are estimated using data through period \( t \), the factor returns at time \( t+1 \) are truly out-of-sample. This recursive estimation procedure is repeated each period, generating a full time series of OOS factor returns for model evaluation.

\section{Clusters}\label{sec:clusters}
This section describes the data used in our analysis and outlines the clustering procedures applied to firm characteristics. We consider two distinct approaches to constructing clusters. The first relies purely on domain knowledge from the asset pricing literature, grouping characteristics according to established economic themes such as value, momentum, profitability, and investment. The second approach integrates both domain knowledge and data-driven insights, leveraging empirical patterns in the data to refine or reallocate characteristics within economically meaningful groups. This hybrid method aims to balance theoretical interpretability with improved empirical performance.

\subsection{Data}\label{subsec:data}
We use the 94 firm characteristics introduced by \cite{gu2020empirical} for the U.S. equity market, covering the period from January 1985 to December 2021.\footnote{The data are publicly available at {https://dachxiu.chicagobooth.edu}, and detailed descriptions of the characteristics are provided in Table A.6 of \cite{gu2020empirical}.} Each characteristic is standardized cross-sectionally in each month by subtracting its mean and dividing by its standard deviation. Monthly stock return data are obtained from the CRSP database. We follow the data cleaning procedures described in \cite{gu2020empirical}.

To ensure that all model evaluations are strictly out-of-sample, we divide the full sample into two sub-periods. The first 15 years (1985:01–1999:12) serve as the training sample for initial parameter estimation. The remaining 22 years (2000:01–2021:12) form the testing sample for evaluating out-of-sample (OOS) performance.\footnote{As a robustness check, we also follow \cite{kelly2019characteristics} by using the first 10 years as the training period; results remain qualitatively unchanged.} At the end of each month in the testing period, model parameters are re-estimated using all data available through that month, ensuring that factor construction relies solely on information known at the time.

\subsection{Intuitive clusters (IC)}
\label{subsec: IC}
A common approach to organizing firm characteristics is to rely on domain knowledge and established economic theory. We refer to this as the \emph{Intuitive Clustering} (IC) method. Under this approach, characteristics are grouped based on their underlying economic concepts—such as momentum, value, or profitability—rather than statistical similarity. (see \cite{hou2015digesting}, \cite{harvey2016and}, \cite{mclean2016does}, \cite{hou2020replicating}, \cite{freyberger2020dissecting}, and \cite{han2024cross}.) The IC approach has been widely applied across various datasets, including the 202 characteristics in \cite{harvey2016and}, the 452 characteristics in \cite{hou2020replicating}, and the 299 characteristics in \cite{han2024cross}.

We apply a similar methodology to the 94 firm characteristics in our dataset. First, following \cite{hou2015digesting}, \cite{hou2020replicating}, \cite{freyberger2020dissecting}, and \cite{han2024cross}, we classify characteristics into six economically motivated groups (with abbreviations in parentheses): Momentum (Mom), Value (Val), Profitability (Prof), Investment (Inv), Intangibles (Int), and Trading Frictions (TFs).

Second, each firm characteristic is assigned to the cluster most closely aligned with its economic meaning. This classification yields six intuitive clusters containing 9, 8, 13, 12, 36, and 16 characteristics, respectively. These IC clusters serve as a benchmark for evaluating the interpretability and empirical performance of our factor models. Table \ref{tab:addlabel} shows the mapping between the ICs and firm characteristics.

\subsection{Data-Driven Clusters}\label{subsec:dc-cluster}

While IC clusters are grounded in economic reasoning, they may overlook important statistical relationships among characteristics which in turn could facilitate our understanding and interpretation. This section introduces the \emph{Data-Driven Cluster} (DC) method, which seeks to combine economic intuition with empirical information to produce clusters that are both interpretable and statistically coherent.

How can we integrate economic priors with empirical signals in a principled way? The Bayesian paradigm provides a natural framework. Under Bayesian inference, one begins with a prior belief - based on existing knowledge - and updates it with data to obtain a posterior belief. Analogously, the DC method treats intuitive clusters as a priori and then refines them based on observed statistical relationships among the characteristics.

In the following subsections, we describe how we combine prior economic classifications with empirical information to form data-driven clusters. Specifically, we proceed in two steps. First, we quantify the statistical similarity between firm characteristics based on their historical behavior, as detailed in Section~\ref{subsec:similarity_def}. Second, in subsection \ref{subsec:cluster}, we describe an agglomerative hierarchical clustering procedure that integrates both sources of information, the prior intuitive clusters and the empirical similarity structure, to produce economically meaningful and statistically coherent clusters.

\subsubsection{Characteristic Similarity}
\label{subsec:similarity_def}

The first step in clustering firm characteristics is to calculate the pairwise similarity or distance between them. In our context, the goal is to ensure that characteristics within the same cluster exhibit high similarity, while those in different clusters exhibit lower similarity.

The similarity between characteristics \(i\) and \(j\) is defined as:
\begin{equation}
s_{ij} = \exp\left(-\left(1 - |\rho_{ij}|\right)\right),
\label{Eq:s_D}
\end{equation}
where \( \rho_{ij} \) is the time-series average of the monthly, value-weighted, cross-sectional rank correlations:
\begin{equation}
\rho_{ij} = \frac{1}{T} \sum_{t=1}^{T} \frac{
\sum_{n=1}^{N_t} w_t^n (x_{it}^n - \bar{x}_{it})(x_{jt}^n - \bar{x}_{jt})
}{
\sqrt{\sum_{n=1}^{N_t} w_t^n (x_{it}^n - \bar{x}_{it})^2}
\sqrt{\sum_{n=1}^{N_t} w_t^n (x_{jt}^n - \bar{x}_{jt})^2}
},
\label{eq:similarity}
\end{equation}
where \( x_{it}^n \) denotes the cross-sectional rank of characteristic \( i \) for stock \( n \) at time \( t \),\( \bar{x}_{it} = \sum_{n=1}^{N_t} w_t^n x_{it}^n \) is the value-weighted average rank for characteristic \( i \) at time \( t \),\( w_t^n \) is the value weight of stock \( n \) at time \( t \), normalized such that \( \sum_{n=1}^{N_t} w_t^n = 1 \),\( N_t \) is the number of stocks at time \( t \). In the appendix \ref{app: EW}, we discuss the results for the equal weights.

The use of value-weighted rank correlations in equation~\eqref{eq:similarity} is motivated by two considerations. First, using ranks helps mitigate the influence of outliers, ensuring that the correlation reflects the underlying economic relationships rather than extreme values. Second, by applying value-weighted ranks, we place greater emphasis on larger firms, consistent with our focus on value-weighted Sharpe ratios as discussed later. This approach aligns with \cite{hou2020replicating}, which argues that anomalies are predominantly driven by micro- and small-cap stocks, which deliver limited information on tangency portfolio in practical portfolio management. By emphasizing value-weighted correlations, we mitigate the concern that our results may be disproportionately influenced by these smaller stocks, whose characteristics may not fully represent broader market trends.

There are several important considerations regarding the transformation in equation~\eqref{Eq:s_D}, which converts rank correlations into pairwise similarities:

\begin{enumerate}
    \item \textbf{Noise Robustness:} The exponential transformation, as used in \cite{saxena2017review} and \cite{von2007tutorial}, further down-weights the similarity for characteristic pairs with low rank correlations. This reduces the influence of noisy or weak correlations that might distort clustering outcomes, ensuring that only the most statistically significant relationships dominate the similarity measure.
    \item \textbf{Correlation Strength:} Following \cite{stambaugh2017mispricing}, we interpret large magnitude of correlations as indicative of strong statistical relationships. This ensures that characteristics with high similarity are meaningful and robust, capturing relationships that truly contribute to the clustering process.
    \item \textbf{Sign Invariance:} By taking the absolute value of the correlations, we ensure that negatively but strongly correlated characteristics (e.g., size and illiquidity, with \( \rho = -0.87 \)) are treated as similar. Despite their opposite signs, these characteristics likely reflect the same underlying economic factor and thus should be grouped together.
\end{enumerate}

We will use the similarity metric to refine the initial IC clusters in the subsequent steps of the Data-Driven Cluster (DC) method. The DC method adjusts the clusters by integrating both the economic intuition of the IC method and the statistical insights captured by the similarity structure between characteristics.

\subsubsection{Posterior Adjustment Process}\label{subsec:cluster}
The DC method refines the initial intuitive cluster by integrating prior economic knowledge with data. We adopt the framework proposed by \cite{karypis1999chameleon}, utilizing a split-and-merge approach to adjust the original clustering (hereafter, we refer the algorithm to as Chameleon).

The core idea is to begin with the initial IC clusters and iteratively partition them into smaller sub-clusters. The goal of this splitting process is to minimize the similarity between clusters while maximizing the similarity within each cluster. This process is controlled by a hyperparameter, denoted as the number of sub-clusters. A key advantage of this approach is that the resulting sub-clusters retain interpretability, as they are derived from the initial intuitive clusters.  Once the sub-clusters are formed, the next step is to merge them back into larger clusters. This allows for the potential combination of sub-clusters from different initial intuitive clusters, resulting in more flexible and potentially more meaningful groupings. The merging process is governed by a hyperparameter, which determines the criteria for merging sub-clusters based on their relative similarity. The iterative nature of this procedure allows the method to adapt to both the economic structure of the data and the statistical patterns uncovered by the data.

To facilitate understanding, it is useful to frame the problem using a graph-based approach, as demonstrated in \cite{guha1998cure} and \cite{karypis1999chameleon}. In this context, the data can be represented as a graph, where each vertex corresponds to a firm characteristic and the edges represent the pairwise similarities between characteristics.

Specifically, the weight of an edge between two vertices is determined by their similarity: the higher the similarity, the larger the edge weight, and the smaller the distance between the vertices. The objective of the clustering method is to partition the graph into distinct groups, such that vertices (firm characteristics) within the same group are highly similar, while those in different groups exhibit lower similarity. For clarity, we use the terms "firm characteristics" and "vertices" interchangeably. Similarly, for convenience, we refer to "low similarity," "small edge weights," and "large distances" as interchangeable concepts throughout the discussion.

Following the Chameleon clustering method \citep{karypis1999chameleon}, we perform a split-and-merge adjustment to the initial IC clusters through a three-step procedure.

\textbf{Step 1: Graph Construction.} The first step involves constructing a sparse graph based on the similarity between firm characteristics. In this process, edges are retained only between characteristics that are among the \(knn\)-nearest neighbors, where \(knn\) is a hyperparameter. This approach not only improves computational efficiency but also ensures that the most significant relationships between characteristics are preserved. Additionally, this step effectively diminishes the impact of small pairwise similarities by shrinking them towards zero. This is beneficial because small similarities could arise from noise in the data, which may distort the clustering process.

\textbf{Step 2: Splitting IC Clusters.} In this step, we partition the initial IC clusters into smaller sub-clusters. The number of sub-clusters, denoted as \( m \), a hyperparameter, which can be fine-tuned during the process. The partitioning is performed iteratively: in each iteration, the largest sub-cluster is selected and split into two smaller sub-clusters. The goal of the splitting process is to minimize the inter-cluster similarity, which is defined as the average similarity between characteristics in the two resulting sub-clusters. By doing so, we ensure that the resulting sub-clusters are as distinct as possible. This iterative process continues until exactly \( m \) sub-clusters are obtained, which we refer to as basic sub-clusters\footnote{{Instead of using the hMetis algorithm as in Karypis et al. (1999) for partitioning, we adopt spectral clustering, which is more computationally efficient and has publicly available source code.}}.

\textbf{Step 3: Merging Sub-clusters.} Starting with the \(m\) basic sub-clusters, we sequentially merge the two most similar sub-clusters. The merging criterion is based on the Relative Inter-Cluster Similarity (RIS), which is defined as:
\begin{equation}
RIS(C_i,C_j) = \frac{INTER(C_i, C_j)}{\frac{|C_i|}{|C_i| + |C_j|} INTRA(C_i) + \frac{|C_j|}{|C_i| + |C_j|} INTRA(C_j)},
\label{Eq:RIS}
\end{equation}
where \( INTER(C_i, C_j) \) is the inter-cluster similarity, measured as the average edge weight between the two clusters, and \( INTRA(C_i) \) and \( INTRA(C_j) \) are the intra-cluster similarities of clusters \(C_i\) and \(C_j\), respectively. The algorithm merges clusters with high inter-cluster similarity but low intra-cluster similarity, ensuring that the final clusters exhibit high internal coherence while minimizing external overlap\footnote{{The merging rule in \cite{karypis1999chameleon} is designed to balance two objectives: robustness to noise and adaptiveness to cluster shape, quantified by the metrics \(RC\) and \(RI\), respectively. Given the low signal-to-noise ratio commonly observed in financial data, we place infinite weight on robustness to noise, effectively disregarding cluster shape adaptability. As a result, our merging rule focuses solely on robustness to noise, and the \(RIS\)
metric introduced in our paper is equivalent to the 
\(RC\) metric defined in \cite{karypis1999chameleon}.}}.

\subsubsection{Advantages of the Data-Driven Cluster Method}
\label{subsubsec: DC_adv}
There are two advantages of our Data-Driven Cluster (DC) method: (1) robustness to data noise, making it well-suited for financial data with a low signal-to-noise ratio, and (2) relaxed assumptions about the data, allowing for clusters of varying shapes and sizes as emphasized by \cite{karypis1999chameleon}.

\textbf{Robustness to Data Noise.} {
The merging rule in equation~\eqref{Eq:RIS} enhances Chameleon’s robustness to data noise. As highlighted by \cite{karypis1999chameleon}, this rule is specifically designed to adapt to datasets where clusters exhibit heterogeneous densities. In other words, it accommodates structures in which some clusters consist of firm characteristics with high similarity (high-density clusters), while others exhibit lower similarity (low-density clusters).
}

{
This versatile is particularly valuable in asset pricing, where noise frequently reduces the density of economically meaningful clusters. To illustrate, recall that firm characteristics within the same cluster are assumed to be noisy proxies for a common latent risk exposure. In the absence of measurement error, each characteristic would perfectly capture the latent exposure, yielding a correlation of one across characteristics within the cluster and maximizing the similarity measure $s_{i,j}$ in equation~(\ref{Eq:s_D}). However, as measurement error increases, observed characteristics deviate from the latent exposure, reducing correlations and thereby lowering both similarity and cluster density. Because the Chameleon algorithm maintains effectiveness even in the presence of low-density clusters, it is particularly well-suited for analyzing financial data subject to substantial noise.
}

\textbf{Relaxed Assumptions on Data Structure.} In the third step of the clustering process (Section~\ref{subsec:cluster}), Chameleon’s strategy of merging sub-clusters, rather than individual firm characteristics, allows it to relax assumptions about the data. Unlike traditional algorithms, such as K-means, which assume clusters to be elliptical and of similar sizes, Chameleon can accommodate clusters of arbitrary shape and size. Many clustering algorithms represent each cluster by a \emph{single vertex} (referred to as the "cluster representative") and compute clustering results based on the distance from each data point to the cluster representative, called \emph{vertex-to-vertex distance}. This approach works well when clusters are roughly spherical and of similar sizes. However, when clusters are concave or vary widely in size, the vertex-to-vertex distance can lead to inaccurate results, as it fails to capture the overall shape and size of the clusters.

In contrast, Chameleon computes the similarity between sub-clusters based on the \emph{sub-cluster-to-sub-cluster} distances, as reflected in the \( INTER(C_i, C_j) \) term of the merging rule in equation~\eqref{Eq:RIS}. This allows the algorithm to adapt to a wider variety of cluster shapes and sizes, ensuring that the final clustering structure better reflects the true underlying data distribution.

\subsubsection{Selection of Hyper-parameters}
\label{subsec: hyper-params}
Our algorithm involves three hyperparameters: (1) the number of nearest neighbors \(knn\) in Step 1, (2) the number of sub-clusters \(m\) in the splitting process (Step 2), and (3) the final number of clusters \(K\) in the merging process (Step 3). This section explains how to select the optimal values for these hyperparameters using a grid search.

We first define a grid of candidate values for each hyperparameter:
\( knn = \{10, 15, \dots, 90\} \), \( m = \{16, 19, 24, 31\} \), and \( K = \{1, 2, \dots, 15\} \), where \( m \) is chosen such that, on average, there are 3, 4, 5, or 6 vertices in each basic sub-cluster after Step 2.

\paragraph{Selection of \(K\).}We begin by selecting the optimal value for \(K\) given \(m\) and \(knn\). This corresponds to determining when to stop the merging process in Step 3 in Section \ref{subsec:cluster}. Following the approach of \cite{barton2019chameleon}, we terminate the merging process when the maximum relative inter-cluster similarity (\(Max\_RIS\)) becomes abnormally low. The \(Max\_RIS\) is defined as the highest RIS value among all cluster pairs, and is calculated as:
\begin{equation}
Max\_RIS = \max_{i,j=1,...,K, i \neq j} RIS(C_i,C_j),
\label{Eq:average RIS}
\end{equation}
where \( RIS(C_i,C_j) \) represents the relative inter-cluster similarity between clusters \(C_i\) and \(C_j\) as in equation \eqref{Eq:RIS}.

The intuition behind using \(Max\_RIS\) is as follows: during the merging process, each pair of sub-clusters merged at a given stage will have the highest RIS at that point. A abnormally low \(Max\_RIS\) indicates that the merger is inappropriate, either because the inter-cluster similarity is very low or the intra-cluster similarity is unusually high, suggesting the clusters are not well-matched. {Therefore,we should stop merging at this point and consider the current number of clusters as the optimal $K$.}

To determine at which point the \(Max\_RIS\) is abnormally low —i.e., identifying the stopping point for merging — we follow the approach of \cite{barton2019chameleon}, which detects a sharp drop in \(Max\_RIS\). In our merging procedure, we always combine the two clusters with the highest \(RIS\), so the value of \(Max\_RIS\) naturally tends to decrease as the number of clusters \(K\)
decreases. However, when this decrease becomes disproportionately large, we take it as evidence that the \(Max\_RIS\) has become abnormally low, marking the point at which we stop merging.

Specifically, the approach to determine the optimal $K$ involves two steps. First, it computes a baseline measure for detecting the abnormal low $Max\_RIS$ values. This baseline measure is defined as the average reciprocal of $Max\_RIS$ within the larger half of
$K$'s range (i.e. the range [$m$/2,$m$] as $K$ ranges from 1 to m )
. We denote this value as $\overline {{Max\_RIS}^{-1}}$. Second, we search in descending order for $K$ within the rest range [1,m/2-1]. Specifically, we look for the first $K$ where the reciprocal $Max\_RIS$ is at least $f$ times greater than the $\overline {{Max\_RIS}^{-1}}$. When found, we take the preceding $K$ as the optimal value. If this condition is
not fulfilled for any $K$, we iteratively relax the threshold by adjusting it from $f$ to $\frac{f}{{\eta}^i}$ times of $\overline {{Max\_RIS}^{-1}}$, where $i$ is an iteration index that starts at 1 and increases by 1 until the condition is fulfilled. Here $f$ is the initial scaling value and $\eta$ controls the rate of decrease in the threshold.
Following \cite{barton2019chameleon}, we set $f$ to $10^3$. The value of $\eta$ depends on the problem, and in this case, we use $\eta=1.3$. Results are robust if $\eta=1.2$ or $\eta=1.4$.

\paragraph{Selection of \(m\) and \(knn\).}Finally, we select the optimal values for \(m\) and \(knn\) by evaluating the performance of the corresponding C-IPCA models and select the hyperparameter with highest Sharpe ratio of the training sample for the tangency portfolio. 
Specifically, we construct the tangency portfolio using all model factors following a purely out-of-sample procedure: for each time \(t\), we estimate the mean and covariance matrix of factor returns using data available up to time \(t\), and then track the realized return of the tangency portfolio in time \(t+1\). 
The tangency portfolio is initially constructed in December 1989. 
{The optimal hyper-parameters are \(m=24, knn=55, K=12\)}.

\section{Data-Driven Clusters vs. Intuitive Clusters}
\label{subsec: DC and IC}

This section provides a comparison of the Data-Driven Cluster (DC) and Intuitive Cluster (IC) methods to highlight the role of empirical data in refining clusters beyond prior domain knowledge. 

\subsection{Connections Between DC and IC}

\begin{figure}[htbp]
\begin{subfigure}[b]{1.0\textwidth} 
        \centering
        \includegraphics[width=\textwidth]{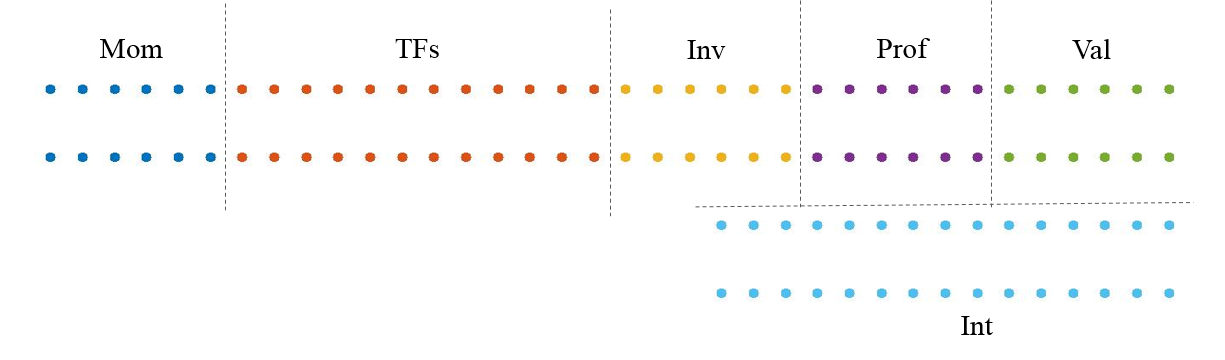} 
        \caption{IC}
        \label{fig:subfig3}
    \end{subfigure}

    \vspace{1em} 
    
    \begin{subfigure}[b]{1.0\textwidth} 
        \centering
        \includegraphics[width=\textwidth]{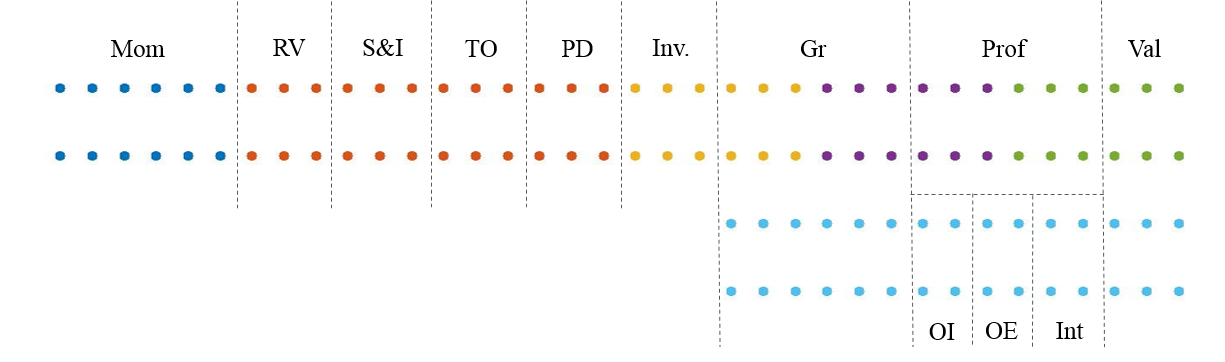} 
        \caption{DC}
        \label{fig:subfig4}
    \end{subfigure}
\caption{Cluster results}
\bigskip
\textit{ Note:} This figure illustrates the relationship between IC and DC clustering structures. Subfigure (a) shows the six IC clusters (Mom = Momentum, TFs = Trading Frictions, Inv = Investment, Prof = Profitability, Val = Value, Int = Intangibles), while Subfigure (b) displays the thirteen DC clusters (Mom = Momentum, RV = Return Volatility, S\&I = Size \& Illiquidity, TO = Turnover, PD = Price Delay, Inv = Investment, Gr = Growth, Prof = Profitability, Val = Value, OI = Operating Illiquidity, OE = Operating Efficiency, Int = Intangibles). Each vertex represents a firm characteristic, with colors indicating IC or DC cluster membership. The number of vertices is chosen for visualization purposes only and does not correspond to the actual count of firm characteristics. Dashed lines depict the clustering boundaries among characteristics.

\vskip 0.1in \label{Fig: DC and IC}
\end{figure}

Figure~\ref{Fig: DC and IC} shows the clustering results for firm characteristics based on both the DC and IC methods. In the figure, the vertices represent firm characteristics, with different colors indicating the clusters assigned by the IC method. The number of vertices is for visualization purposes only and does not reflect the actual number of firm characteristics in the dataset. The dashed lines in the figure represent the clustering results for each method. For the details on the mapping between characteristics and clusters, please see the table in Appendix \ref{tab:addlabel}.

Panel A shows the clustering results based on the IC method, while Panel B shows the results from the DC method. In our data, the optimal number of clusters for the DC method is 12, with the following abbreviations: Momentum (Mom), Return Volatility (RV), Size and Illiquidity (S\&I), Turnover (TO), Price Delay (PD), Investment (Inv), Growth (Gr), Profitability (Prof), Operating Illiquidity (OI), Operating Efficiency (OE), Intangibles (Int), and Value (Val).

There are several key observations when comparing the clustering results between Panel A (IC) and Panel B (DC).

1. \textbf{Momentum Cluster (Mom)}: The Mom cluster in the IC method is fully aligned with the DC method, with the same firm characteristics grouped together, demonstrating a complete overlap.

2. \textbf{Trading Frictions (TFs) Cluster}: The TFs cluster in the IC method is further divided into four distinct clusters in the DC method: RV, S\&I, TO, and PD. The RV cluster includes characteristics such as idiosyncratic return volatility and overall return volatility. The S\&I cluster contains characteristics like size and Amihud illiquidity. The TO cluster includes characteristics such as share turnover and the volatility of share turnover. The PD cluster includes characteristics like abnormal earnings announcement volume and price delay.

3. \textbf{Intangibles Cluster (Int)}: The Int cluster in the IC method is split into three distinct clusters in the DC method: OI, OE, and Int. OI includes characteristics such as the quick ratio and current ratio, OE contains characteristics like the accruals, sales-to-inventory ratio, and Int includes characteristics such as R\&D investment.

4. \textbf{Investment, Profitability, and Value Clusters}: In the IC method, the characteristics related to Investment (Inv), Profitability (Prof), and Value (Val) are re-clustered into four separate clusters in the DC method: Gr (growth), Inv (investment), Prof (profitability) and Val (value).

These observations highlight the flexibility of the DC method in refining the initial intuitive clusters by leveraging empirical data. Although the IC method relies solely on prior economic knowledge on the characteristics of the group, the DC method adapts these groupings exploit the underlying statistical relationships between the characteristics, offering a more nuanced clustering structure.

\subsection{Evaluating the Effectiveness of DC Clustering}
\label{subsec: DCClusteringSimilarity}

A core goal of clustering is to group together firm characteristics that are highly similar while separating those that are dissimilar. This subsection evaluates whether our Data-Driven Clustering (DC) method effectively achieves this goal. Specifically, we assess whether characteristics grouped together under DC exhibit high pairwise similarity and whether characteristics placed in different clusters exhibit low similarity.

We approach this evaluation from two complementary perspectives. First, we directly examine the pairwise similarity between firm characteristics within and across clusters. Second, we visualize the clustering structure by mapping characteristics into a low-dimensional space that preserves pairwise distances, revealing how well the clusters are separated geometrically.

\paragraph{Pair-wise Similarity.} The DC method incorporates empirical information via the similarity measure \( s_{ij} \). If the clustering structure is informed by data, we expect within-cluster similarity to be high and between-cluster similarity to be low.

\noindent\paragraph{Distance-Based Visualization via MDS.} 
To provide an alternative perspective, we transform the similarity matrix into a distance matrix using the following transformation:
\begin{equation}
d_{ij} = \frac{1}{s_{ij}} - 1,
\label{Eq:distance}
\end{equation}
where \( s_{ij} \) is the similarity between characteristics \(i\) and \(j\) as defined in equation~\eqref{eq:similarity}, with \(d_{ij} \geq 0\) and \(d_{ii} = 0\). This transformation preserves the inverse relationship between similarity and distance: higher similarity implies shorter distance.

To visualize the clustering structure implied by these distances in a two-dimensional space to enhance our intuition, we apply Multidimensional Scaling (MDS) , a statistical technique that maps objects to points in a low-dimensional space such that pairwise distances are preserved as closely as possible \citep{borg2011multidimensional}. Specifically, given a set of \(N\) firm characteristics and their pairwise distances \(d_{ij}\), MDS map each characteristic into one point in a two-dimensional space by solving:
\begin{equation}
\text{Stress} = \sqrt{ \frac{ \sum_{i<j} (d_{ij} - \hat{d}_{ij})^2 }{ \sum_{i<j} d_{ij}^2 } },
\label{Eq:stress}
\end{equation}
where \( \hat{d}_{ij} \) is the Euclidean distance between the mapped points of characteristics \(i\) and \(j\). The objective is to minimize the stress function, thereby ensuring that the projected distances in the two-dimensional space approximate the original distances.

\begin{figure}[htbp] 
    \begin{subfigure}[b]{1.0\textwidth} 
        \centering
        \includegraphics[width=\textwidth]{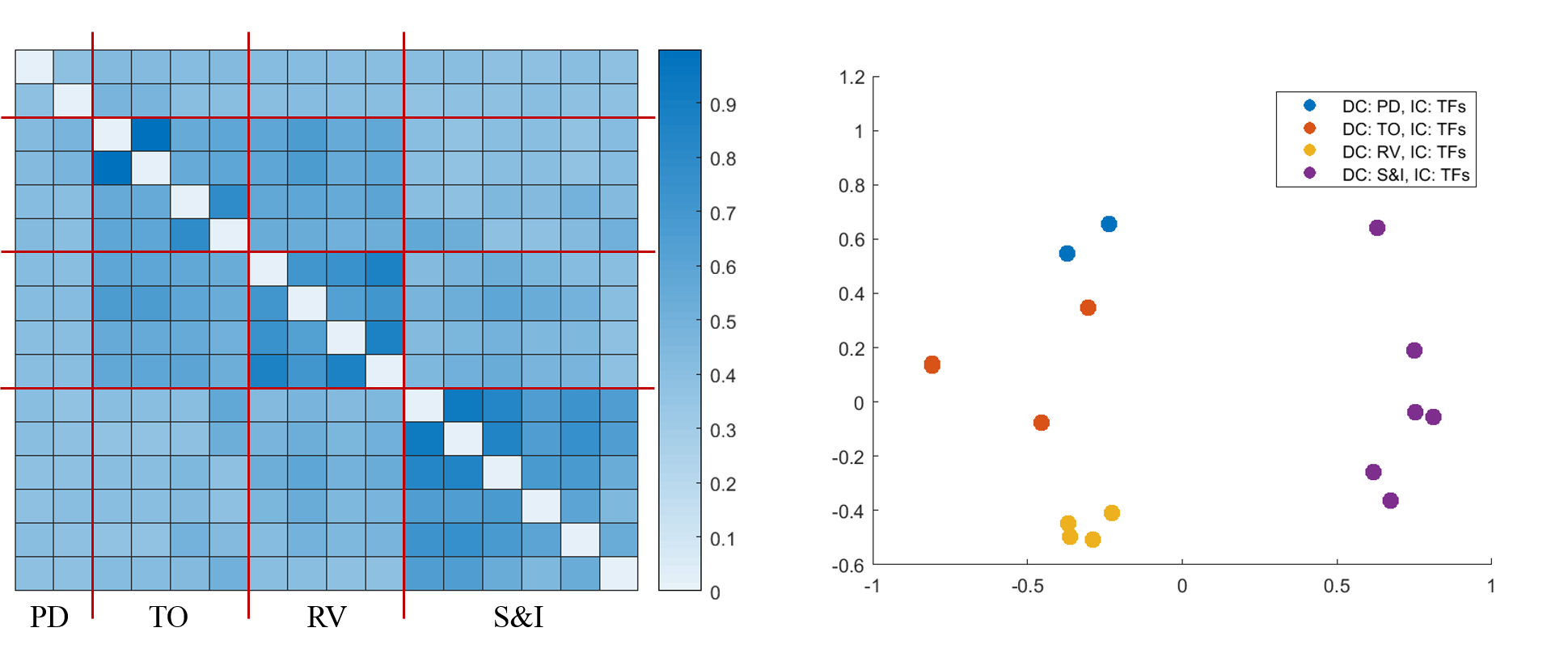} 
        \caption{RV, S\&I, TO and PD clusters in DC}
        \label{fig:subfig1}
    \end{subfigure}

    \vspace{1em} 
    
    \begin{subfigure}[b]{1.0\textwidth} 
        \centering
        \includegraphics[width=\textwidth]{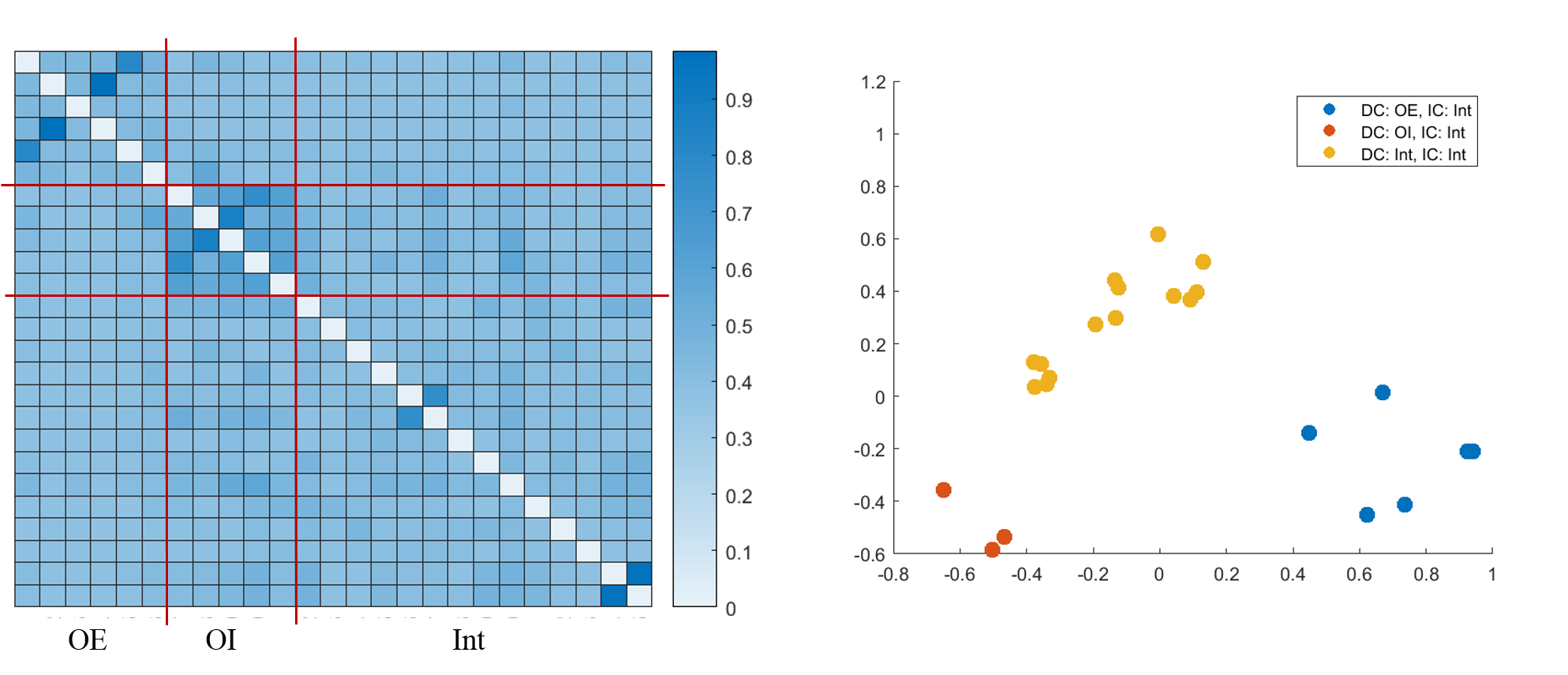} 
        \caption{OE, OI and Int clusters in DC}
        \label{fig:subfig2}
    \end{subfigure}
\caption{\textbf{Role of Data Information in Clustering.} This figure illustrates the role of data information in clustering. Subfigure (a) focuses on four DC trading-friction clusters: RV, S\&I, TO, and PD. Subfigure (b) highlights three DC intangibles-related clusters: OE, OI, and Int. (See Figure~\ref{Fig: DC and IC} for cluster abbreviations.) In each subfigure, the left panel shows the similarity matrix \(s_{i,j}\), where rows and columns represent firm characteristics, and cell shading indicates pairwise similarity (darker cells denote higher similarity). Red lines delineate DC clusters. The right panel visualizes firm characteristics in two dimensions space using Multidimensional Scaling (MDS) algorithm following \cite{borg2011multidimensional}, based on distances defined as \(d_{i,j} = 1/s_{i,j} - 1\).  The MDS maps each firm characteristic into a point in a two dimensional space while keep the distance between characteristics roughly unchanged. Vertices represent firm characteristics, with colors denoting DC clusters and shapes indicating IC clusters, enabling direct comparison between clustering approaches..
}
\label{Fig: DataRole1}
\end{figure}

\begin{figure}[htbp] 
    
    \begin{subfigure}[b]{1.0\textwidth} 
        \centering
        \includegraphics[width=\textwidth]{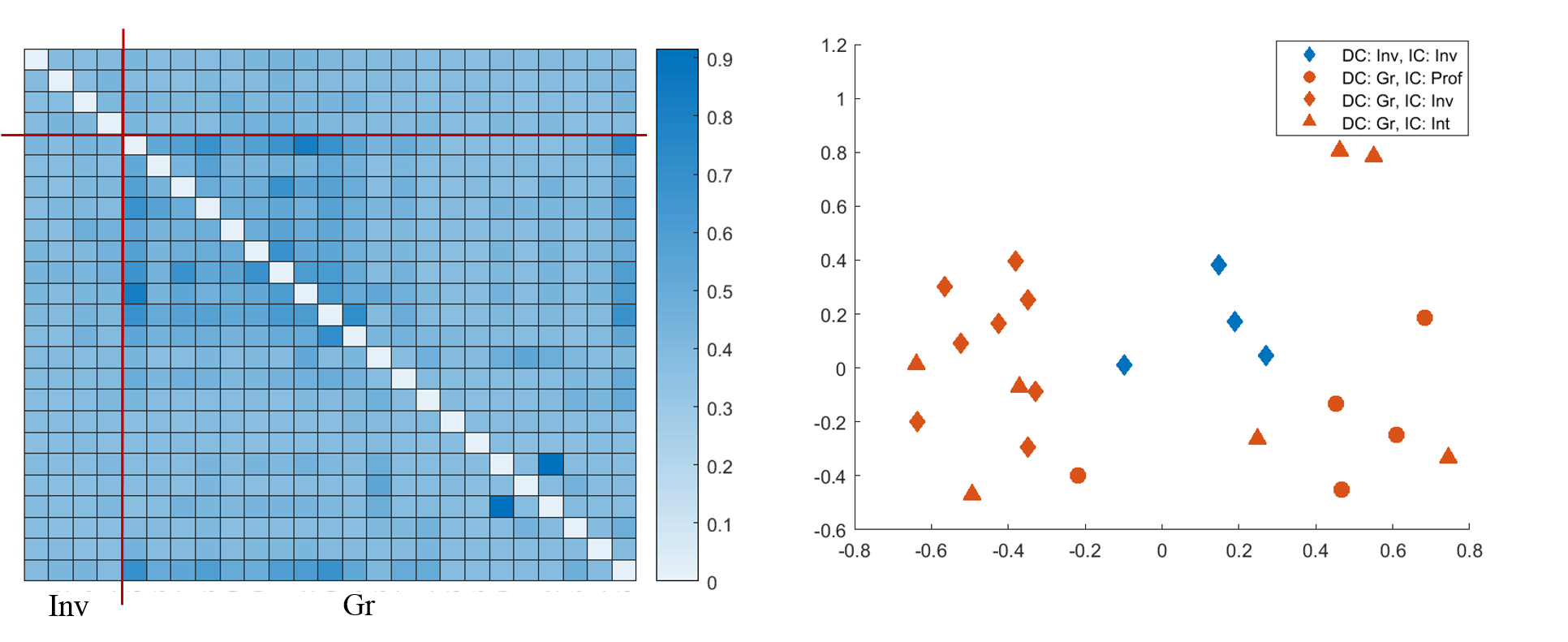} 
        \caption{Inv and Gr clusters in DC}
        \label{fig:subfig3}
    \end{subfigure}

    \vspace{1em} 
    
    \begin{subfigure}[b]{1.0\textwidth} 
        \centering
        \includegraphics[width=\textwidth]{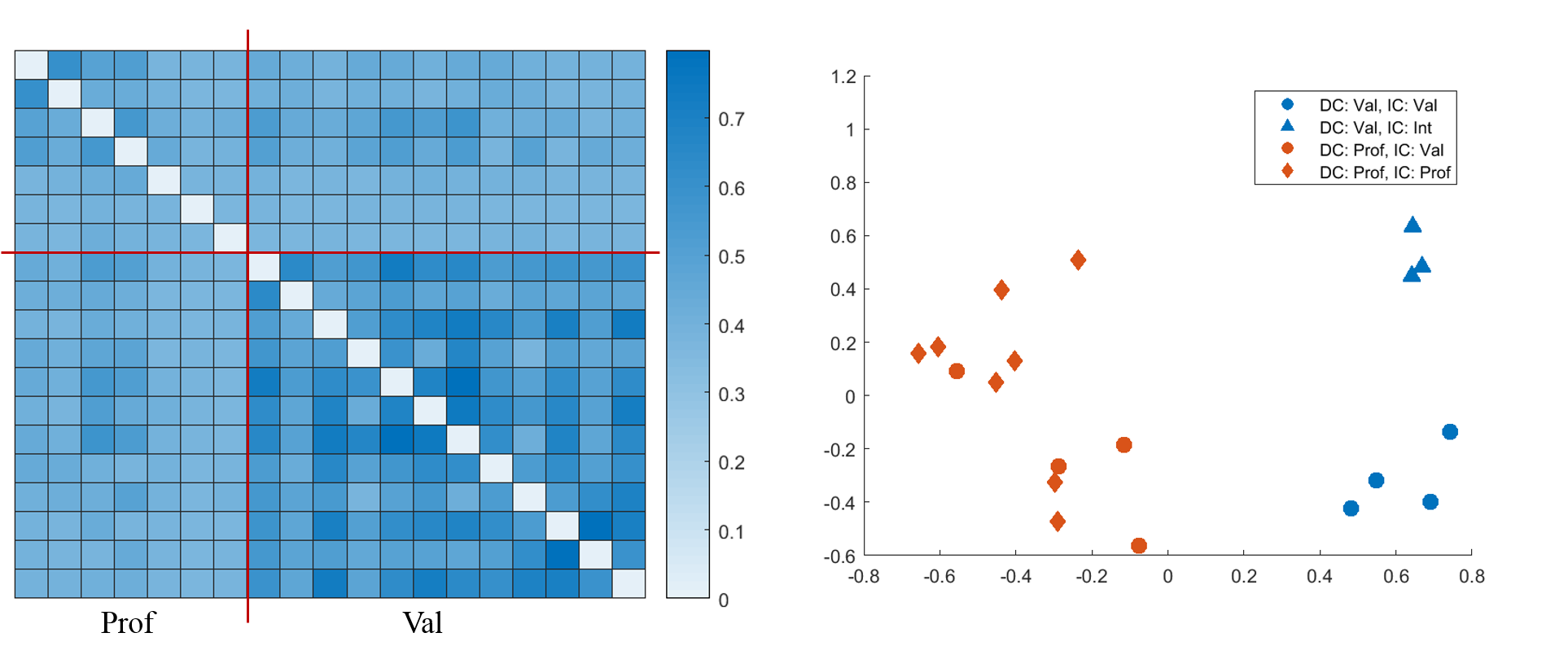} 
        \caption{Val and Prof clusters in DC}
        \label{fig:subfig4}
    \end{subfigure}

    \caption{\textbf{Role of Data Information in Clustering (cont).} This figure examines how data information influences clustering outcomes. Subfigure (a) focuses on the Inv and Growth clusters in DC, while Subfigure (b) highlights the Val and Prof clusters in DC. (See Figure~\ref{Fig: DC and IC} for cluster abbreviations.) In each panel, the left subfigure displays the similarity matrix \(s_{i,j}\), where rows and columns correspond to firm characteristics, and cell shading reflects similarity (darker colors indicate higher similarity). The right subfigure visualizes the spatial distribution of characteristics using Multidimensional Scaling (MDS) following \cite{borg2011multidimensional}, based on distances defined as \(d_{i,j} = 1/s_{i,j} - 1\). The MDS maps each firm characteristic into a point in a two dimensional space while keep the distance between characteristics roughly unchanged. Vertices represent firm characteristics, with colors denoting DC clusters and shapes indicating IC clusters, enabling direct comparison between clustering approaches.}
    \label{Fig: DataRole2}
\end{figure}

Given the large number of firm characteristics, it is inherently difficult to preserve all pairwise distances accurately in two-dimensional space. To enhance interpretability, we present the results in four separate panels in Figures~\ref{Fig: DataRole1} and~\ref{Fig: DataRole2}, with each panel corresponding to a subset of DC clusters.




Figure~\ref{Fig: DataRole1}(a) presents the results for the four trading frictions clusters in the DC method: Return Volatility (RV), Turnover (TO), Size \& Illiquidity (S\&I), and Price Delay (PD), which together include 15 firm characteristics.

The left panel displays a heatmap of the pairwise similarity matrix among these 15 characteristics. Darker colors indicate higher similarity. To enhance interpretability, the characteristics are ordered by DC cluster membership, and red grid lines partition the matrix into 16 submatrices, corresponding to within- and between-cluster similarities. The visual contrast in shading clearly reveals that within-cluster similarities are substantially higher than between-cluster similarities, consistent with the goal of the DC method to group statistically similar characteristics.

The right panel of Figure~\ref{Fig: DataRole1}(a) shows the two-dimensional embedding of these characteristics using MDS. Each point represents a  characteristic, and points are colored according to their DC cluster membership. The plot demonstrates that characteristics within the same cluster tend to be located near one another, while those from different clusters are well-separated. This spatial pattern provides additional evidence that the DC clustering method effectively reduces between-cluster similarity while enhancing within-cluster coherence.

Figure \ref{Fig: DataRole1}(b) shows the similarity between firm characteristics in the three intangible clusters in DC: OI, OE and Int. From the heatmap on the left, we can see that darker grid cells appear more frequently in the diagonal matrices enclosed by the red lines, indicating that firm characteristics with higher similarity are likely to be in the same cluster. Besides, there is an interesting finding in firm characteristics in the third diagonal matrix enclosed by the red lines. Their similarity with each other is not very high, and their similarity to other characteristics is also low. This reflects the advantage of the Chameleon clustering method: low similarity may be due to high noise, and these firm characteristics should belong to the same cluster rather than being treated as separate clusters. This is also evident from the scatter plot on the right: some yellow vertices are far from all other vertices, which may be due to high noise. Each of these vertices should belong to the same cluster as other yellow vertices rather than forming a separate cluster.

Figure~\ref{Fig: DataRole2}(a) displays the similarity structure among firm characteristics in the DC Investment (Inv) and Growth (Gr) clusters. The right panel shows the two-dimensional embedding of firm characteristics based on their pairwise distances, with colors representing DC cluster membership and shapes indicating IC cluster assignments. Several key patterns emerge. First, consider the IC Investment cluster (represented by diamond-shaped points). The DC method splits this group into two distinct sub-clusters, shown as yellow and blue diamonds. These two groups are well-separated in the embedded space, indicating that they are statistically dissimilar despite sharing a common economic label under the IC scheme. The DC method, therefore, refines the initial classification by recognizing and separating these empirically distinct subgroups.

Second, the DC method merges firm characteristics from the IC Profitability, Intangibles, and Investment clusters. For instance, yellow circular (IC Profitability), triangle (IC Intangibles), and diamond (IC Investment) points are all positioned close to one another, suggesting that they capture a shared empirical signal. Accordingly, the DC algorithm groups them into a single cluster, highlighting its ability to uncover cross-cutting relationships not reflected in the IC taxonomy.

Finally, while some points within the same DC cluster appear dispersed - such as the yellow circular points on the far left and right—this reflects the influence of prior economic classifications. These points belong to the same IC cluster and share a common economic interpretation (e.g., profitability), even though their empirical similarity \(s_{ij}\) is low. The DC method incorporates this prior structure and retains them in the same cluster, demonstrating its flexibility in balancing data-driven evidence with domain knowledge.

Figure~\ref{Fig: DataRole2}(b) presents the similarity structure of firm characteristics within the DC Profitability (Prof) and Value (Val) clusters. The heatmap on the left shows that the diagonal blocks enclosed by red lines—representing within-cluster similarities—are distinctly darker than the off-diagonal blocks, indicating strong intra-cluster similarity and weak inter-cluster similarity. This visual evidence suggests that the DC method effectively groups together characteristics with high empirical similarity while separating those that are dissimilar.

The right panel shows the two-dimensional embedding of firm characteristics based on the pairwise distance matrix. As in earlier panels in Figure \ref{Fig: DataRole2}(a), colors denote DC cluster membership, while shapes indicate IC classifications. The figure clearly shows that characteristics assigned to the DC Value and Profitability clusters form two well-separated groups in the embedded space. This result reinforces the core strength of the DC methodology: after incorporating data-driven similarity measures, firm characteristics with strong empirical affinity are more likely to be grouped together, even when their intuitive classification is ambiguous or overlapping. These findings are consistent with the patterns observed in Figures~\ref{Fig: DataRole1} and~\ref{Fig: DataRole2}(a), further validating the effectiveness of our approach.

\section{Performance of the C-IPCA model}
\label{sec: Data and Perf_IC-IPCA}
Given the characteristic clusters constructed in Section~\ref{sec:clusters}, we proceed to estimate the C-IPCA model, which imposes cluster-specific restrictions on the characteristic coefficients in the factor loadings, as described in Section~\ref{subsec:CIPCA}. In this section, we evaluate the empirical performance of the C-IPCA model under different clustering schemes for firm characteristics. This comparison allows us to assess the impact of clustering methodology on both the interpretability and pricing performance of the resulting factor models.

\subsection{ Interpretable factors}
\label{subsec: Perf_factors_IC-IPCA}
Given the estimated characteristic coefficient matrix \(\bm{\hat{\Gamma}}\), recall that each latent factor in the C-IPCA model can be expressed as a linear combination of individual stock returns, with weights determined by the projection formula:
\[
\hat{\bm{f}}_t = \left( \hat{\bm{\Gamma}}' \bm{X}_{t-1}' \bm{W}_{t-1} \bm{X}_{t-1} \hat{\bm{\Gamma}} \right)^{-1} \hat{\bm{\Gamma}}' \bm{X}_{t-1}' \bm{W}_{t-1} \hat{\bm{r}}_t.
\]
This formulation enables us to construct factor-mimicking portfolios directly from the estimated model parameters. Each factor is interpretable by design, as it is extracted from a distinct cluster of firm characteristics.

Figure~\ref{Fig: Gamma} illustrates the interpretability of C-IPCA factors through an example. The figure displays the absolute values of the \(\hat{\Gamma}\) matrix from equation (\ref{Eq: IPCA_beta}) for an IPCA model with 13 factors, estimated using the full sample. Each column corresponds to a factor, while each row represents a characteristic. The red solid lines partition the 94 characteristics into 13 clusters based on the DC structure. The cell at row \(i\) and column \(j\) represents the absolute value of the loading of the \(i\)-th characteristic on the \(j\)-th factor exposure, reflecting the importance of that characteristic for the corresponding factor exposure. Darker shading indicates larger absolute loadings, signifying higher importance, as characteristics have been standardized to have zero mean and unit variance. To enhance visibility of within-column variation, each column is scaled so that the sum of squared elements equals one.

Panel (a) of Figure~\ref{Fig: Gamma} demonstrates that in the standard IPCA model, the loadings for each factor are dispersed across multiple clusters, complicating factor interpretation. For example, in the second column, both market beta and momentum exhibit large loadings, making it difficult to classify the corresponding factor as primarily a market or momentum factor. In contrast, Panel (b) shows that under the C-IPCA model, loadings for each factor are highly concentrated within a single cluster of characteristics. For instance, the first column exhibits significant loadings exclusively on operating efficiency (OE)-related characteristics, indicating that the associated risk exposure is driven predominantly by OE. This clustering enables a clear and economically meaningful interpretation of each factor.

Table~\ref{Table: Perf_factors} shows the summary statistics on the properties of interpretable factors under both the IC-IPCA (intuitive clustering) and DC-IPCA (data-driven clustering) models. Panel A reports results for IC-IPCA, while Panel B presents the corresponding outcomes for DC-IPCA. Each row represents a latent factor, with the final column indicating the economic label or abbreviation of its associated characteristic cluster. To aid interpretation, factors are sorted by their out-of-sample Sharpe ratios over the testing period (2000:01–2021:12), except for the market factor, which is listed in the final row.

The second through fifth columns report standard performance metrics, including the time-series mean (Mean), standard deviation (S.D.), annualized Sharpe ratio (Sharpe), and maximum drawdown (MDD) of monthly factor returns. The Sharpe ratio is computed as the ratio of the time-series mean to the standard deviation, and annualized by multiplying the monthly Sharpe by \(\sqrt{12}\). MDD is defined as the largest cumulative loss from peak to trough over the sample period and serves as an indicator of downside risk.

\vspace{0.2cm}
\noindent \textbf{Panel A: Performance of IC-IPCA Factors.}
Among the seven factors estimated under IC-IPCA, the two with highest Sharpe ratio are the Investment (Inv) and Intangibles (Int) factors, with annualized Sharpe ratios of 0.62 and 0.44, respectively.  In terms of downside risk, the Momentum (Mom) factor exhibits the largest maximum drawdown, consistent with the findings of \cite{daniel2016momentum}.

\vspace{0.2cm}
\noindent \textbf{Panel B: Performance of DC-IPCA Factors.}
Results from DC-IPCA corroborate and refine the insights from Panel A. Among the thirteen factors, the top two performers are the Operating Illiquidity (OI) and Return Volatility (RV), with Sharpe ratios of 0.73 and 0.56, respectively. These results are consistent with Panel A’s strong performance of the Intangibles factor, whose cluster includes OI. In terms of risk, the S\&I and Mom factors have the highest drawdowns, mirroring the risk pattern in Panel A.

Several DC-IPCA factors exhibit relatively low Sharpe ratios - namely, Price Delay (PD), Investment (Inv), and Value (Val)—with values below 0.15 (0.13, 0.08, and 0.02, respectively). This suggests that not all DC-IPCA factors are equally important for pricing cross-sectional returns, and a parsimonious factor model may retain only a subset of the constructed factors.

\vspace{0.2cm}
\noindent \textbf{Panel C: Market Factor Correlation.}
Panel C reports the correlations between market factors (MktRf) and zero-correlation factors (ZC) of various models, including IC-IPCA and DC-IPCA. For comparison, we estimate two versions of the standard IPCA model, one with seven factors (IPCA7) and another with 13 (IPCA13) that match the number of factors in the IC-IPCA and DC-IPCA models, respectively. In each case, we identify the IPCA factor most correlated with the market return and denote it as MF(IPCA7) or MF(IPCA13). The benchmark market factor (MktRf) is the value-weighted excess return on the aggregate stock market.

Across all models, we find that the estimated factors are highly correlated with the true market factor, with correlations near 1. This result confirms that the ZC factor in C-IPCA can be interpreted as a market-mimicking portfolio. It also supports the use of restriction~\eqref{Eq:res_CIPCA2} in the C-IPCA model. Although standard IPCA does not impose the existence of a market factor, one naturally emerges from the data. This reinforces the necessity of explicitly incorporating a market factor in the C-IPCA specification to ensure robust empirical performance and interpretability.

\subsection{Comparison with Traditional Factors}

A natural and interesting question is how our clustering-based factors relate to traditional asset pricing factors. To address this, Panels A and B of Table~\ref{Table: Perf_factors} report the alphas of our factors using Fama–French regressions. Specifically, columns six through eight present alphas from regressions of each factor on standard factor models, including the Fama–French three-factor model (FF3; \citeauthor{fama1993common}, \citeyear{fama1993common}), the Fama–French five-factor model (FF5; \citeauthor{fama2015five}, \citeyear{fama2015five}), and the q-factor model (Q4; \citeauthor{hou2015digesting}, \citeyear{hou2015digesting}).

Two main findings emerge. First, while traditional factors explain some of our clustering-based factors, this explanatory power is concentrated among factors with the lowest Sharpe ratios. In contrast, factors with the highest Sharpe ratios from our models deliver economically and statistically significant alphas relative to all traditional models. For IC-IPCA, three of the top four factors (ranked by Sharpe ratio) exhibit significant alphas across FF3, FF5, and Q4. For example, the first and second IC-IPCA factors earn monthly alphas of 0.17\% and 0.15\%, respectively, in the FF5 model, as shown in Panel A of Table \ref{Table: Perf_factors}.

The results are even stronger for DC-IPCA. The top two factors—Operating Illiquidity (OI) and Return Volatility (RV)—which play a central role in the tangency portfolios of our best-performing models (see Sections~\ref{subsec:ordered_cipca_results} and~\ref{subsec:emp_B_C_IPCA}), generate substantial abnormal returns. For instance, OI earns a monthly alpha of 0.34\% in the FF5 model, while RV earns 0.17\%, both statistically significant at 1\% as shown in Panel B of Table \ref{Table: Perf_factors}.

Taken together, these findings underscore that our clustering-based approach extracts economically meaningful and distinct sources of risk not captured by standard factor models \citep{fama1993common,fama2015five,hou2015digesting}. By combining economic intuition with data-driven clustering, our method delivers interpretable factors that enhance pricing performance and provide new insights into the structure of expected returns.

\subsection{ Mean-Variance Efficiency}
\label{subsec: MaxPort_IC-IPCA}
This section evaluates the mean-variance efficiency of various factor models by examining the out-of-sample performance of their corresponding tangency portfolios estimated using historical data. Specifically, we construct the tangency portfolio using model-implied factors, following a purely out-of-sample procedure: for each month \(t\), we estimate the mean and covariance matrix of factor returns using data available up to time \(t\), and then track the realized return of the tangency portfolio in month \(t+1\).

Formally, let \(\bm{f}_t\) denote the vector of factor returns at time \(t\), and let \(\bm{\mu}_t\) and \(\bm{\Sigma}_t\) represent the sample mean vector and covariance matrix of \(\bm{f}_t\), estimated using data up to time \(t\). The weights of the tangency portfolio are given by:
\[
\bm{w}_t^{*} = c_t \bm{\Sigma}_t^{-1} \bm{\mu}_t,
\]
where \(c_t\) is a scalar that normalizes the portfolio's volatility. Following \cite{kelly2019characteristics}, we scale the weights \(c_t\) each period to make sure an ex-ante volatility of 1\% per month. Specifically, \(c_t\) is chosen such that the historical volatility of the tangency portfolio's returns, \(\{\bm{w}_t^{*'} \bm{f}_\tau\}_{\tau=1}^{t-1}\), equals 1\%. This scaling ensures comparability across models while preserving the tangency portfolio’s Sharpe-optimal composition. The realized return on the tangency portfolio in month \(t+1\) is then:
\[
R_{t+1}^{\text{TP}} = \bm{w}_t^{*'} \bm{f}_{t+1}.
\]

Table~\ref{Table: Perf_factors} reveals that not all factors extracted by the C-IPCA model command economically meaningful risk premia. As such, we implement a model selection procedure to retain only a subset of the most informative factors. We consider two approaches to factor selection.

\subsubsection{Ordered Model Selection.} \label{subsub:ordered_factor_selection}
Our first approach to model specification selects factors based on their out-of-sample Sharpe ratios in the training period. Following \cite{stambaugh2017mispricing}, we treat the market factor as a baseline and sequentially augment it with the top \(J-1\) C-IPCA factors ranked by Sharpe performance. This procedure yields a family of \(J\)-factor models, which we refer to as the \textit{Ordered C-IPCA} models (O-C-IPCA). We provide the methodology in this subsection and the Subsection \ref{subsec:ordered_cipca_results} shows the empirical results.

For the IC-based C-IPCA specification, we generate seven O-C-IPCA models corresponding to \(J = 1, 2, \ldots, 7\), reflecting the total number of interpretable factors identified in the IC clustering. Similarly, for the DC-based C-IPCA specification, we construct thirteen O-C-IPCA models for \(J = 1, 2, \ldots, 13\), based on the richer set of clusters derived from the data-driven clustering procedure.

To evaluate the incremental value of the O-C-IPCA model, we compare its performance against two benchmark models, each designed to isolate the role of (i) factor interpretability and clustering structure, and (ii) ex-ante Sharpe-based factor selection. Both benchmarks are constructed with the same number of factors as the O-C-IPCA model to ensure comparability in model complexity.

\textbf{Benchmark 1: Standard IPCA.} The first benchmark is the canonical IPCA model proposed by \cite{kelly2019characteristics}, estimated using the full set of firm characteristics without imposing any structural restrictions. For comparability, we retain the same number of factors \(J\) as in the corresponding O-C-IPCA specification. This benchmark serves as a baseline for evaluating the effectiveness of our two-step refinement—first imposing cluster-based interpretability constraints through the C-IPCA framework, and then selecting factors based on their Sharpe ratios to construct the O-C-IPCA model.

\textbf{Benchmark 2: Ordered IPCA (O-IPCA).} The second benchmark isolates the effect of Sharpe-based factor selection while holding the IPCA estimation method fixed. Specifically, we estimate a standard IPCA model with either 7 or 13 factors—matching the number of factors in the IC-IPCA and DC-IPCA models, respectively. We then construct an \textit{Ordered IPCA} (O-IPCA) model by selecting the 
top \(J\) factors.
based on their Sharpe ratios in the training sample
This yields a sequence of O-IPCA models indexed by \(J = 1, 2, \ldots, 7\) for comparison with the O-IC-IPCA models, and \(J = 1, 2, \ldots, 13\) for comparison with the O-DC-IPCA models.

In summary, to ensure fair comparisons across specifications, we construct O-IPCA benchmarks that are directly aligned with their C-IPCA counterparts. The \textit{O-IC-IPCA} model, which relies on 7 interpretable factors from intuitive clustering, is benchmarked against the \textit{O-IPCA7} model constructed from a 7-factor IPCA estimation. Similarly, the \textit{O-DC-IPCA} model, based on 13 data-driven interpretable factors, is benchmarked against the \textit{O-IPCA13} specification.

This benchmarking framework allows us to separately evaluate the economic value added by interpretable factor construction through clustering and the statistical value added by Sharpe-based factor selection within each modeling approach.



\subsubsection{Bayesian Model Selection}\label{subsec:bayesian_model_selection}
In contrast to the ordered factor selection approach discussed in Subsection~\ref{subsub:ordered_factor_selection}, which selects the top \(J\) factors with the highest Sharpe ratio from the training set to form the final factor model, this section introduces a Bayesian model selection method, the Bayesian C-IPCA (B-C-IPCA) model. This approach includes two variants: B-IC-IPCA, which selects models from the IC-IPCA framework where clusters are intuitive, and B-DC-IPCA, which selects models from the DC-IPCA framework where clusters are DC-based. The technical details of the model selection process are outlined here, while empirical results are presented in Subsection~\ref{subsec:emp_B_C_IPCA}.

\paragraph{Bayesian Model Framework.} We adopt the Bayesian framework developed by \cite{chib2020comparing} for model comparison across different factor sets. Specifically, consider a C-IPCA model with \(J\) factors (\(J = 7\) for IC-IPCA and \(J = 13\) for DC-IPCA). The \(J\) factors results in \(L = 2^{J} - 1\) candidate models, each corresponding to a subset of the \(J\) factors. As shown in \cite{chib2020comparing}, this framework enables the estimation of the posterior probability for each potential model, based on observed data \(D\). We select the top 10 models with the highest posterior probabilities. The basic idea of this framework is outlined below, with further details available in \cite{chib2020comparing}.

Formally, let \( M_1, M_2, \dots, M_L \) represent the \(L\) candidate models, where each model \( M_l \) consists of a subset of factors from the full set of C-IPCA factors. We begin by assigning each model an equal prior probability, reflecting a non-informative prior over the model space:
\[
\Pr(M_l) = \frac{1}{L}, \quad \forall l = 1, \dots, L.
\]
This assignment implies that, prior to observing the data, all models are equally likely.

\paragraph{Posterior Probability Calculation.} Upon observing the data \( D \), we update the prior using Bayes' theorem to compute the posterior probability of each model:
\[
\Pr(M_l | D) = \frac{\Pr(M_l) \Pr(D | M_l)}{\sum_{i=1}^{L} \Pr(M_i) \Pr(D | M_i)} \propto \Pr(D | M_l),
\]
where \( \Pr(M_l) \) is the prior probability of model \( M_l \), and \( \Pr(D | M_l) \) is the likelihood of the data given that model \( M_l \) is correct.

\paragraph{Likelihood of the Data.} The likelihood \( \Pr(D | M_l) \) represents the probability of observing the data \( D \) given that model \( M_l \) is the true model. The data \( D \) consists of two components: the C-IPCA factors and the test-assets returns. As noted in \cite{chib2020comparing}, the relative posterior probability between models with various subsets of factors is invariant to the choice of test-assets returns. For simplicity, we assume an empty set for the test-assets returns in this context.

For each model \( M_l \), let \(f_l\) denote the set of factors included in the model, and \(f_l^*\) the set of factors excluded from the model. The factors \( f_l \) included in the model are assumed to be linear combinations of the corresponding characteristics, with an intercept term \( \alpha_l \) and an error term \( \epsilon_l \):
\[
f_l = \alpha_l + \epsilon_l,
\]
where \( \epsilon_l \sim N(0, \Sigma_l) \) which can be estimated from the data. For the factors not included in model \( M_l \), we write
\begin{equation}\label{eq:factor}
f_l^* = \alpha_l^* + \beta_l^* f_l + \epsilon_l^*,
\end{equation}
where \( \epsilon_l^* \sim N(0, \Sigma_l^*) \). Following \cite{chib2020comparing}, if the factors \( f_l \) are mean-variance efficient, then \( \alpha_l^* = 0 \). The relative posterior probability across all models that include a subset of factors is proportional to the probability of explaining \( f_l^* \) given the factors \( f_l \) included in the model.

The prior distribution for the intercept \( \alpha_l \) is assumed to be normal:
\[
\alpha_l | \Sigma_l \sim N(\alpha_{l0}, k_l \Sigma_l),
\]
where \( \alpha_{l0} \) will be estimated from the data, and \( k_l \) is a scaling factor related to the maximum achievable squared Sharpe ratio in the market:
\[
k_l = \frac{\text{Sh}^2_{\text{max}}}{J},
\]
where \( J \) is the number of factors in model \( M_l \), and \( \text{Sh}^2_{\text{max}} \) is the maximum achievable squared Sharpe ratio.

\paragraph{Estimation of Parameters.} To estimate the parameters \( \alpha_{l0} \), \( \Sigma_l \), and \( k_l \), we split the training sample into two sub-samples. The first sub-sample consists of the first \( tr \times T \) months, where \( T \) is the total length of the training sample and \( tr \) is the proportion used for prior estimation. This sub-sample is used to estimate the prior parameters \( \alpha_{l0} \) and \( \Sigma_l \), with \( \hat \alpha_l \) and \( \hat \Sigma_l \) representing the mean and standard deviation of the factors. The second sub-sample consists of the remaining months, used to compute the posterior probability \( \Pr(M_l | D) \).

\paragraph{Model Selection.} For each model \( M_l \), based on a subset of C-IPCA factors, we calculate the posterior probability \( \Pr(M_l | D) \) following \cite{chib2020comparing}. The 10 models with the highest posterior probabilities are selected to form the B-C-IPCA model.

\paragraph{Benchmark Models.} To evaluate the relative performance of the C-IPCA model, we compare it against two benchmark models, both derived from IPCA factors but employing different model selection methods.

\textbf{Benchmark 1: B-IPCA Models (Method of \cite{chib2024winners}).} The first benchmark follows the same Bayesian model selection approach as B-C-IPCA. We begin by estimating an IPCA model with 7 or 13 factors, corresponding to the IC-IPCA and DC-IPCA models, respectively. We then evaluate all possible subsets of IPCA factors based on their posterior probabilities and select the top 10 models. These models are referred to as B-IPCA models, using the method outlined by \cite{chib2024winners}.

\textbf{Benchmark 2: B-IPCA Models (Method of \cite{kelly2019characteristics}).} The second benchmark adopts the approach used by \cite{kelly2019characteristics}, where the number of factors is treated as a hyperparameter. This method estimates a series of IPCA models with various numbers of factors, ranging from 1 to 13, and reports the Sharpe ratio for each model. These models are referred to as KPS-IPCA models. Although this method is not part of the Bayesian framework, it provides a useful comparison by treating the number of factors as a hyperparameter.

\subsection{Performance of Ordered C-IPCA Models}
\label{subsec:ordered_cipca_results}

This subsection presents the empirical results on the performance of C-IPCA models using a subset of factors selected according to the ordered factor selection procedure described in Subsection~\ref{subsub:ordered_factor_selection}. Specifically, we evaluate the out-of-sample Sharpe performance of tangency portfolios constructed from the top \(J-1\) C-IPCA factors—ranked by their Sharpe ratios in the training sample—together with the market factor.

Table~\ref{Table: MaxPort_ordered} reports the annualized Sharpe ratios of these tangency portfolios for both the O-IC-IPCA and O-DC-IPCA specifications, across different values of \(J\). For each model, we also provide results for two benchmark specifications: the standard IPCA model and the O-IPCA model with the same number of factors. The final column of the table indicates the economic clusters associated with the selected C-IPCA factors, offering interpretability of the underlying factor structure.

Panel A reports how the Sharpe ratios of the O-IC-IPCA models evolve as the number of included factors increases, using the top \(J\) factors ranked by their Sharpe ratios in the training sample. For comparison, we also include results for the standard IPCA models with the same number of factors, as well as for the O-IPCA7 model, constructed from a 7-factor IPCA model using the same factor-selection criteria as the O-IC-IPCA.

Panel B provides analogous results for the O-DC-IPCA specification, reporting Sharpe ratios as a function of \(J\), alongside the corresponding IPCA and O-IPCA13 benchmarks. This design enables a direct assessment of both the value added by clustering (IC vs. DC) and the efficacy of factor selection.

There are several things worth mentioning via comparing the Panels A and B. \textit{First}, Panel A demonstrates that the O-IC-IPCA model generally underperforms both the standard IPCA and O-IPCA models. Except for the smallest model sizes ($J = 1$ and $J = 2$), the O-IC-IPCA consistently exhibits lower Sharpe ratios. For instance, at $J = 3$, the Sharpe ratio of the O-IC-IPCA model is 0.67, compared to 0.86 for IPCA and 0.83 for O-IPCA—gaps of 0.19 and 0.16, respectively. The performance deficit is most pronounced at $J = 6$, where the O-IC-IPCA’s Sharpe ratio is 0.77 versus 1.19 for IPCA (gap of 0.42), and at $J = 5$, with a gap of 0.45 relative to O-IPCA. One likely explanation for this underperformance is the rigidity introduced by the intuitive clustering constraints: although economically motivated, such constraints may miss important cross-sectional correlations among firm characteristics that the data-driven clustering method is better suited to capture.

\vspace{0.3em}
 \textit{Second}, Panel B shows that the O-DC-IPCA model consistently and significantly outperforms its O-IC-IPCA counterpart across all values of \(J\). This performance dominance underscores the value of leveraging data-driven clustering to extract more accurate and informative groupings of characteristics. The DC clustering method captures latent correlation structures that are essential for identifying priced factors, thereby enhancing both model precision and empirical performance.

\vspace{0.3em}
\textit{Third}, focusing on the within-model performance of the O-DC-IPCA specification, we observe that the Sharpe ratio of the tangency portfolio improves steadily with \(J\) until peaking at \(J = 7\), beyond which performance plateaus. This pattern suggests that the first seven factors contain the majority of pricing-relevant information, and adding more factors yields diminishing returns. This aligns with results in Table~\ref{Table: Perf_factors}, which shows that only a subset of the factors have substantial risk premia, while others contribute little to cross-sectional pricing.

\vspace{0.3em}
 \textit{Finally}, contrasting the O-DC-IPCA model with its benchmarks provides further insights. Across nearly all model sizes \(J = 1,2,\ldots,13\), the O-DC-IPCA outperforms the O-IPCA13 model, although the margin of improvement narrows as \(J\) increases. This suggests that the DC-IPCA framework not only improves interpretability but also delivers stronger pricing performance than a similarly specified IPCA model. Similarly, the O-DC-IPCA model generally outperforms the standard IPCA benchmark, except for a marginal underperformance at \(J = 11\).

\vspace{0.3em}
\noindent Taken together, these findings highlight the empirical and practical value of combining economic structure with statistical discipline. The data-driven C-IPCA framework offers a compelling approach to constructing interpretable factor models that are also empirically efficient.

\subsection{Performance of Bayesian C-IPCA models}\label{subsec:emp_B_C_IPCA}
This subsection presents the empirical results comparing model performance based on the Bayesian model selection procedure described in \ref{subsec:bayesian_model_selection}.

\begin{figure}[htbp] 
    \begin{subfigure}[b]{1.0\textwidth} 
        \centering
        \includegraphics[width=\textwidth]{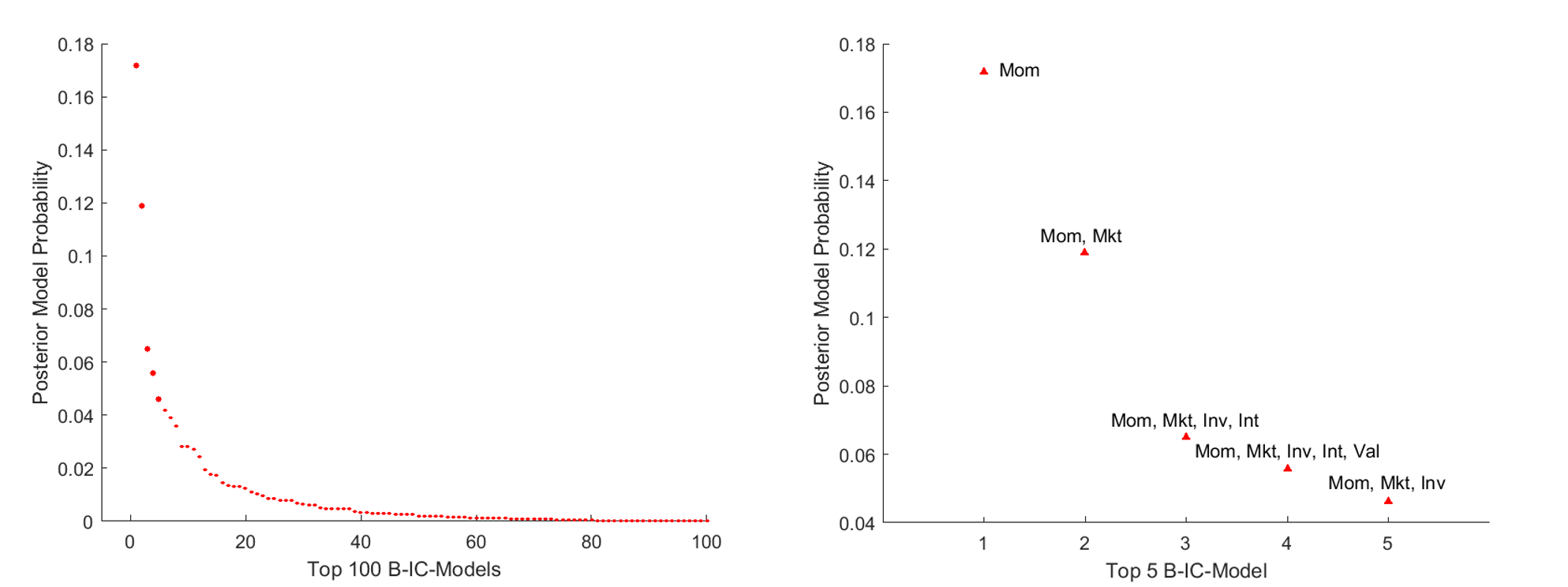} 
        \caption{B-IC-IPCA models}
        \label{fig:subfig1}
    \end{subfigure}

    \vspace{1em} 
    
    \begin{subfigure}[b]{1.0\textwidth} 
        \centering
        \includegraphics[width=\textwidth]{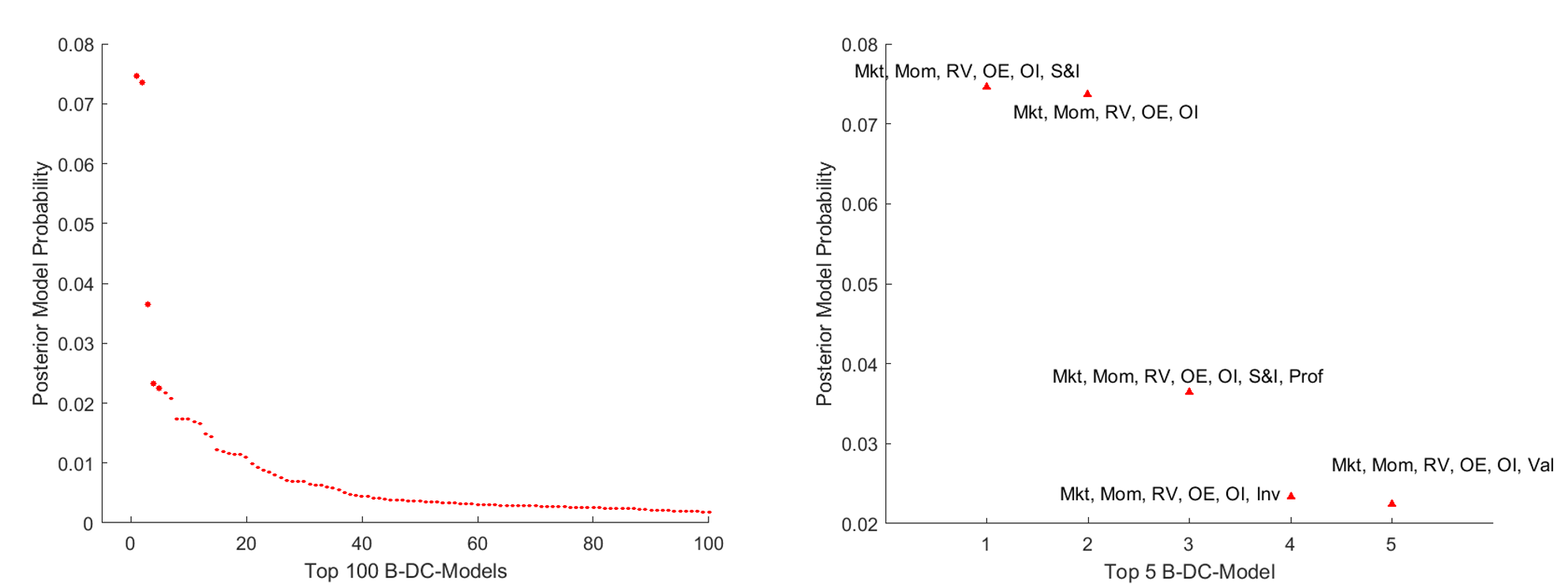} 
        \caption{B-DC-IPCA models}
        \label{fig:subfig2}
    \end{subfigure}
\caption{\textbf{Posterior Probability of B-C-IPCA Models.} This figure displays the posterior probabilities of B-C-IPCA models. For all candidate models—each representing a subset of factors from a given C-IPCA specification—we compute and rank posterior probabilities. The ranked probabilities are shown. Subfigures (a) and (b) correspond to two versions of B-C-IPCA models based on different clustering schemes: B-IC-IPCA and B-DC-IPCA, respectively. In each subfigure, the left panel reports the most likely 100 B-C-IPCA models, where the x-axis indicates the posterior probability rank and the y-axis shows the posterior probability. The right panel highlights the most likely five B-C-IPCA models along with their associated clusters. (See Figure~\ref{Fig: DC and IC} for cluster abbreviations; “Mkt” denotes the zero-correlation market factor (ZC).)}
    \label{Fig: Pr_C}
\end{figure}

Figure \ref{Fig: Pr_C} displays the posterior probabilities of the top 100 models with the highest posterior probabilities. Panel A of Figure \ref{Fig: Pr_C} presents results for models based on the intuitive clusters. Within Panel A, the left subfigure shows the posterior probabilities of the top 100 models, while the right subfigure depicts the associated models. For clarity, we only display the top 5 most likely models in the right subfigure. Among the \(2^7 - 1\) possible models, the most likely 5 models include (1) only one factor Momentum, (2) Momentum and Market, (3) Momentum, Market, Investment, and Intangible, (4) Momentum, Market, Investment, Intangible, and Value,  and (5) Momentum, Market, and Investment.

Panel B of Figure \ref{Fig: Pr_C} shows the results for models based on DC clusters. The left subfigure shows the posterior probabilities of the top 100 models, while the right subfigure highlights the top 5 models. Models include (1) Market, Momentum, Return Volatility, Operating Efficiency, Operating Illiquidity, and Size \& Illiquidity, (2)Market, Momentum, Return Volatility, Operating Efficiency, and Operating Illiquidity, (3)Market, Momentum, Return Volatility, Operating Efficiency, Operating Illiquidity, Size \& Illiquidity, and Profitability, (4)Market, Momentum, Return Volatility, Operating Efficiency, Operating Illiquidity, Investment, and (5)Market, Momentum, Return Volatility, Operating Efficiency, Operating Illiquidity, Value. 

It is important to emphasize that the left subfigures in both Panel A and Panel B show that the most likely models exhibit much higher posterior probabilities than the remaining models. For instance, in the B-IC-IPCA models, the top model delivers a posterior probability of 18\%, followed by 12\%, 6.5\%, 5.8\%, and 4.2\% for the top 5 models. Similarly, in the B-DC-IPCA models, the top two models deliver posterior probabilities of approximately 7.5\% and 7.4\%, which is significantly high considering that there are over 8,000 models. The following models have probabilities of 3.7\%, 2.4\%, and 2.3\%. Overall, the posterior distribution indicates that the posterior probability is highly concentrated among the top models, thereby lending confidence to the selection of these models.

A natural and interesting question is the performance of the most likely models. Table \ref{Table: MaxPort_best} shows the performance, measured as the annualized Sharpe ratio of the tangency portfolio for each model. Specifically, Panel A in Table \ref{Table: MaxPort_best} presents the performance of the top 10 models based on the DC and IC clusters, respectively. As a comparison, Panel B shows results for the corresponding IPCA models, estimated as discussed in Section \ref{subsec:bayesian_model_selection}.  We begin by estimating IPCA models with 7 and 13 factors, respectively. We then evaluate and rank these models using the Bayesian framework, selecting the top 10 most likely models.
Panel C  reports the performance of KPS-IPCA models, constructed following the method of \cite{kelly2019characteristics}. The first row presents the Sharpe ratios, while the second row shows the number of factors included in each model.

There are several key points to note. First, when comparing the performance of the tangency portfolios between B-IC-IPCA and B-DC-IPCA, we observe that the B-DC-IPCA models consistently deliver higher Sharpe ratios. Specifically, excluding the sixth model, the Sharpe ratios of the B-DC-IPCA models are all above 1.4, whereas the Sharpe ratios of the B-IC-IPCA models remain below 0.80. This evidence suggests that incorporating data-driven information significantly boosts performance without compromising model interpretability.

When comparing the most likely models derived from the IPCA, we find that B-DC-IPCA continues to deliver much higher Sharpe ratios than the IPCA models, which do not impose any restrictions on the characteristics coefficients of factor exposures.

Finally, when comparing KPS-IPCA, which simply treats the number of factors as a hyperparameter and estimates models with varying numbers of factors, we still find that our B-DC-IPCA model delivers comparable or superior out-of-sample Sharpe ratios. In most cases, our model outperforms the KPS-IPCA model, except in one instance with 11 factors (where the KPS-IPCA model achieves a Sharpe ratio of 1.51).
\footnote{{One important point  is that \cite{kelly2019characteristics} examines IPCA models with the number of factors \(J=1,2,...,6\), and find that the model with \(J=5\)  achieves the highest Sharpe ratio. In contrast, in Panel B of our Table \ref{Table: MaxPort_best}, the IPCA model with \(J=5\) rank only 10 in terms of Sharpe ratio. The main reason for this discrepancy lies in the weighting scheme used during estimation. As shown in Equation (\ref{Eq:obj_fun}), we employ value weights in estimating the IPCA model, whereas \cite{kelly2019characteristics} uses equal weights. Table \ref{Table: MaxPort_best_EW} in Appendix \ref{app: EW} reports the results based on equal weighting using our dataset. In the testing sample, the Sharpe ratio of the \(J=5\) IPCA model under equal weights is 1.67, smaller only than the models with \(J=8, 9, 10\), which yield Sharpe ratios of 1.78, 1.86, and 1.93, respectively. Among the models with \(J=1,2,...,6\), the \(J=5\) model achieves the highest Sharpe ratio, consistent with the findings of \cite{kelly2019characteristics}.}}
As mentioned, the KPS-IPCA model faces challenges in interpretability.

\section{Mechanism and Robustness}
\label{sec: Machanism_and_robustness}
Sections \ref{subsec:ordered_cipca_results} and \ref{subsec:emp_B_C_IPCA} demonstrate that the DC-IPCA models perform at least as well as and even better than the IPCA models no matter whether we use the ordered model selection or the Bayesian model selection to select a subset of factors (see details in Subsection~\ref{subsub:ordered_factor_selection} and \ref{subsec:bayesian_model_selection}). This section delves into the underlying mechanisms driving this model performance.

There are several factors which could boost the out-performance of our DC-IPCA models: information implied by the data, domain knowledge, and the parameter restriction. We discuss each of them in detail.

\subsection{Value of Data} Our DC clustering method combines both the domain knowledge (using the intuitive cluster as the starting point to make an initial partition as discussed in Section~\ref{subsec:CIPCA}) and the characteristics similarity information implied by the data. The comparison between the DC-IPCA and IC-IPCA models indicates that our DC-IPCA model consistently and significantly out-performs that of IC-IPCA in the out-of-sample Sharpe ratio, which indicates that beyond the domain knowledge, the information implied by the data provides valuable insights for the clustering.

\subsection{Value of Domain Knowledge}
Another potential explanation for the out-performance of our DC-IPCA method is the domain knowledge implied by the initial intuitive clusters. To evaluate the role of economic intuition, we construct alternative clusters using a purely data-driven approach - referred to as Purely Data-Driven Clusters (PDC). We compare the performance of PDC-IPCA, which is based solely on data-derived clusters, to DC-IPCA, which combines both economic intuition and data. This comparison allows us to assess the marginal value of incorporating economic information into the clustering process.

Specifically, the PDC is constructed by removing the economic information incorporated in the DC method, relying solely on data-derived clusters. In the DC method, economic information is incorporated by requiring that, within each sub-cluster, formed by splitting the set of firm characteristics, the characteristics must share the same economic explanation, or equivalently, belong to the same IC cluster. These economically coherent sub-clusters are then merged to form the final DC clusters in Section \ref{subsec:cluster}.  In contrast, PDC removes the requirement that characteristics in each sub-cluster is within the same IC cluster. We obtain sub-clusters by partitions the whole set of firm characteristics. This comparison allows us to assess the marginal value of incorporating economic information into the clustering process.

The optimal hyper-parameters for PDC are \(m =19, knn=85, K=10\). This means that PDC has 10 clusters and the corresponding PDC-IPCA model is constructed following Section \ref{subsec:CIPCA}, with 11 factors (10 factors corresponding to 10 clusters and 1 ZC or market factor). To compare the model performance between PDC-IPCA and DC-IPCA, we follow the procedures outlined in Sections \ref{subsub:ordered_factor_selection} and \ref{subsec:bayesian_model_selection} for factor selection to find the mean-variance efficiency portfolio which may only include a subsect of factors. Specifically, we evaluate the out-of-sample Sharpe performance of tangency portfolios constructed from a subset of factors selected according to the ordered factor selection procedure or Bayesian model selection procedure, respectively.

Panel A of Table \ref{Table: Mechanism} presents the model performance of the PDC-IPCA. The first two columns report the performance of tangency portfolios based on the ordered factor selection approach. Specifically, factors are first ranked by their Sharpe ratio in the training sample, and then gradually included based on their Sharpe ratio to construct \(J\)-factor O-PDC-IPCA models. The first column indicates the number of factors included, while the second column shows the Sharpe ratio of the tangency portfolios associated with the corresponding \(J\)-factor model.

The right panel (B-C-IPCA) in Panel A displays the Sharpe ratios of the tangency portfolios based on the Bayesian model selection procedure (the B-PDC-IPCA model). This procedure selects the 11 models with the highest posterior probability, ranking them by likelihood. The final column reports the number of factors in each model.

There are several important findings worth mentioning in Panel A. First, the Sharpe ratios of the O-PDC-IPCA models are generally lower than those of the O-DC-IPCA models (see Table \ref{Table: MaxPort_ordered}). Except for the cases where \(J=1\) or \(J=4\), the O-DC-IPCA model consistently delivers higher Sharpe ratios. For instance, when \(J=2\), the Sharpe ratio of the O-PDC-IPCA model is 0.74, which is 0.11 lower than the 0.85 achieved by the O-DC-IPCA model. As \(J\) increases beyond 4, the gap in Sharpe ratios between the O-PDC-IPCA and O-DC-IPCA models becomes even more pronounced. The highest Sharpe ratio for the O-PDC-IPCA model is 1.09 (when \(J=11\)), whereas the O-DC-IPCA model attains Sharpe ratios exceeding 1.4 when \(J\geq4\).

Second, similar trends are observed when comparing the Sharpe ratios based on the Bayesian model selection procedure. The Sharpe ratios of the B-PDC-IPCA models are consistently lower than those of the B-DC-IPCA models (see Panel A in Table \ref{Table: MaxPort_best}). Specifically, among all B-PDC-IPCA models, the highest Sharpe ratio is 1.07 (Rank = 5). In contrast, the lowest Sharpe ratio among the B-DC-IPCA models is 1.09 (Rank = 6), which is still higher. Moreover, the second lowest Sharpe ratio from the B-DC-IPCA models is 1.41 (Rank = 9), significantly higher than the highest Sharpe ratio from the B-PDC-IPCA models (1.07).

Taken together, these results indicate that models based on PDC consistently underperform those based on DC, underscoring the significant role that economic information plays in enhancing the performance of the DC-IPCA model.

\subsection{Coefficient Restrictions}

We have demonstrated that both data and economic intuition contribute to the out-performance of the DC-IPCA model. However, one might argue that this improved performance is not solely due to the clustering structure itself but rather to the reduced number of parameters estimated. The risk exposure of the associated factor has non-q-fac coefficients on characteristics within the cluster. Specifically, the DC-IPCA model is constructed by imposing parameter restrictions derived from DC, effectively setting some parameters to zero and reducing the number of parameters to be estimated.

To assess whether the out-performance of the DC-IPCA model comes from its clustering structure rather than merely the smaller number of parameters, we employ a placebo approach. Specifically, we generate random clusterings with the same number of clusters as the DC-IPCA model and construct a corresponding factor model denoted as RC-IPCA. RC is generated without any reference to data or economic intuition, thus carrying no informational content implied by the data and domain knowledge. Instead, it takes a clustering structure that allows non-zero coefficients on characteristics within the cluster. As a result, the RC-IPCA model benefits from a similarly reduced number of parameters but lacks any meaningful clustering structure implied by the data and domain knowledge. If the out-performance of the DC-IPCA model is indeed attributable to its clustering structure driven by the data and intuition rather than parameter sparsity alone, we would expect the RC-IPCA to deliver a much inferior performance to the DC-IPCA model.

For each RC, we use the ordered and Bayesian model selection to select a subset of factors as in \ref{subsub:ordered_factor_selection} and \ref{subsec:bayesian_model_selection}. For our analysis, to mitigate concerns about the randomness inherent in the RC-IPCA  affecting our results, we bootstrap 100 different RC-IPCAs.

Panel B of Table \ref{Table: Mechanism} presents the results. The first three columns report the performance of tangency portfolios constructed using an ordered model selection approach, where factors are progressively added based on their in-sample Sharpe ratios (the O-RC-IPCA model). The first column indicates the number of factors included, while the second and third columns display the mean and median Sharpe ratios, respectively, derived from 100 bootstrap samples. The right panel of Panel B presents the results from the Bayesian model selection procedure (hereafter referred to as B-RC-IPCA). Models are ranked according to their posterior likelihood, with the 13 most likely models shown. The "Rank" column represents the rank of each model based on its posterior probability. The columns "Sharpe - mean" and "Sharpe - median" report the average and median Sharpe ratios, respectively, from the 100 bootstrap samples. The "J - mean" and "J - median" columns show the average and median number of factors, respectively, across the 100 independently generated B-RC-IPCA models.

There are several things worth mentioning in Panel B. First, generally speaking, the Sharpe ratios of the O-RC-IPCA models are lower than those of the O-DC-IPCA models (see Table \ref{Table: MaxPort_ordered}). Specifically, with the exception of $J=1$ and $J=4$, the O-DC-IPCA model consistently outperforms the O-RC-IPCA model in terms of Sharpe ratios. For instance, at $J=2$, the O-RC-IPCA model yields an average and median Sharpe ratio of 0.66, which is 0.19 points below the 0.85 achieved by the O-DC-IPCA model. As $J$ exceeds 4, the performance disparity between the O-RC-IPCA and O-DC-IPCA models becomes even more pronounced. Although the O-RC-IPCA model reaches its highest average and median Sharpe ratios of 1.19 (at $J=12$) and 1.21 (at $J=13$), respectively, the O-DC-IPCA model surpasses 1.4 in both metrics once the number of factors exceeds 4.

Second, similar patterns emerge when comparing Sharpe ratios based on subsets of factors selected via the Bayesian procedure. The Sharpe ratios of the B-RC-IPCA models are consistently lower than those of the B-DC-IPCA models (see Table \ref{Table: MaxPort_best}). Regardless of the rank of the B-RC-IPCA model, its average (or median) Sharpe ratio never exceeds 0.6. In contrast, the lowest Sharpe ratio of the B-DC-IPCA model is 1.09, and the second-lowest is 1.41 — both significantly higher than the average level achieved by the B-RC-IPCA models.

Overall, models based on DC consistently outperform those based on RC, suggesting that the out-performance of the DC-IPCA model is due to its clustering structure rather than parameter sparsity alone.

\subsection{Robustness test}
\subsubsection{Training Sample}
Our baseline analysis used the first 180 months as the training sample. To assess robustness, we extend the training window to 240, 300, 360, and 420 months. Table~\ref{Table: O_C_IPCA_Train} reports the out-of-sample performance of O-DC-IPCA models alongside two benchmarks: (i) the IPCA model with the same number of factors and (ii) the O-IPCA models. For each training length, models are estimated in-sample, and tangency portfolio Sharpe ratios are computed out-of-sample. The first column lists the number of factors (\(J\)), and the remaining columns report annualized Sharpe ratios for the three model classes. The accompanying figure summarizes these results: the horizontal axis shows the number of factors, the vertical axis shows annualized Sharpe ratios, and bars with different colors correspond to different models.

Table~\ref{Table: O_C_IPCA_Train} indicates that O-DC-IPCA generally outperforms both benchmarks. With a 240-month training sample, O-DC-IPCA consistently delivers higher Sharpe ratios than O-IPCA across all factor dimensions. Its performance relative to IPCA depends on model size: for larger \(J\), O-DC-IPCA dominates, whereas for smaller \(J\) the difference is minimal. For example, with \(J=2\), O-DC-IPCA achieves a Sharpe ratio of 0.65 versus 0.70 for IPCA. Results for the 300-month sample are similar. As the training window expands to 360 or 420 months, the advantage of O-DC-IPCA becomes more pronounced: across all factor counts, O-DC-IPCA achieves higher Sharpe ratios than both O-IPCA and IPCA.

\begin{landscape} 
\begin{figure}[htbp]
\centering
\includegraphics[height=0.5\textheight,width=1.3\textwidth]{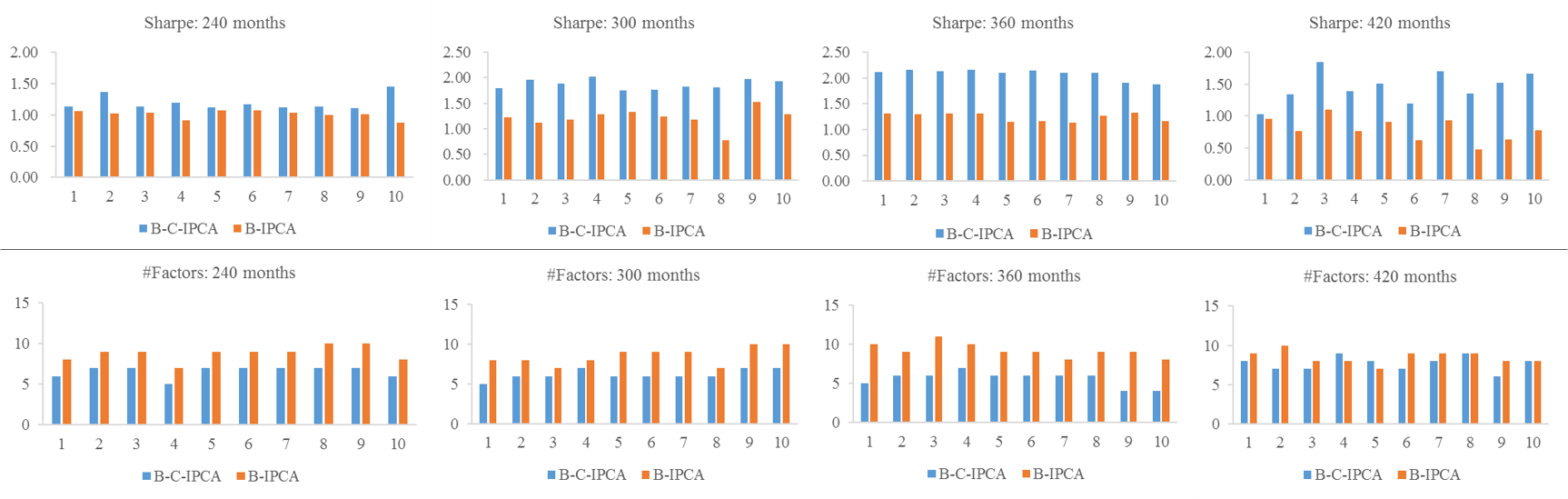} 
\captionsetup{justification=justified, singlelinecheck=false} 
\caption{
\textbf{Different Training Samples: Mean-Variance Efficiency of the B-C-IPCA and B-IPCA Models.}
This figure presents the out-of-sample Sharpe ratios of tangency portfolios for the top 10 B-C-IPCA models across varying training sample lengths. The horizontal axis indicates model rank. In the first row of subfigures, the vertical axis reports annualized Sharpe ratios, while in the second row, it shows the number of factors in each model. Bars with different colors represent different models.
}
\label{Figure: B_C_IPCA_Train}
\end{figure}
\end{landscape}

Table~\ref{Table: O_B_IPCA_Train} and Figure~\ref{Figure: B_C_IPCA_Train} report the performance of the top 10 B-DC-IPCA models with the highest posterior probabilities and compare them with the B-IPCA benchmarks. Models are estimated using different training sample lengths, and we evaluate the out-of-sample tangency portfolio Sharpe ratios in the corresponding testing periods.

In each panel of Table~\ref{Table: O_B_IPCA_Train}, models are ranked by posterior probability. The first column reports the model rank, the second and third columns report the annualized Sharpe ratios, and the fourth and fifth columns report the number of factors in each model. Figure~\ref{Figure: B_C_IPCA_Train} provides a graphical representation of the same results. The horizontal axis denotes model rank. In the top row of subfigures, the vertical axis represents annualized Sharpe ratios, while in the bottom row it represents the number of factors. Bars with different colors correspond to different models.

Both Table~\ref{Table: O_B_IPCA_Train} and Figure~\ref{Figure: B_C_IPCA_Train} indicate that, for the same posterior probability rank, B-DC-IPCA models achieve higher Sharpe ratios than B-IPCA models while using fewer factors. This suggests that B-DC-IPCA delivers superior predictive performance with a more parsimonious model structure.

Table~\ref{Table: MF_Train} examines the performance of interpretable factors in DC-IPCA models across different training samples. The first column reports the factor IDs (MF), the second column lists annualized Sharpe ratios, and the third column shows the associated clusters. An asterisk (*) indicates that the factor is included in the top B-DC-IPCA model as selected by \cite{chib2024winners}. Across all training samples, momentum-related factors—both fundamental momentum and return momentum—are consistently included in the top model with highest posterior probability. Factor selection varies with the training sample length: for the first 240 months, return volatility, growth of equity, and market beta are selected; for 300 months, R\&D and illiquidity; for 360 months, operating illiquidity and return volatility; and for 420 months, market beta, profitability, size, financial stability, growth, and value.

\subsubsection{Model with Equal Weights}
\label{subsec: EW}
In the previous sections, we estimate the C-IPCA model by minimizing the \emph{value-weighted} mean squared error, as defined in Equation~\eqref{Eq:obj_fun}, to mitigate the dominance of micro-cap stocks, as suggested by \cite{hou2020replicating}. Under this specification, the estimated model factors correspond to the returns of value-weighted portfolios, as shown in Equation~\eqref{Eq:f_hat}.

To assess the robustness of our results with respect to portfolio weighting schemes, we consider an alternative specification in which the model is estimated by minimizing the \emph{equal-weighted} mean squared error. The empirical results for this specification are reported in Appendix~\ref{app: EW}. The key message is that our main conclusions remain unchanged: combining economic intuition with data-driven clustering significantly enhances both model performance and interpretability. The objective function for the equal-weighted case is given by:
\begin{equation}
\min_{\bm{\Gamma}, \{\bm{f}_t\}_{t=1}^T} \sum_{t=1}^T \left( \bm{r}_t - \bm{X}_{t-1} \bm{\Gamma} \bm{f}_t \right)' \left( \bm{r}_t - \bm{X}_{t-1} \bm{\Gamma} \bm{f}_t \right),
\label{Eq: obj_fun_EW}
\end{equation}
where all notations follow those in Equation~\eqref{Eq:obj_fun}. The resulting model factors correspond to the returns of equal-weighted portfolios, as expressed in:
\begin{equation}
\hat{\bm{f}}_t = \left( \hat{\bm{\Gamma}}' \bm{X}_{t-1}' \bm{X}_{t-1} \hat{\bm{\Gamma}} \right)^{-1} \hat{\bm{\Gamma}}' \bm{X}_{t-1}' \bm{r}_t,
\label{Eq:f_hat_EW}
\end{equation}
where notations are consistent with those in Equation~\eqref{Eq:f_hat}.

To mitigate the disproportionate influence of low-priced stocks, we exclude observations with closing prices below \$5 when constructing equal-weighted portfolios, as such stocks tend to exhibit high volatility and limited liquidity, which can distort factor estimation. To maintain consistency between factor estimation and clustering, we also compute characteristic similarity using equal-weighted correlations during the clustering stage:
\begin{equation}
    \rho_{ij} = \frac{1}{T} \sum_{t=1}^{T} \frac{
\sum_{n=1}^{N_t}(x_{it}^n - \bar{x}_{it})(x_{jt}^n - \bar{x}_{jt})
}{
\sqrt{\sum_{n=1}^{N_t} (x_{it}^n - \bar{x}_{it})^2}
\sqrt{\sum_{n=1}^{N_t} (x_{jt}^n - \bar{x}_{jt})^2}
},
\end{equation}
where \(\bar{x}_{jt} = \frac{1}{N_t} \sum_{n=1}^{N_t}x_{jt}^n\), and all other notations follow those in Equation~\eqref{Eq:s_D}. All other methodological procedures remain identical to those described in the main text.


\section{Conclusion}
\label{sec: conclusion}
A vast literature documents numerous firm characteristics that explain the cross-section of stock returns. Treating each characteristic as a separate factor, however, leads to severe over-identification. Existing approaches—based on latent factor models or machine learning methods—reduce dimensionality by constructing factors as linear combinations of all characteristics. While these methods deliver a plausible number of factors, the resulting factors often lack clear economic interpretation.

This paper introduces a new framework that achieves both parsimony and interpretability in factor construction. Our approach proceeds in two steps. First, we cluster firm characteristics by combining economic intuition with data-driven similarity, producing statistically and economically coherent groups. Second, we extract a representative factor from each cluster that captures its underlying economic driving force. Applying this method to 94 characteristics from \cite{gu2020empirical}, we identify at most nine factors, each with a clear economic interpretation.

We further extend the influential IPCA model of \cite{kelly2019characteristics} by incorporating clustering structure, yielding the C-IPCA model.
Clusters can be pre-specified based on economic theory or formed adaptively from the data. Empirically, C-IPCA retains the strong pricing performance
of IPCA while delivering substantial gains in interpretability. Compared to the five latent factors in the original IPCA, our approach identifies
 factors such as illiquidity, long-run momentum, short-term reversal, investment, and market exposure—factors that not only perform as well or better,
  but are also economically transparent. Moreover, our interpretable factors can explain the IPCA factors
   in a standard regression, whereas the reverse does not hold. These results underscore that combining economic structure with data-driven clustering can significantly improve both predictive performance and interpretability.

Because factor models are foundational to asset pricing, portfolio management, and risk analysis, the proposed framework has broad implications. By bridging the gap between statistical efficiency and economic interpretability, our approach offers a practical and theoretically grounded tool for understanding the fundamental sources of cross-sectional variation in expected returns.

\clearpage
\newgeometry{top=3cm, bottom=3cm, left=2cm, right=2cm}
\setstretch{1.5} 

\setlength{\bibsep}{0pt plus 0.3ex}
\makeatletter
\patchcmd{\thebibliography}
  {\settowidth}
  {\setlength{\itemsep}{0.5\baselineskip} \setlength{\parskip}{0pt} \settowidth}
  {}{}
\makeatother

\bibliographystyle{elsarticle-harv}

\begin{thebibliography}{}

\bibitem[Avramov et~al., 2023]{avramov2023integrating}
Avramov, D., Cheng, S., Metzker, L., and Voigt, S. (2023).
\newblock Integrating factor models.
\newblock {\em The Journal of Finance}, 78(3):1593--1646.

\bibitem[Barton et~al., 2019]{barton2019chameleon}
Barton, T., Bruna, T., and Kordik, P. (2019).
\newblock Chameleon 2: an improved graph-based clustering algorithm.
\newblock {\em ACM Transactions on Knowledge Discovery from Data (TKDD)}, 13(1):1--27.

\bibitem[Borg, 2011]{borg2011multidimensional}
Borg, I. (2011).
\newblock Multidimensional scaling.
\newblock In {\em International encyclopedia of statistical science}, pages 875--878. Springer.

\bibitem[B{\"u}chner and Kelly, 2022]{buchner2022factor}
B{\"u}chner, M. and Kelly, B. (2022).
\newblock A factor model for option returns.
\newblock {\em Journal of Financial Economics}, 143(3):1140--1161.

\bibitem[Chen and Zimmermann, 2022]{chen2022publication}
Chen, A.~Y. and Zimmermann, T. (2022).
\newblock Publication bias in asset pricing research.
\newblock {\em arXiv preprint arXiv:2209.13623}.

\bibitem[Chen et~al., 2024]{chen2024deep}
Chen, L., Pelger, M., and Zhu, J. (2024).
\newblock Deep learning in asset pricing.
\newblock {\em Management Science}, 70(2):714--750.

\bibitem[Chib et~al., 2020]{chib2020comparing}
Chib, S., Zeng, X., and Zhao, L. (2020).
\newblock On comparing asset pricing models.
\newblock {\em The Journal of Finance}, 75(1):551--577.

\bibitem[Chib et~al., 2024]{chib2024winners}
Chib, S., Zhao, L., and Zhou, G. (2024).
\newblock Winners from winners: A tale of risk factors.
\newblock {\em Management Science}, 70(1):396--414.

\bibitem[Chordia et~al., 2020]{chordia2020anomalies}
Chordia, T., Goyal, A., and Saretto, A. (2020).
\newblock Anomalies and false rejections.
\newblock {\em The Review of Financial Studies}, 33(5):2134--2179.

\bibitem[Cochrane, 2011]{cochrane2011presidential}
Cochrane, J.~H. (2011).
\newblock Presidential address: Discount rates.
\newblock {\em The Journal of finance}, 66(4):1047--1108.

\bibitem[Cong et~al., 2024]{cong2024textual}
Cong, L.~W., Liang, T., Zhang, X., and Zhu, W. (2024).
\newblock Textual factors: A scalable, interpretable, and data-driven approach to analyzing unstructured information.
\newblock Technical report, National Bureau of Economic Research.

\bibitem[Cong et~al., 2021]{cong2021alphaportfolio}
Cong, L.~W., Tang, K., Wang, J., and Zhang, Y. (2021).
\newblock Alphaportfolio: Direct construction through deep reinforcement learning and interpretable ai.
\newblock {\em Available at SSRN}, 3554486.

\bibitem[Connor and Korajczyk, 1986]{connor1986performance}
Connor, G. and Korajczyk, R.~A. (1986).
\newblock Performance measurement with the arbitrage pricing theory: A new framework for analysis.
\newblock {\em Journal of financial economics}, 15(3):373--394.

\bibitem[Daniel and Moskowitz, 2016]{daniel2016momentum}
Daniel, K. and Moskowitz, T.~J. (2016).
\newblock Momentum crashes.
\newblock {\em Journal of Financial economics}, 122(2):221--247.

\bibitem[Daniel et~al., 2020]{daniel2020cross}
Daniel, K., Mota, L., Rottke, S., and Santos, T. (2020).
\newblock The cross-section of risk and returns.
\newblock {\em The Review of Financial Studies}, 33(5):1927--1979.

\bibitem[DeMiguel et~al., 2020]{demiguel2020transaction}
DeMiguel, V., Martin-Utrera, A., Nogales, F.~J., and Uppal, R. (2020).
\newblock A transaction-cost perspective on the multitude of firm characteristics.
\newblock {\em The Review of Financial Studies}, 33(5):2180--2222.

\bibitem[Fama and French, 1993]{fama1993common}
Fama, E.~F. and French, K.~R. (1993).
\newblock Common risk factors in the returns on stocks and bonds.
\newblock {\em Journal of financial economics}, 33(1):3--56.

\bibitem[Fama and French, 2015]{fama2015five}
Fama, E.~F. and French, K.~R. (2015).
\newblock A five-factor asset pricing model.
\newblock {\em Journal of financial economics}, 116(1):1--22.

\bibitem[Feng et~al., 2020]{feng2020taming}
Feng, G., Giglio, S., and Xiu, D. (2020).
\newblock Taming the factor zoo: A test of new factors.
\newblock {\em The Journal of Finance}, 75(3):1327--1370.

\bibitem[Freyberger et~al., 2020]{freyberger2020dissecting}
Freyberger, J., Neuhierl, A., and Weber, M. (2020).
\newblock Dissecting characteristics nonparametrically.
\newblock {\em The Review of Financial Studies}, 33(5):2326--2377.

\bibitem[Grullon et~al., 2012]{grullon2012real}
Grullon, G., Lyandres, E., and Zhdanov, A. (2012).
\newblock Real options, volatility, and stock returns.
\newblock {\em The Journal of Finance}, 67(4):1499--1537.

\bibitem[Gu et~al., 2020]{gu2020empirical}
Gu, S., Kelly, B., and Xiu, D. (2020).
\newblock Empirical asset pricing via machine learning.
\newblock {\em The Review of Financial Studies}, 33(5):2223--2273.

\bibitem[Guha et~al., 1998]{guha1998cure}
Guha, S., Rastogi, R., and Shim, K. (1998).
\newblock Cure: An efficient clustering algorithm for large databases.
\newblock {\em ACM Sigmod record}, 27(2):73--84.

\bibitem[Han et~al., 2019]{han2019firm}
Han, Y., He, A., Rapach, D., and Zhou, G. (2019).
\newblock Firm characteristics and expected stock returns.
\newblock {\em Available at SSRN}, 3185335.

\bibitem[Han et~al., 2024]{han2024cross}
Han, Y., He, A., Rapach, D.~E., and Zhou, G. (2024).
\newblock Cross-sectional expected returns: new fama--macbeth regressions in the era of machine learning.
\newblock {\em Review of Finance}, 28(6):1807--1831.

\bibitem[Harvey et~al., 2016]{harvey2016and}
Harvey, C.~R., Liu, Y., and Zhu, H. (2016).
\newblock … and the cross-section of expected returns.
\newblock {\em The Review of Financial Studies}, 29(1):5--68.

\bibitem[Hou et~al., 2015]{hou2015digesting}
Hou, K., Xue, C., and Zhang, L. (2015).
\newblock Digesting anomalies: An investment approach.
\newblock {\em The Review of Financial Studies}, 28(3):650--705.

\bibitem[Hou et~al., 2020]{hou2020replicating}
Hou, K., Xue, C., and Zhang, L. (2020).
\newblock Replicating anomalies.
\newblock {\em The Review of financial studies}, 33(5):2019--2133.

\bibitem[Karypis et~al., 1999]{karypis1999chameleon}
Karypis, G., Han, E.-H., and Kumar, V. (1999).
\newblock Chameleon: Hierarchical clustering using dynamic modeling.
\newblock {\em computer}, 32(8):68--75.

\bibitem[Kelly et~al., 2023]{kelly2023modeling}
Kelly, B., Palhares, D., and Pruitt, S. (2023).
\newblock Modeling corporate bond returns.
\newblock {\em The Journal of Finance}, 78(4):1967--2008.

\bibitem[Kelly et~al., 2019]{kelly2019characteristics}
Kelly, B.~T., Pruitt, S., and Su, Y. (2019).
\newblock Characteristics are covariances: A unified model of risk and return.
\newblock {\em Journal of Financial Economics}, 134(3):501--524.

\bibitem[Kozak et~al., 2020]{kozak2020shrinking}
Kozak, S., Nagel, S., and Santosh, S. (2020).
\newblock Shrinking the cross-section.
\newblock {\em Journal of Financial Economics}, 135(2):271--292.

\bibitem[McLean and Pontiff, 2016]{mclean2016does}
McLean, R.~D. and Pontiff, J. (2016).
\newblock Does academic research destroy stock return predictability?
\newblock {\em The Journal of Finance}, 71(1):5--32.

\bibitem[Newey and West, 1986]{newey1986simple}
Newey, W.~K. and West, K.~D. (1986).
\newblock A simple, positive semi-definite, heteroskedasticity and autocorrelationconsistent covariance matrix.

\bibitem[Palazzo, 2012]{palazzo2012cash}
Palazzo, B. (2012).
\newblock Cash holdings, risk, and expected returns.
\newblock {\em Journal of Financial Economics}, 104(1):162--185.

\bibitem[Patton and Weller, 2020]{patton2020you}
Patton, A.~J. and Weller, B.~M. (2020).
\newblock What you see is not what you get: The costs of trading market anomalies.
\newblock {\em Journal of Financial Economics}, 137(2):515--549.

\bibitem[Saxena et~al., 2017]{saxena2017review}
Saxena, A., Prasad, M., Gupta, A., Bharill, N., Patel, O.~P., Tiwari, A., Er, M.~J., Ding, W., and Lin, C.-T. (2017).
\newblock A review of clustering techniques and developments.
\newblock {\em Neurocomputing}, 267:664--681.

\bibitem[Sloan, 1996]{sloan1996stock}
Sloan, R.~G. (1996).
\newblock Do stock prices fully reflect information in accruals and cash flows about future earnings?
\newblock {\em Accounting review}, pages 289--315.

\bibitem[Stambaugh and Yuan, 2017]{stambaugh2017mispricing}
Stambaugh, R.~F. and Yuan, Y. (2017).
\newblock Mispricing factors.
\newblock {\em The review of financial studies}, 30(4):1270--1315.

\bibitem[Von~Luxburg, 2007]{von2007tutorial}
Von~Luxburg, U. (2007).
\newblock A tutorial on spectral clustering.
\newblock {\em Statistics and computing}, 17:395--416.

\bibitem[Ziegel, 2003]{ziegel2003elements}
Ziegel, E.~R. (2003).
\newblock The elements of statistical learning.

\end{thebibliography}

\restoregeometry

\section{Figures \& Tables}
\clearpage

\clearpage
\begin{figure}[htbp]
  \centering
  \includegraphics[width=0.85\textwidth]{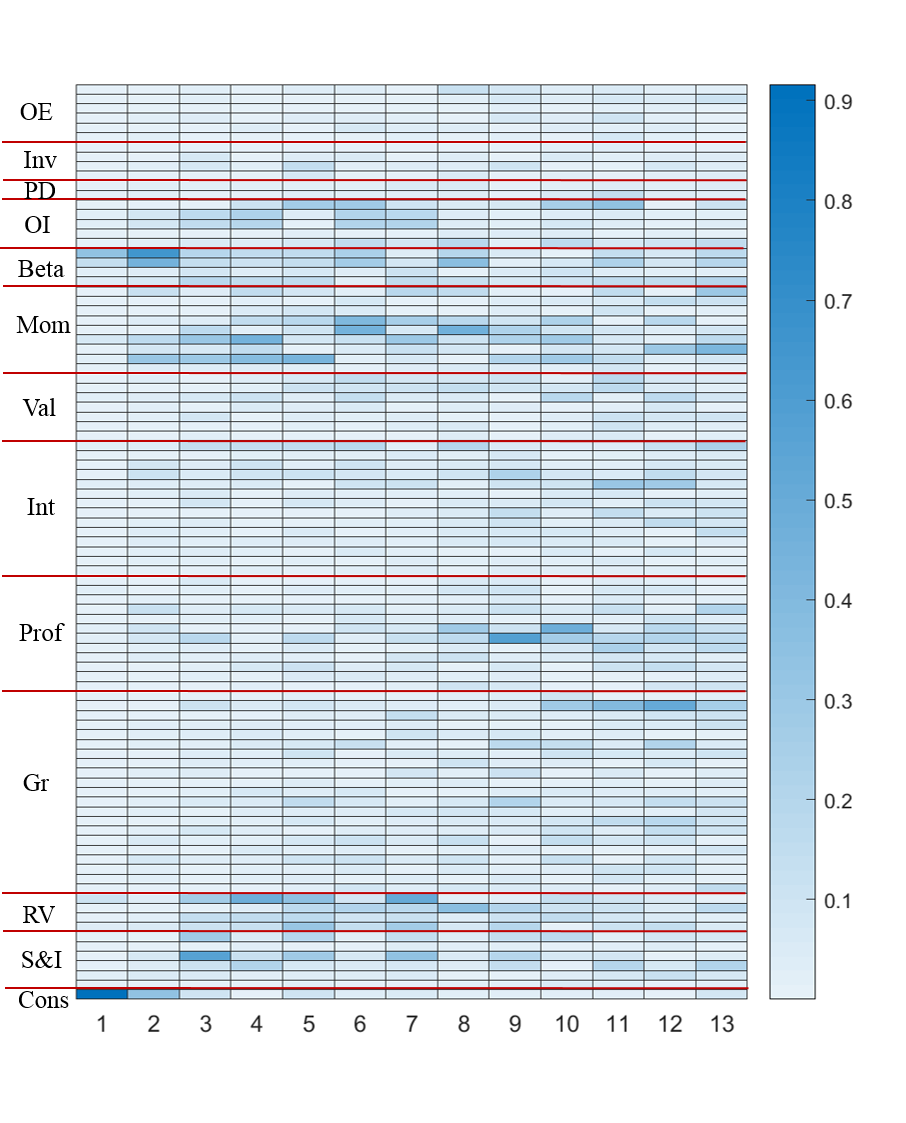}
  \begin{center}{(a) IPCA}\end{center}
  \caption{\textbf{The Gamma matrix \(\hat \Gamma\) for models}}
  \label{Fig: Gamma}
\end{figure}


\clearpage
\begin{figure}[htbp]
  \ContinuedFloat
  \centering
  \includegraphics[width=0.85\textwidth]{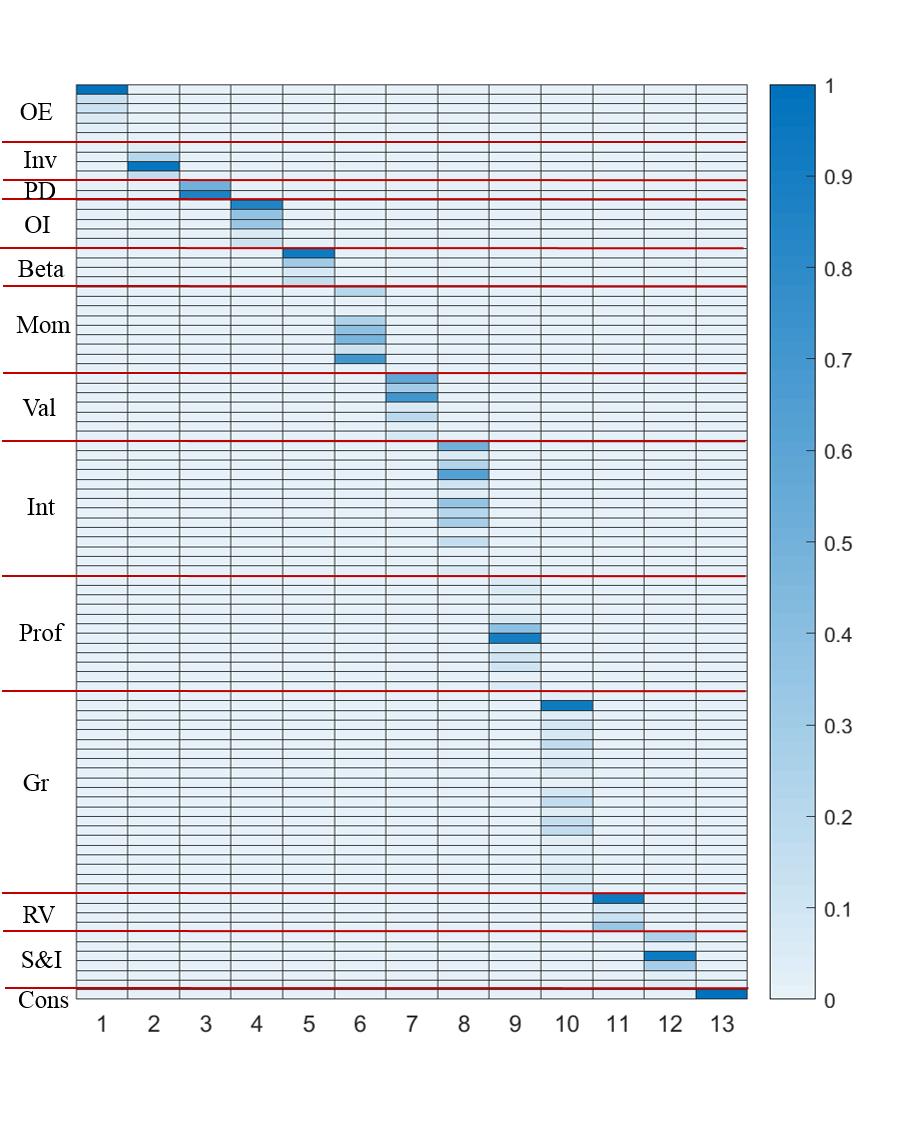}
  \begin{center}(b) C-IPCA\end{center}
  \captionsetup{justification=justified, singlelinecheck=false} 
  \caption{
    \textbf{The Gamma matrix \(\hat{\Gamma}\) for models (cont.).} This figure presents the absolute values of the \(\hat{\Gamma}\) matrix from equation (\ref{Eq: IPCA_beta}) for the IPCA model with 13 factors (Panel (a)) and the DC-IPCA model (Panel (b)), estimated using the full sample. Each column corresponds to a factor in \(f_t\) (or an exposure in \(\beta_t\)) in equation (\ref{eq:ret}), while each row represents a firm characteristic. Red lines partition the rows into 12 clusters based on DC and one constant. Each cell represents the absolute loading of characteristic \(i\) on exposure \(j\), reflecting the characteristic’s importance for that factor. Darker shading indicates higher absolute values and thus greater importance. To enhance comparability across factors, each column is scaled so that the sum of squared elements equals one.
  }
\end{figure}

\clearpage
\clearpage

\clearpage

\begin{table}[htbp]
\footnotesize
  \caption{Performance of interpretable factors}
This table reports the performance of interpretable factors from the IC-IPCA model (Panel A) and the DC-IPCA model (Panel B). Each row corresponds to an individual factor. The first column lists factor IDs. To facilitate comparison, factors are ranked by their out-of-sample Sharpe ratios over the testing period (2000:01–2021:12), except for the market factor, which is reported in the final row. Columns two through five present performance metrics of each factor portfolio: sample mean (Mean), sample standard deviation (S.D.), annualized Sharpe ratio (Sharpe), and maximum drawdown (MDD) of monthly returns. 
Columns six through eight present the alpha of regressing model factors on traditional factor models, including Fama French 3  factor model(FF3, \citeauthor{fama1993common}, \citeyear{fama1993common}), Fama French 5 factor model(FF5, \citeauthor{fama2015five}, \citeyear{fama2015five}) and q-factor factor model(Q4, \citeauthor{hou2015digesting}, \citeyear{hou2015digesting}). *, **, and *** indicate statistical significance of alpha at the 10\%, 5\%, and 1\% confidence levels, respectively, based on Newey-West adjusted standard errors \citep{newey1986simple}.
The final column indicates the associated clusters (see Figure~\ref{Fig: DC and IC} for abbreviations; “Mkt” refers to the zero-correlation market factor (ZC)). Panel C reports correlations between the market factor (MktRf) and the following: (i) the ZC factors from the IC-IPCA and DC-IPCA models, and (ii) the factor from the IPCA model — of the same number of factors as IC-IPCA and DC-IPCA — that exhibits the highest correlation with the market factor (denoted as MF(IPCA7) and MF(IPCA13), respectively). MktRf is computed as the excess return on the value-weighted portfolio. The sample period spans January 2000 to December 2021.
\vskip 0.1in

  \centering
    \begin{tabular}{lcccccccc}
    \toprule
    \multicolumn{1}{c}{MFs} & Mean (\%) & S.D. (\%) & Sharpe & MDD   & alpha-FF3 & alpha-FF5 & alpha-Q4 & Econ. Interp. \\
    \midrule
    \multicolumn{9}{l}{Panel A. IC-IPCA} \\
    \midrule
    \multicolumn{1}{c}{1} & 0.19  & 1.06  & 0.62  & 7.93  & 0.17*** & 0.17*** & 0.14** & Inv \\
    \multicolumn{1}{c}{2} & 0.11  & 0.87  & 0.44  & 10.02  & 0.13** & 0.15** & 0.17*** & Int \\
    \multicolumn{1}{c}{3} & 0.09  & 0.87  & 0.35  & 8.77  & 0.06  & -0.02 & 0.02  & Val \\
    \multicolumn{1}{c}{4} & 0.08  & 1.11  & 0.25  & 10.81  & 0.18*** & 0.11** & 0.15*** & TFs \\
    \multicolumn{1}{c}{5} & 0.07  & 1.04  & 0.23  & 16.96  & 0.1   & -0.01 & 0.09  & Prof \\
    \multicolumn{1}{c}{6} & 0.07  & 1.15  & 0.20  & 18.70  & 0.11  & 0.05  & 0.15** & Mom \\
    \multicolumn{1}{c}{7} & 0.14  & 1.03  & 0.46  & 15.12  & 0.00  & 0.00  & 0.00  & Mkt \\
    \midrule
    \multicolumn{9}{l}{Panel B. DC-IPCA} \\
    \midrule
    \multicolumn{1}{c}{1} & 0.24  & 1.14  & 0.73  & 8.59  & 0.25*** & 0.34*** & 0.29*** & OI \\
    \multicolumn{1}{c}{2} & 0.21  & 1.27  & 0.56  & 8.80  & 0.30*** & 0.17*** & 0.24*** & RV \\
    \multicolumn{1}{c}{3} & 0.10  & 0.98  & 0.37  & 11.67  & 0.09* & 0.06  & 0.09  & OE \\
    \multicolumn{1}{c}{4} & 0.08  & 0.93  & 0.30  & 23.06  & -0.08 & -0.05 & -0.09 & S\&I \\
    \multicolumn{1}{c}{5} & 0.07  & 0.86  & 0.29  & 8.96  & 0.08  & 0.06  & 0.05  & Int \\
    \multicolumn{1}{c}{6} & 0.08  & 0.95  & 0.29  & 9.80  & 0.07  & 0.08  & 0.05  & Gr \\
    \multicolumn{1}{c}{7} & 0.06  & 1.12  & 0.20  & 17.86  & 0.11  & 0.05  & 0.15** & Mom \\
    \multicolumn{1}{c}{8} & 0.05  & 1.00  & 0.17  & 17.15  & -0.10* & 0.01  & -0.12* & Prof \\
    \multicolumn{1}{c}{9} & 0.05  & 1.07  & 0.16  & 10.13  & 0.03  & 0.08  & 0.04  & Turn \\
    \multicolumn{1}{c}{10} & 0.04  & 1.12  & 0.13  & 16.78  & -0.04 & -0.02 & -0.03 & PD \\
    \multicolumn{1}{c}{11} & 0.02  & 0.98  & 0.08  & 11.78  & 0.02  & 0.02  & 0.02  & Inv \\
    \multicolumn{1}{c}{12} & 0.00  & 0.92  & 0.02  & 12.92  & 0.01  & 0.01  & 0.00  & Val \\
    \multicolumn{1}{c}{13} & 0.14  & 1.03  & 0.46  & 15.12  & 0.00  & 0.00  & 0.00  & Mkt \\
    \midrule
    \multicolumn{9}{l}{Panel C. corr. of market factors} \\
    \midrule
    ZC(IC-IPCA)    & 1.00  &       &       &       &       &       &       &  \\
    ZC(DC-IPCA)    & 1.00  & 1.00  &       &       &       &       &       &  \\
    MF(IPCA7) & 0.99  & 0.99  & 1.00  &       &       &       &       &  \\
    MF(IPCA13) & 0.99  & 0.99  & 1.00  & 1.00  &       &       &       &  \\
    MktRf & 1.00  & 1.00  & 0.99  & 0.99  & 1.00  &       &       &  \\
    \bottomrule
    \end{tabular}%
    \label{Table: Perf_factors}%
\end{table}%

\clearpage
\begin{table}[htbp]
\small
  \caption{Mean-variance efficiency of the O-C-IPCA, IPCA and O-IPCA models}
This table reports the out-of-sample Sharpe ratios of tangency portfolios for O-C-IPCA models and two benchmarks: (i) the IPCA model with the same number of factors, and (ii) the O-IPCA models. The \(J\) factor O-C-IPCA model consists of the first \(J-1\) factors from the corresponding C-IPCA specification and the market factor, with factors ordered by their Sharpe ratios in the training sample (1985:01–1999:12). The O-IPCA benchmark for a given O-C-IPCA model includes the first \(J\) factors from an IPCA model. Specifically, the O-IPCA model paired with O-IC-IPCA (Panel A) is based on an IPCA model with 7 factors (O-IPCA7), while the benchmark for O-DC-IPCA (Panel B) is based on an IPCA model with 13 factors (O-IPCA13).  Tangency portfolios are constructed entirely out-of-sample by estimating the mean and covariance matrix of model factors using data up to time \(t\) and computing the portfolio return at \(t+1\). The first column lists the number of factors (\(J\)). Columns two through four report the annualized Sharpe ratios for O-C-IPCA, IPCA, and O-IPCA models, respectively. The final column identifies the cluster associated with the newly added factor in the O-C-IPCA model. (See Figure~\ref{Fig: DC and IC} for cluster abbreviations; “Mkt” refers to the zero-correlation market factor (ZC).) The sample period spans January 2000 through December 2021.

\vskip 0.1in
  \centering
    \begin{tabular}{ccccc}
    \toprule
    \multicolumn{5}{l}{Panel A. IC} \\
    \midrule
    J     & O-IC-IPCA & IPCA  & O-IPCA7 & Clusters \\
    \midrule
    1     & 0.43  & 0.31  & 0.37  & Mkt \\
    2     & 0.50  & 0.59  & 0.44  & Mom \\
    3     & 0.67  & 0.86  & 0.83  & Inv \\
    4     & 0.77  & 0.88  & 0.90  & Int \\
    5     & 0.75  & 0.86  & 1.20  & Prof \\
    6     & 0.77  & 1.19  & 1.19  & Val \\
    7     & 0.88  & 1.23  & 1.23  & TFs \\
    \midrule
    \multicolumn{5}{l}{Panel B. DC} \\
    \midrule
    J     & O-DC-IPCA & IPCA  & O-IPCA13 & Clusters \\
    \midrule
    1     & 0.43  & 0.31  & 0.21  & Mkt \\
    2     & 0.85  & 0.59  & 0.34  & OI \\
    3     & 0.84  & 0.86  & 0.45  & OE \\
    4     & 0.83  & 0.88  & 0.77  & Mom \\
    5     & 1.46  & 0.86  & 0.73  & RV \\
    6     & 1.44  & 1.19  & 1.00  & Inv \\
    7     & 1.49  & 1.23  & 1.31  & Gr \\
    8     & 1.49  & 1.24  & 1.32  & PD \\
    9     & 1.46  & 1.20  & 1.29  & Prof \\
    10    & 1.47  & 1.28  & 1.27  & S\&I \\
    11    & 1.46  & 1.51  & 1.26  & Int \\
    12    & 1.47  & 1.41  & 1.38  & TO \\
    13    & 1.45  & 1.35  & 1.35  & Val \\
    \bottomrule
    \end{tabular}%
  \label{Table: MaxPort_ordered}%
\end{table}%

\clearpage
\begin{table}[htbp]
\small
  \caption{Mean-variance efficiency of the B-C-IPCA and B-IPCA models}
This table reports the out-of-sample Sharpe ratios of tangency portfolios for the top 10 B-C-IPCA models with the highest posterior probabilities (Panel A) and two benchmarks: B-IPCA models following \cite{chib2024winners} (Panel B) and \cite{kelly2019characteristics} (Panel C). Panel A presents results for two variants of B-C-IPCA models: B-DC-IPCA and B-IC-IPCA. Panel B reports B-IPCA models corresponding to IPCA specifications with 7 and 13 factors (denoted as IPCA7 and IPCA13, respectively), enabling comparison with IC-IPCA and DC-IPCA models. For each IPCA specification, Bayesian model selection is applied to identify the 10 most probable models. Panel C summarizes two metrics: the first row shows annualized Sharpe ratios, and the second row indicates the number of factors (\(J\)) in the B-IPCA models. All tangency portfolios are constructed on a purely out-of-sample basis, using all observations up to time \(t\) to estimate the mean and covariance matrix, with portfolio returns evaluated at \(t+1\). The sample period spans January 2000 through December 2021.

\vskip 0.1in
  \centering
    \begin{tabular}{lcccccccccc}
    \toprule
          & Top1     & 2     & 3     & 4     & 5     & 6     & 7     & 8     & 9     & 10 \\
    \midrule
    \multicolumn{11}{l}{Panel A. B-C-IPCA} \\
    \midrule
    DC-IPCA    & 1.44  & 1.46  & 1.42  & 1.44  & 1.42  & 1.09  & 1.45  & 1.45  & 1.41  & 1.43  \\
    IC-IPCA    & 0.21  & 0.50  & 0.77  & 0.79  & 0.67  & 0.39  & 0.63  & 0.48  & 0.33  & 0.62  \\
    \midrule
    \multicolumn{11}{l}{Panel B. B-IPCA(\cite{chib2024winners} method)} \\
    \midrule
    IPCA7 & 0.83  & 0.44  & 0.37  & 0.73  & 0.74  & 0.90  & 0.32  & 0.88  & 1.16  & 0.82  \\
    IPCA13 & 0.54  & 0.82  & 0.96  & 0.85  & 0.52  & 0.34  & 1.10  & 0.80  & 0.64  & 1.06  \\
    \midrule
    \multicolumn{11}{l}{Panel C. KPS-IPCA(\cite{kelly2019characteristics} method)} \\
    \midrule
    IPCA  & 1.35  & 1.41  & 1.28  & 1.51  & 1.20  & 0.86  & 1.24  & 0.31  & 0.59  & 0.86  \\
    J     & 13    & 12    & 10    & 11    & 9     & 3     & 8     & 1     & 2     & 5  \\
    \bottomrule
    \end{tabular}%
  \label{Table: MaxPort_best}%
\end{table}%

\renewcommand{\arraystretch}{0.8}
\begin{table}[htbp]
\small
  \caption{Mean-variance efficiency of the (O)B-PDC-IPCA and (O)B-RC-IPCA models}
This table reports the mean-variance efficiency of the (O)B-PDC-IPCA and (O)B-RC-IPCA models. \emph{PDC} (Pure Data-Driven Clustering) refers to clusters derived exclusively from data, without incorporating economic information from the IC prior. Specifically, to construct PDC, we modify Step 2 of the split-and-merge procedure in Section~\ref{subsec:dc-cluster} by removing the restriction that subclusters remain within intuitive clusters during splitting. \emph{RC} (Random Clustering) denotes clusters generated entirely at random, without using any economic or data-driven inputs. Panel A presents results for the (O)B-PDC-IPCA models, while Panel B reports results for the (O)B-RC-IPCA models. Within each panel, the left set of columns shows results based on ordered model selection, and the right set shows results based on Bayesian model selection. Column \emph{J} denotes the number of model factors; \emph{Sharpe} reports the tangency portfolio Sharpe ratio; and \emph{Rank} gives the B-model’s ranking. For the (O)B-RC-IPCA models, \emph{Mean} and \emph{Median} report the average and median results across 100 random clustering iterations.
  \vskip 0.1in
  \centering
    
    \begin{tabular}{cccccccccrcc}
    \toprule
    \multicolumn{12}{l}{Panel A. Performance of PDC-IPCA} \\
    \midrule
    \multicolumn{4}{c}{O-PDC-IPCA}  &       & \multicolumn{7}{c}{B-PDC-IPCA} \\
\cmidrule{1-4}\cmidrule{6-12}    J     &       & \multicolumn{2}{c}{Sharpe} &       & Rank  &       & \multicolumn{2}{c}{Sharpe} &       & \multicolumn{2}{c}{J} \\
\cmidrule{1-1}\cmidrule{3-4}\cmidrule{6-6}\cmidrule{8-9}\cmidrule{11-12}    1     &       & \multicolumn{2}{c}{0.43 } &       & 1     &       & \multicolumn{2}{c}{0.87 } &       & \multicolumn{2}{c}{5} \\
    2     &       & \multicolumn{2}{c}{0.74 } &       & 2     &       & \multicolumn{2}{c}{0.69 } &       & \multicolumn{2}{c}{4} \\
    3     &       & \multicolumn{2}{c}{0.69 } &       & 3     &       & \multicolumn{2}{c}{0.76 } &       & \multicolumn{2}{c}{3} \\
    4     &       & \multicolumn{2}{c}{0.94 } &       & 4     &       & \multicolumn{2}{c}{0.83 } &       & \multicolumn{2}{c}{6} \\
    5     &       & \multicolumn{2}{c}{0.87 } &       & 5     &       & \multicolumn{2}{c}{1.07 } &       & \multicolumn{2}{c}{6} \\
    6     &       & \multicolumn{2}{c}{0.83 } &       & 6     &       & \multicolumn{2}{c}{0.94 } &       & \multicolumn{2}{c}{4} \\
    7     &       & \multicolumn{2}{c}{0.86 } &       & 7     &       & \multicolumn{2}{c}{0.72 } &       & \multicolumn{2}{c}{4} \\
    8     &       & \multicolumn{2}{c}{0.88 } &       & 8     &       & \multicolumn{2}{c}{0.65 } &       & \multicolumn{2}{c}{5} \\
    9     &       & \multicolumn{2}{c}{0.85 } &       & 9     &       & \multicolumn{2}{c}{0.89 } &       & \multicolumn{2}{c}{6} \\
    10    &       & \multicolumn{2}{c}{0.90 } &       & 10    &       & \multicolumn{2}{c}{0.84 } &       & \multicolumn{2}{c}{6} \\
    11    &       & \multicolumn{2}{c}{1.09 } &       & 11    &       & \multicolumn{2}{c}{0.89 } &       & \multicolumn{2}{c}{6} \\
    \midrule
    \multicolumn{12}{l}{Panel B. average performance of RC-IPCA} \\
    \midrule
    \multicolumn{4}{c}{O-RC-IPCA}  &       & \multicolumn{7}{c}{B-RC-IPCA} \\
\cmidrule{1-4}\cmidrule{6-12}    \multirow{2}[4]{*}{J} &       & \multicolumn{2}{c}{Sharpe} &       & \multirow{2}[4]{*}{Rank} &       & \multicolumn{2}{c}{Sharpe} &       & \multicolumn{2}{c}{J} \\
\cmidrule{3-4}\cmidrule{8-9}\cmidrule{11-12}          &       & Mean  & Median &       &       &       & Mean  & Median &       & Mean  & Median \\
    \midrule
    1     &       & 0.43  & 0.43  &       & 1     &       & 0.54  & 0.54  &       & 4.83  & 4 \\
    2     &       & 0.66  & 0.66  &       & 2     &       & 0.49  & 0.50  &       & 4.61  & 4.5 \\
    3     &       & 0.80  & 0.78  &       & 3     &       & 0.52  & 0.53  &       & 4.71  & 5 \\
    4     &       & 0.90  & 0.89  &       & 4     &       & 0.52  & 0.52  &       & 4.71  & 5 \\
    5     &       & 0.98  & 0.96  &       & 5     &       & 0.53  & 0.53  &       & 5.01  & 5 \\
    6     &       & 1.02  & 1.00  &       & 6     &       & 0.57  & 0.57  &       & 5.16  & 5 \\
    7     &       & 1.05  & 1.02  &       & 7     &       & 0.56  & 0.55  &       & 5.15  & 5 \\
    8     &       & 1.08  & 1.07  &       & 8     &       & 0.55  & 0.56  &       & 5.13  & 5 \\
    9     &       & 1.10  & 1.11  &       & 9     &       & 0.53  & 0.55  &       & 5.24  & 5 \\
    10    &       & 1.12  & 1.11  &       & 10    &       & 0.56  & 0.56  &       & 5.13  & 5 \\
    11    &       & 1.13  & 1.12  &       & 11    &       & 0.52  & 0.53  &       & 5.12  & 5 \\
    12    &       & 1.16  & 1.16  &       & 12    &       & 0.55  & 0.52  &       & 5.28  & 5 \\
    13    &       & 1.19  & 1.21  &       & 13    &       & 0.56  & 0.56  &       & 5.20   & 5 \\
    \bottomrule
    \end{tabular}%
  \label{Table: Mechanism}%
\end{table}%

\begin{landscape}
\begin{table}[htbp]
  \caption{Different training samples: Mean-variance efficiency of the O-DC-IPCA, IPCA and O-IPCA models}
This table reports the out-of-sample Sharpe ratios of tangency portfolios for O-DC-IPCA models and two benchmarks: (i) the IPCA model with the same number of factors, and (ii) the O-IPCA models. The length of the training sample varies from 240 to 420 months. The first column lists the number of factors (\(J\)), while the remaining columns report annualized Sharpe ratios for O-DC-IPCA, IPCA, and O-IPCA models in the testing sample (the full sample excluding the training period). The bottom figure illustrates these results: the horizontal axis represents the number of factors, and the vertical axis represents annualized Sharpe ratios. Bars with different colors indicate different models.
\vspace{1em}

\resizebox{1.3\textwidth}{!}{%
\begin{tabular}{ccccrcccrcccrccc}
    \toprule
          & \multicolumn{3}{c}{Train: 240 months} &       & \multicolumn{3}{c}{Train: 300 months} &       & \multicolumn{3}{c}{Train: 360 months} &       & \multicolumn{3}{c}{Train: 420 months} \\
\cmidrule{2-4}\cmidrule{6-8}\cmidrule{10-12}\cmidrule{14-16}          & O-DC-IPCA & IPCA  & O-IPCA &       & O-DC-IPCA & IPCA  & O-IPCA &       & O-DC-IPCA & IPCA  & O-IPCA &       & O-DC-IPCA & IPCA  & O-IPCA \\
    \midrule
    1     & 0.59  & 0.53  & 0.17  &       & 0.92  & 0.81  & 0.23  &       & 0.84  & 0.78  & 0.36  &       & 1.01  & 1.05  & -0.08  \\
    2     & 0.65  & 0.70  & 0.26  &       & 0.95  & 1.09  & 0.15  &       & 1.14  & 0.87  & 0.17  &       & 0.83  & 0.69  & -0.75  \\
    3     & 0.83  & 0.95  & 0.29  &       & 0.85  & 1.34  & 0.55  &       & 1.47  & 1.14  & 0.46  &       & 1.57  & 1.09  & -0.82  \\
    4     & 0.84  & 1.07  & 0.56  &       & 1.27  & 1.53  & 0.65  &       & 1.79  & 1.42  & 0.65  &       & 2.09  & 1.46  & -0.57  \\
    5     & 0.98  & 1.13  & 0.88  &       & 1.44  & 1.35  & 0.51  &       & 2.21  & 1.38  & 1.10  &       & 2.04  & 1.73  & -0.06  \\
    6     & 0.97  & 1.22  & 0.77  &       & 1.52  & 1.60  & 0.55  &       & 2.26  & 1.30  & 1.82  &       & 2.41  & 1.05  & 0.69  \\
    7     & 1.44  & 1.13  & 0.91  &       & 1.47  & 1.61  & 0.61  &       & 2.27  & 1.14  & 2.02  &       & 1.85  & 0.88  & 0.91  \\
    8     & 1.42  & 0.97  & 0.87  &       & 1.99  & 1.45  & 1.10  &       & 2.21  & 1.14  & 1.92  &       & 1.70  & 0.73  & 0.76  \\
    9     & 1.41  & 1.12  & 1.01  &       & 1.96  & 1.54  & 1.32  &       & 2.21  & 1.42  & 1.86  &       & 1.78  & 1.74  & 0.62  \\
    10    & 1.33  & 1.11  & 1.01  &       & 1.94  & 1.81  & 1.28  &       & 2.17  & 1.81  & 1.83  &       & 1.42  & 1.51  & 0.76  \\
    \bottomrule
    \end{tabular}%
}

\vspace{1em} 

\includegraphics[width=1.3\textwidth]{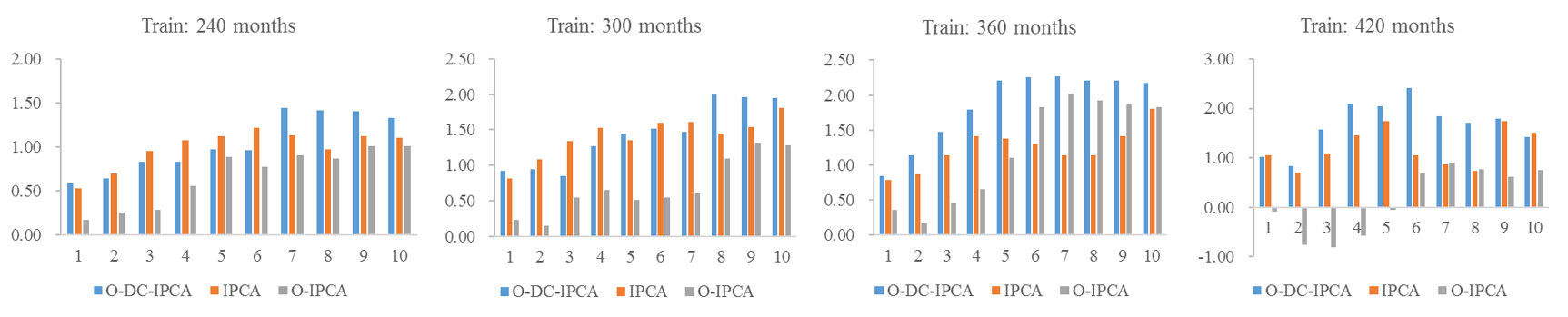} 

  \label{Table: O_C_IPCA_Train}%
\end{table}%
\end{landscape}

\begin{landscape}
\begin{table}[htbp]
\small
  \caption{Different training samples: Mean-variance efficiency of the B-C-IPCA and B-IPCA models}
This table reports the out-of-sample Sharpe ratios of tangency portfolios for the top 10 B-C-IPCA models with the highest posterior probabilities, estimated using different training sample lengths. Panels A through D correspond to training periods of 240, 300, 360, and 420 months, respectively. Within each panel, the first column indicates the model rank, the second and third columns report the annualized Sharpe ratios in the testing sample, and the fourth and fifth columns show the number of factors included in each model.
\vskip 0.1in
    \begin{tabular}{ccccccrcccccc}
\cmidrule{1-6}\cmidrule{8-13}    \multicolumn{6}{l}{Panel A. Training Sample: 240 months} &       & \multicolumn{6}{l}{Panel B. Training Sample: 300 months} \\
\cmidrule{1-6}\cmidrule{8-13}    \multirow{2}[4]{*}{Rank} & \multicolumn{2}{c}{Sharpe} &       & \multicolumn{2}{c}{\#Factors} &       & \multirow{2}[4]{*}{Rank} & \multicolumn{2}{c}{Sharpe} &       & \multicolumn{2}{c}{\#Factors} \\
\cmidrule{2-3}\cmidrule{5-6}\cmidrule{9-10}\cmidrule{12-13}          & B-C-IPCA & B-IPCA &       & B-C-IPCA & B-IPCA &       &       & B-C-IPCA & B-IPCA &       & B-C-IPCA & B-IPCA \\
\cmidrule{1-6}\cmidrule{8-13}    1     & 1.13  & 1.05  &       & 6     & 8     &       & 1     & 1.80  & 1.23  &       & 5     & 8 \\
    2     & 1.37  & 1.01  &       & 7     & 9     &       & 2     & 1.96  & 1.12  &       & 6     & 8 \\
    3     & 1.13  & 1.03  &       & 7     & 9     &       & 3     & 1.89  & 1.17  &       & 6     & 7 \\
    4     & 1.20  & 0.91  &       & 5     & 7     &       & 4     & 2.02  & 1.28  &       & 7     & 8 \\
    5     & 1.12  & 1.08  &       & 7     & 9     &       & 5     & 1.74  & 1.32  &       & 6     & 9 \\
    6     & 1.16  & 1.06  &       & 7     & 9     &       & 6     & 1.76  & 1.24  &       & 6     & 9 \\
    7     & 1.12  & 1.04  &       & 7     & 9     &       & 7     & 1.82  & 1.18  &       & 6     & 9 \\
    8     & 1.13  & 0.99  &       & 7     & 10    &       & 8     & 1.80  & 0.77  &       & 6     & 7 \\
    9     & 1.10  & 1.01  &       & 7     & 10    &       & 9     & 1.98  & 1.53  &       & 7     & 10 \\
    10    & 1.45  & 0.87  &       & 6     & 8     &       & 10    & 1.92  & 1.28  &       & 7     & 10 \\
\cmidrule{1-6}\cmidrule{8-13}          &       &       &       &       &       &       &       &       &       &       &       &  \\
          &       &       &       &       &       &       &       &       &       &       &       &  \\
\cmidrule{1-6}\cmidrule{8-13}    \multicolumn{6}{l}{Panel C. Training Sample: 360 months} &       & \multicolumn{6}{l}{Panel D. Training Sample: 420 months} \\
\cmidrule{1-6}\cmidrule{8-13}    \multirow{2}[4]{*}{Rank} & \multicolumn{2}{c}{Sharpe} &       & \multicolumn{2}{c}{\#Factors} &       & \multirow{2}[4]{*}{Rank} & \multicolumn{2}{c}{Sharpe} &       & \multicolumn{2}{c}{\#Factors} \\
\cmidrule{2-3}\cmidrule{5-6}\cmidrule{9-10}\cmidrule{12-13}          & B-C-IPCA & B-IPCA &       & B-C-IPCA & B-IPCA &       &       & B-C-IPCA & B-IPCA &       & B-C-IPCA & B-IPCA \\
\cmidrule{1-6}\cmidrule{8-13}    1     & 2.11  & 1.31  &       & 5     & 10    &       & 1     & 1.03  & 0.95  &       & 8     & 9 \\
    2     & 2.16  & 1.30  &       & 6     & 9     &       & 2     & 1.34  & 0.76  &       & 7     & 10 \\
    3     & 2.13  & 1.32  &       & 6     & 11    &       & 3     & 1.85  & 1.10  &       & 7     & 8 \\
    4     & 2.16  & 1.30  &       & 7     & 10    &       & 4     & 1.38  & 0.76  &       & 9     & 8 \\
    5     & 2.09  & 1.15  &       & 6     & 9     &       & 5     & 1.50  & 0.91  &       & 8     & 7 \\
    6     & 2.14  & 1.16  &       & 6     & 9     &       & 6     & 1.20  & 0.62  &       & 7     & 9 \\
    7     & 2.09  & 1.13  &       & 6     & 8     &       & 7     & 1.70  & 0.93  &       & 8     & 9 \\
    8     & 2.10  & 1.27  &       & 6     & 9     &       & 8     & 1.35  & 0.47  &       & 9     & 9 \\
    9     & 1.90  & 1.33  &       & 4     & 9     &       & 9     & 1.52  & 0.63  &       & 6     & 8 \\
    10    & 1.87  & 1.16  &       & 4     & 8     &       & 10    & 1.67  & 0.78  &       & 8     & 8 \\
\cmidrule{1-6}\cmidrule{8-13}   
 \end{tabular}%
  \label{Table: O_B_IPCA_Train}%
\end{table}%
\end{landscape}

\begin{table}[htbp]
\small
\caption{Different Training Samples: Performance of DC-IPCA Model Factors}
This table shows the performance of interpretable factors from the DC-IPCA models across different training sample lengths. The first column lists the IDs of model factors (MFs). To facilitate interpretation, factors are ranked by their Sharpe ratios in the testing sample, except for the market factor, which is reported in the final row. The second column presents the annualized Sharpe ratios (Sharpe), while the third column indicates the corresponding clusters. An asterisk (*) denotes that the factor is included in the top B-C-IPCA model selected using the method by \cite{chib2024winners}. Cluster abbreviations are as follows: Beta = beta; FS = financial stability; FunMom = fundamental momentum; FunVol = fundamental volatility; GA = growth of assets; GE = growth of equity; Gr = growth; Illiq = illiquidity; Inv = investment; Mom = momentum; OE = operating efficiency; OI = operating illiquidity; Prof = profitability; PD = price delay; RV = return volatility; RetMom = return momentum; RD = R\&D; Size = size; Turn = turnover; Val = value.
\bigskip
  \vskip 0.1in
        \begin{tabular}{cclrrrrrrrcl}
    \toprule
    \multirow{2}[4]{*}{MFs} & \multicolumn{2}{c}{Train: 240 months} &       & \multicolumn{2}{c}{Train: 300 months} &       & \multicolumn{2}{c}{Train: 240 months} &       & \multicolumn{2}{c}{Train: 360 months} \\
\cmidrule{2-3}\cmidrule{5-6}\cmidrule{8-9}\cmidrule{11-12}          & \multicolumn{1}{l}{Sharpe} & Clusters &       & \multicolumn{1}{l}{Sharpe} & \multicolumn{1}{l}{Clusters} &       & \multicolumn{1}{l}{Sharpe} & \multicolumn{1}{l}{Clusters} &       & \multicolumn{1}{l}{Sharpe} & Clusters \\
    \midrule
    1     & 0.77  & OI    &       & \multicolumn{1}{c}{0.82 } & \multicolumn{1}{l}{OI} &       & \multicolumn{1}{c}{1.18 } & \multicolumn{1}{l}{OI*} &       & 1.45  & OE \\
    2     & 0.55  & RV*   &       & \multicolumn{1}{c}{0.61 } & \multicolumn{1}{l}{Mom} &       & \multicolumn{1}{c}{0.69 } & \multicolumn{1}{l}{FundMom*} &       & 1.19  & Beta* \\
    3     & 0.48  & Size  &       & \multicolumn{1}{c}{0.56 } & \multicolumn{1}{l}{FunVol} &       & \multicolumn{1}{c}{0.57 } & \multicolumn{1}{l}{Inv} &       & 1.16  & Prof* \\
    4     & 0.44  & GE*   &       & \multicolumn{1}{c}{0.55 } & \multicolumn{1}{l}{Val} &       & \multicolumn{1}{c}{0.52 } & \multicolumn{1}{l}{RV*} &       & 1.06  & Size* \\
    5     & 0.41  & RetMom* &       & \multicolumn{1}{c}{0.51 } & \multicolumn{1}{l}{FunMom*} &       & \multicolumn{1}{c}{0.50 } & \multicolumn{1}{l}{RetMom*} &       & 0.79  & Mom* \\
    6     & 0.33  & OE    &       & \multicolumn{1}{c}{0.46 } & \multicolumn{1}{l}{RD*} &       & \multicolumn{1}{c}{0.49 } & \multicolumn{1}{l}{Size} &       & 0.71  & FS* \\
    7     & 0.28  & GA    &       & \multicolumn{1}{c}{0.30 } & \multicolumn{1}{l}{Size} &       & \multicolumn{1}{c}{0.45 } & \multicolumn{1}{l}{RD} &       & 0.66  & Gr* \\
    8     & 0.24  & FunMom* &       & \multicolumn{1}{c}{0.15 } & \multicolumn{1}{l}{Prof} &       & \multicolumn{1}{c}{0.34 } & \multicolumn{1}{l}{Turn} &       & 0.49  & Inv \\
    9     & 0.17  & Beta* &       & \multicolumn{1}{c}{0.06 } & \multicolumn{1}{l}{Illiq*} &       & \multicolumn{1}{c}{0.19 } & \multicolumn{1}{l}{PD} &       & 0.34  & Val* \\
    10    & 0.03  & SI    &       & \multicolumn{1}{c}{0.01 } & \multicolumn{1}{l}{PD} &       & \multicolumn{1}{c}{0.18 } & \multicolumn{1}{l}{FS} &       & 0.17  & FunMom \\
    11    & 0.03  & ChgOI &       & \multicolumn{1}{c}{0.93 } & \multicolumn{1}{l}{Market*} &       & \multicolumn{1}{c}{0.85 } & \multicolumn{1}{l}{Market*} &       & 0.03  & RD \\
    12    & 0.02  & Inv   &       &       &       &       &       &       &       & 0.01  & SI \\
    13    & 0.61  & Market* &       &       &       &       &       &       &       & 1.04  & Market* \\
    \bottomrule
    \end{tabular}%
  \label{Table: MF_Train}%
\end{table}%

\appendix
\section*{Appendices}
\addcontentsline{toc}{section}{Appendices}
\counterwithin{table}{section}
\renewcommand{\thetable}{\thesection\arabic{table}}
\counterwithin{figure}{section}
\renewcommand{\thefigure}{\thesection\arabic{figure}}

\section{Characteristics in IC and DC}\label{app: reasons_algorithm}
\begin{table}[htbp]
  \caption{Characteristics in IC and DC}
This table presents the characteristics of each IC and DC cluster. The first column outlines the economic meanings of the six clusters in the IC, while the second column describes the economic meanings of the thirteen clusters in the DC (see Figure~\ref{Fig: DC and IC} for cluster abbreviations). The final column lists the characteristics associated with each IC and DC cluster (see Table A.6 of \cite{gu2020empirical} for characteristic abbreviations). We use the same set of characteristics as in \cite{gu2020empirical}, adopting their abbreviations accordingly.
\bigskip

        \begin{tabular}{ccp{28.065em}}
    \toprule
    IC    & DC    & \multicolumn{1}{c}{characteristics} \\
    \midrule
    Mom   & Mom   & \multicolumn{1}{l}{chmom, chtx, ear, indmom, mom12m, mom1m, mom36m, mom6m, nincr} \\
    \midrule
    \multirow{4}[8]{*}{TFs} & RV    & \multicolumn{1}{l}{baspread, idiovol, maxret, retvol} \\
\cmidrule{2-3}          & S\&I  & \multicolumn{1}{l}{dolvol, ill, mve, mve\_ia, std\_dolvol, zerotrade} \\
\cmidrule{2-3}          & PD    & \multicolumn{1}{l}{aeavol, pricedelay} \\
\cmidrule{2-3}          & TO    & \multicolumn{1}{l}{beta, betasq, std\_turn, turn} \\
    \midrule
    \multirow{2}[4]{*}{Inv} & Inv   & \multicolumn{1}{l}{cinvest, depr, pchcapx\_ia, pchdepr} \\
\cmidrule{2-3}          & \multirow{2}[4]{*}{Gr} & \multicolumn{1}{l}{agr, chcsho, chinv, egr, grcapx, grltnoa, invest, lgr} \\
\cmidrule{1-1}\cmidrule{3-3}    \multirow{2}[4]{*}{Prof} &       & \multicolumn{1}{l}{chatoia, chpmia, ps, rsup, tb} \\
\cmidrule{2-3}          & \multirow{2}[4]{*}{Prof} & \multicolumn{1}{l}{cashpr, gma, lev, ms, operprof, roaq, roeq, roic} \\
\cmidrule{1-1}\cmidrule{3-3}    \multirow{2}[4]{*}{Val} &       & \multicolumn{1}{l}{bm, bm\_ia, cashdebt, sp} \\
\cmidrule{2-3}          & \multirow{2}[4]{*}{val} & \multicolumn{1}{l}{cfp, cfp\_ia, dy, ep} \\
\cmidrule{1-1}\cmidrule{3-3}    \multirow{5}[10]{*}{Int} &       & \multicolumn{1}{l}{absacc, divi, divo} \\
\cmidrule{2-3}          & Int   & age, convind, herf, orgcap, rd\_mve, rd\_sale, realestate, roavol, salerec, secured, securedind, sin, stdacc, stdcf \\
\cmidrule{2-3}          & OI    & \multicolumn{1}{l}{cash, currat, quick, salecash, tang} \\
\cmidrule{2-3}          & OE    & \multicolumn{1}{l}{acc, pchsale\_pchinvt, pchsale\_pchrect, pchsaleinv, pctacc, saleinv} \\
\cmidrule{2-3}          & Gr    & chempia, hire, pchcurrat, pchgm\_pchsale, pchquick, pchsale\_pchxsga, rd, sgr \\
    \bottomrule
    \end{tabular}%
  \label{tab:addlabel}%
\end{table}%

\clearpage
\section{{Model Performance with Equal Weights}}\label{app: EW}
The empirical results suggest that the optimal clustering hyper-parameters are \{m=19, knn=15, K=7\}, with the resulting clustering denoted as \(DC^{EW}\). This clustering yields seven distinct clusters, labeled as: Return momentum(RetMom), Size\&illiquidity(S\&I), Return volatility(RV) Turnover(TO), Operating illiquitidy(OI), Intangibles(Int), and Short-run reversal(SR)\footnote{The SR cluster includes firm characteristics originally clustered in the Value (Val), Profitability (Prof), and Investment (Inv) clusters in the IC. However, among these characteristics, the estimated loading on characteristic 1-month momentum (mom1m) in \(\Gamma\) in equation~\eqref{Eq: IPCA_beta} is substantially higher than for the other characteristics. Since \(\Gamma\) reflects the sensitivity of firm characteristics to risk exposures (\(\beta\) in equation~\eqref{Eq: IPCA_beta}), this suggests that the risk exposure in this cluster is primarily driven by short-run reversal. Accordingly, we label the cluster as SR.}. A summary of the firm characteristics clustered in each cluster under both IC and \(DC^{EW}\) is provided in Table \ref{Table: DC_IC_EW}.

Consistent with Section \ref{subsec: MaxPort_IC-IPCA} in the main text, we construct the \(DC^{EW}\)-IPCA model based on the seven clusters of firm characteristics. We then compare the performance of \(DC^{EW}\)-IPCA, IC-IPCA, and the standard IPCA by selecting a subset of factors using the ordered model selection and Bayesian model selection approaches, as described in Subsections \ref{subsub:ordered_factor_selection} and \ref{subsec:bayesian_model_selection}. The comparison results are reported in Tables \ref{Table: MaxPort_ordered_EW} and \ref{Table: MaxPort_best_EW}.

Similar to the value-weighted results presented in the main text, these tables demonstrate that even under an equal-weighted specification, the DC-IPCA model performs at least as well as the standard IPCA model in most cases. This indicates that our main findings are robust to the choice of weighting scheme. Overall, the use of equal weights still allows our approach to improve model interpretability without materially sacrificing performance.

\begin{table}[htbp]
  \caption{Characteristics in IC and \(DC^{EW}\)}
This table presents the characteristics in each IC and \(DC^{EW}\) cluster. The first column outlines the economic meanings of the six clusters in the IC, while the second column describes the economic meanings of the seven clusters in the \(DC^{EW}\). Cluster abbreviations are as follows: Return momentum(RetMom), Size\&illiquidity(S\&I), Return volatility(RV) Turnover(TO), Operating illiquitidy(OI), Intangibles(Int), Short-run reversal(SR), Trading frictions(TFs), Value(Val), Profitability(Prof), Investment(Inv), Momentum(Mom). The final column lists the characteristics associated with each IC and \(DC^{EW}\) cluster (see Table A.6 of \cite{gu2020empirical} for characteristic abbreviations). We use the same set of characteristics as in \cite{gu2020empirical}, adopting their abbreviations accordingly.
\bigskip

        \begin{tabular}{ccp{28.065em}}
    \toprule
    IC    & \(DC^{EW}\)    & \multicolumn{1}{c}{characteristics} \\
    \midrule
    \multirow{3}[6]{*}{TFs} & RV    & \multicolumn{1}{l}{baspread, idiovol, maxret, retvol} \\
\cmidrule{2-3}          & S\&I  & \multicolumn{1}{l}{dolvol, ill, mve, mve\_ia, std\_dolvol, zerotrade} \\
\cmidrule{2-3}          & TO    & \multicolumn{1}{l}{aeavol, beta, betasq, pricedelay, std\_turn, turn} \\
    \midrule
    \multirow{3}[6]{*}{Int} & OI    & \multicolumn{1}{l}{cash, currat, quick, salecash, tang} \\
\cmidrule{2-3}          & Int   & age, convind, divo, herf, orgcap, rd, rd\_mve, rd\_sale, realestate, saleinv, salerec, secured, securedind, sin, stdacc, stdcf \\
\cmidrule{2-3}          & \multirow{5}[10]{*}{SR} & absacc, acc, chempia, divi, hire, pchcurrat, pchgm\_pchsale, pchquick, pchsale\_pchinvt, pchsale\_pchrect, pchsale\_pchxsga, pchsaleinv, pctacc, roavol, sgr \\
\cmidrule{1-1}\cmidrule{3-3}    Val   &       & \multicolumn{1}{l}{bm, bm\_ia, cashdebt, cfp, cfp\_ia, dy, ep, sp} \\
\cmidrule{1-1}\cmidrule{3-3}    Prof  &       & \multicolumn{1}{l}{cashpr, chatoia, chpmia, gma, lev, ms, operprof, ps, roaq, roeq, roic, rsup, tb} \\
\cmidrule{1-1}\cmidrule{3-3}    Inv   &       & \multicolumn{1}{l}{agr, chcsho, chinv, cinvest, depr, egr, grcapx, grltnoa, invest, lgr, pchcapx\_ia, pchdepr} \\
\cmidrule{1-1}\cmidrule{3-3}    \multirow{2}[4]{*}{Mom} &       & \multicolumn{1}{l}{chtx, ear, mom1m, mom36m, nincr} \\
\cmidrule{2-3}          & RetMom & chmom, indmom, mom12m, mom6m \\
    \bottomrule
    \end{tabular}%
  \label{Table: DC_IC_EW}%
\end{table}%

\begin{table}[htbp]
\small
  \caption{Mean-variance efficiency of the O-C-IPCA, IPCA and O-IPCA models under an equal-weighted scheme}
This table reports the out-of-sample Sharpe ratios of tangency portfolios using an equal-weighted scheme for O-C-IPCA models and two benchmarks: (i) the IPCA model with the same number of factors, and (ii) the O-IPCA models. The \(J\) factor O-C-IPCA model consists of the first \(J-1\) factors from the corresponding C-IPCA specification and the market factor, with factors ordered by their Sharpe ratios in the training sample (1985:01–1999:12). The O-IPCA benchmark for a given O-C-IPCA model includes the first \(J\) factors from an IPCA model. Specifically, the O-IPCA model paired with O-IC-IPCA (Panel A) is based on an IPCA model with 7 factors (O-IPCA7), while the benchmark for O-DC-IPCA (Panel B) is based on an IPCA model with 8 factors (O-IPCA8).  Tangency portfolios are constructed entirely out-of-sample by estimating the mean and covariance matrix of model factors using data up to time \(t\) and computing the portfolio return at \(t+1\). The first column lists the number of factors (\(J\)). Columns two through four report the annualized Sharpe ratios for O-C-IPCA, IPCA, and O-IPCA models, respectively. The final column identifies the cluster associated with the newly added factor in the O-C-IPCA model. (See Table~\ref{Table: DC_IC_EW} for cluster abbreviations; “Mkt” refers to the zero-cost market factor.) The sample period spans January 2000 through December 2021.

\vskip 0.1in
  \centering
    \begin{tabular}{ccccc}
    \toprule
    \multicolumn{5}{l}{Panel A. IC} \\
    \midrule
    J     & O-IC-IPCA & IPCA  & O-IPCA7 & Econ. Interp. \\
    \midrule
    1     & 0.37  & 0.30  & 0.46  & Mkt \\
    2     & 0.51  & 1.05  & 0.70  & Inv \\
    3     & 0.75  & 1.23  & 0.66  & Prof \\
    4     & 1.18  & 1.25  & 0.65  & Mom \\
    5     & 1.25  & 1.67  & 0.70  & Val \\
    6     & 1.55  & 1.56  & 0.72  & Int \\
    7     & 1.50  & 1.57  & 1.57  & TFs \\
    \midrule
    \multicolumn{5}{l}{Panel B. \(DC^{EW}\)} \\
    \midrule
    J     & O-\(DC^{EW}\)-IPCA & IPCA  & O-IPCA8 & Econ. Interp. \\
    \midrule
    1     & 0.19  & 0.30  & 0.62  & Mkt \\
    2     & 0.85  & 1.05  & 0.54  & RetMom \\
    3     & 0.75  & 1.23  & 0.72  & Int \\
    4     & 1.54  & 1.25  & 0.77  & SR \\
    5     & 1.84  & 1.67  & 0.81  & RV \\
    6     & 1.87  & 1.56  & 1.79  & OI \\
    7     & 1.86  & 1.57  & 1.79  & TO \\
    8     & 1.90  & 1.78  & 1.78  & S\&I \\
    \bottomrule
    \end{tabular}%
  \label{Table: MaxPort_ordered_EW}%
\end{table}%

\begin{table}[htbp]
\small
  \caption{Mean-variance efficiency of the B-C-IPCA and B-IPCA models under an equal-weighted scheme}
This table reports the out-of-sample Sharpe ratios of tangency portfolios using an equal-weighted scheme for the top 10 B-C-IPCA models with the highest posterior probabilities (Panel A) and two benchmarks: B-IPCA models following \cite{chib2024winners} (Panel B) and \cite{kelly2019characteristics} (Panel C). Panel A presents results for two variants of B-C-IPCA models: B-\(DC^{EW}\)-IPCA and B-IC-IPCA. Panel B reports B-IPCA models corresponding to IPCA specifications with 7 and 8 factors (denoted as IPCA7 and IPCA8, respectively), enabling comparison with IC-IPCA and \(DC^{EW}\)-IPCA models. For each IPCA specification, Bayesian model selection is applied to identify the 10 most probable models. Panel C summarizes two metrics: the first row shows annualized Sharpe ratios, and the second row indicates the number of factors (\(J\)) in the B-IPCA models. All tangency portfolios are constructed on a purely out-of-sample basis, using all observations up to time \(t\) to estimate the mean and covariance matrix, with portfolio returns evaluated at \(t+1\). The sample period spans January 2000 through December 2021.

\vskip 0.1in
  \centering
    \begin{tabular}{lcccccccccc}
    \toprule
          & Top1     & 2     & 3     & 4     & 5     & 6     & 7     & 8     & 9     & 10 \\
    \midrule
    \multicolumn{11}{l}{Panel A. B-C-IPCA} \\
    \midrule
    \(DC^{EW}\)-IPCA    & 1.76  & 1.78  & 1.76  & 1.73  & 1.84  & 1.78  & 1.87  & 1.86  & 1.79  & 1.82  \\
    IC-IPCA    & 0.85  & 1.10  & 0.88  & 1.12  & 0.93  & 1.09  & 0.84  & 1.21  & 0.33  & 0.62  \\
    \midrule
    \multicolumn{11}{l}{Panel B. B-IPCA(\cite{chib2024winners} method))} \\
    \midrule
    IPCA7 & 0.68  & 0.65  & 0.72  & 0.70  & 1.57  & 1.51  & 1.57  & 1.51  & 0.00  & 0.00  \\
    IPCA8 & 0.81  & 0.79  & 1.79  & 0.77  & 0.81  & 1.78  & 1.81  & 0.79  & 0.00  & 0.00  \\
    \midrule
    \multicolumn{11}{l}{Panel C. KPS-IPCA(\cite{kelly2019characteristics} method)} \\
    \midrule
    IPCA  & 1.56  & 1.67  & 1.57  & 1.78  & 1.25  & 1.86  & 1.93  & 1.23  & 1.05  & 0.30  \\
    J     & 6     & 5     & 7     & 8     & 4     & 10    & 9     & 3     & 2     & 1  \\
    \bottomrule
    \end{tabular}%
  \label{Table: MaxPort_best_EW}%
\end{table}%

\clearpage
\section{Rationality of Chameleon}
{Section \ref{subsubsec: DC_adv} argues that Chameleon is robust to data noise because of the merging rule in equation~\eqref{Eq:RIS}. This section details this argument.}

We first recall that firm characteristics within the same cluster are assumed to be noisy measurements of a common latent risk exposure. In the absence of measurement errors, each characteristic would perfectly measure the latent exposure, resulting in a correlation of 1 between characteristics within the same cluster, thus maximizing the similarity \(s_{ij}\) in equation~\eqref{Eq:s_D}. However, as measurement errors increase, the observed characteristics deviate more from the latent exposure, reducing their correlation and lowering their similarity.

\begin{figure}[htbp]
\centering
\includegraphics[width=12cm]{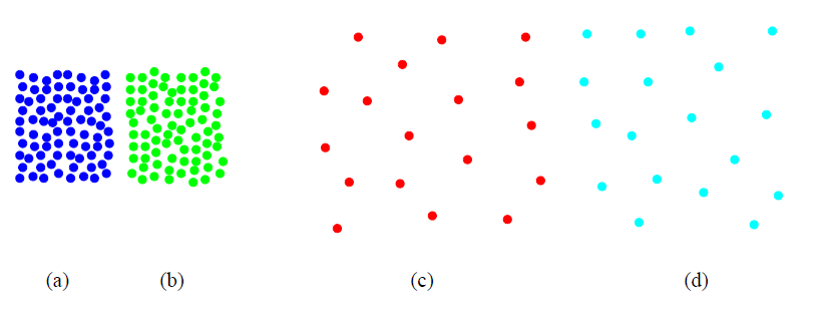}
\caption{Clustering with Noise}\vskip 0.1in \label{fig:cluster_noise}
\end{figure}

The inclusion of intra-cluster similarity, \(INTRA(C_i)\), in the merging rule (\ref{Eq:RIS}) allows the algorithm to detect clusters with high measurement errors and prevents them from being erroneously split into multiple sub-clusters. For illustration, consider four clusters of firm characteristics. The first two clusters, shown in panels (a) and (b) of Figure~\ref{fig:cluster_noise}(copied from \cite{karypis1999chameleon}), exhibit low measurement errors and thus high intra-cluster similarity (depicted by small intra-cluster distances). In contrast, the other two clusters, shown in panels (c) and (d), have high measurement errors and low intra-cluster similarity (depicted by large intra-cluster distances).

The task is to determine whether to merge clusters (a) and (b), or (c) and (d). Intuitively, merging (c) and (d) seems more reasonable, as their intra-cluster distance appears comparable to their inter-cluster distance, indicating high homogeneity between them. If the algorithm only considers inter-cluster similarity (or distance), it would incorrectly merge (a) and (b) due to their smaller inter-cluster distance. However, by incorporating both intra-cluster and inter-cluster distances, as in the Chameleon algorithm, the algorithm correctly merges (c) and (d), aligning with intuition. Specifically, Chameleon avoids merging (a) and (b) because such a merger would increase the intra-cluster distance and reduce the Relative Inter-Cluster Similarity (RIS) in equation~\eqref{Eq:RIS}. Thus, the inclusion of intra-cluster similarity ensures that high-noise sub-clusters, such as (c) and (d), are merged into a single high-noise cluster, preventing the erroneous splitting of a high-noise cluster into multiple sub-clusters.

\end{document}